\documentclass[twocolumn]{aastex631}
\shorttitle{The $\gamma$-Ray Radiation of High-redshift Blazars}
\shortauthors{Wu et al.}
\graphicspath{{./}{figures/}}
\begin{document}
\title{Study of the $\gamma$-Ray Radiation Properties of High-redshift Blazars at $z>2.5$}
\email{bzhdai@ynu.edu.cn}

\author[0000-0002-6292-057X]{Fan Wu}
\affiliation{School of Physics and Astronomy, Key Laboratory of Astroparticle Physics of Yunnan Province, Yunnan University, Kunming 650091, P. R. China\\}
\author[0000-0001-7908-4996]{Benzhong Dai}
\affiliation{School of Physics and Astronomy, Key Laboratory of Astroparticle Physics of Yunnan Province, Yunnan University, Kunming 650091, P. R. China\\}

\begin{abstract}

We study a sample of 30 high-redshift blazars ($z>2.5$) by means of spectra and the radiation mechanism with Fermi Large Area Telescope $\gamma$-ray observations spanning 15 years. Three models---the power law, power law with an exponential cutoff, and log-parabola---are employed to analyze the spectral properties, and most sources exhibit significant curvature. The high-redshift blazars exhibit higher $\gamma$-ray luminosities and softer spectral indices compared with their low-redshift counterparts, where B3~1343+451 has the highest integrated flux, $\rm 1.13 \times 10^{-7} \mathrm{\ ph \ cm^{-2} s^{-1}}$. We use a standard one-zone leptonic emission model to reproduce the spectral energy distributions of 23 sources with multiwavelength observations. We find that modeling with infrared seed photons is systematically better than with broad-line region (BLR) photons based on a $\chi^2$ test, which suggests that the $\gamma$-ray-emitting regions are most likely located outside the BLR. The fit results show that high-redshift blazars exhibit higher energy density, jet power, kinetic power, and accretion disk luminosities, along with lower synchrotron and inverse Compton (IC) peak frequencies, relative to their lower-redshift counterparts. We find that blazars with higher accretion disk luminosities tend to have lower IC peak frequencies, leading to more efficient cooling of high-energy electrons. The positive correlation between jet power and accretion disk luminosity further supports the possibility of an accretion--jet connection in these high-redshift sources.
\end{abstract}

\section{Introduction\label{sec1}}

Blazars are a subclass of active galactic nuclei (AGNs). They have a relativistic jet oriented at a small angle with respect to the line of the sight \citep{1995PASP..107..803U}. Based on the absence or presence of broad emission lines, blazars are divided into two subclasses: BL~Lacertae objects and flat-spectrum radio quasars (FSRQs) \citep{1991ApJ...374..431S}. BL~Lacertae objects exhibit weak or undetected emission lines, while FSRQs display prominent emission lines.

Blazar spectral energy distributions (SEDs) span from the radio to $\gamma$-ray bands, exhibiting a double-peaked structure: a low-energy (LE) peak between infrared (IR) and X-rays, and a high-energy (HE) peak at MeV--TeV $\gamma$-rays. The LE component is generally attributed to synchrotron radiation from highly relativistic electrons in the jet \citep{1986ApJ...308...78L, 1989MNRAS.241P..43G}, while the HE component is thought to result from inverse Compton (IC) scattering, either via synchrotron self-Compton (SSC) \citep[e.g.,][]{1996ApJ...461..657B, 1985A&A...146..204G} or external Compton (EC) processes. In BL~Lacertae objects, SSC dominates the $\gamma$-ray emission, whereas EC is more important in FSRQs. The source of external photons is debated, with likely candidates being IR emission from the molecular torus (MT) \citep[e.g.,][]{2000ApJ...545..107B} or optical emission from the broad-line region (BLR) \citep[e.g.,][]{1994ApJ...421..153S, 2009MNRAS.397..985G, 2018MNRAS.477.4749C}.

The high-redshift blazars typically show flat or increasing X-ray spectra, especially in the hard X-ray band, have soft $\gamma$-ray spectral indices, and possess $\gamma$-ray luminosities $> 10^{48}\ \rm ergs\ s^{-1}$. High-redshift blazars are believed to contain supermassive black holes (SMBHs) with masses often $> 10^9 M_\odot$ and possess disk luminosities $> 10^{46}\ \rm ergs\ s^{-1}$ \citep[e.g.,][]{2010MNRAS.405..387G, Paliya2020ApJ, 2020MNRAS.498.2594S, Marcotulli2020ApJ}.

High-redshift AGN black holes (BHs) are believed to possess abundant accretable gas, making them more luminous than their low-redshift counterparts \citep{2004ApJ...612..724Y} and able to produce powerful jets \citep{1982Natur.295...17R}. Their prominent emission lines may originate from various sources, including the jet, hot accretion flows, disk structures, or cold clumps within the hot flow \citep{2000ARA&A..38..521S, 2012ApJ...759...65T, 2004ApJ...612..724Y, 2022MNRAS.514..780W, 2017MNRAS.469.2997N, 1999ApJ...515L..69N, 2010MNRAS.403.1102C}.

Studying the common properties of high-redshift blazars is important. First, these distant and powerful objects help us understand the physics of relativistic jets and the accretion--jet connection in the early Universe \citep{Ghisellini2013MNRAS, 2011MNRAS.416..216V}. Second, they provide valuable constraints on the density of extragalactic background light (EBL) \citep[e.g.,][]{2007ApJ...667L..29S,2008A&A...487..837F, Finke2010ApJ}, which improves our understanding of cosmological evolution \citep[e.g.,][]{Ackermann2012Sci, Fermi2018Sci, Desai2019ApJ, Finke2022ApJ}.

Combining $\gamma$-ray data with archival multiwavelength observations helps us to study the physical processes producing their $\gamma$-ray emission. It also allows us to understand the properties of relativistic jets in early Universe blazars. This study uses SED theoretical modeling to determine the location of the HE emission region and to understand the radiation properties of the jet. The discussion focuses on the jet dynamics and radiation mechanisms of these high-redshift blazars, and how these findings align with existing radiation models and astrophysical theories.

Given the distant nature of high-redshift blazars, the observational data quality tends to be lower, leading to significant uncertainties in individual SED fitting parameters. However, constructing and analyzing average SEDs from many sources reduces these uncertainties. This approach allows us to compensate for the limitations of individual observations through the power of statistical averaging across larger samples.

The rest of paper is organized as follows. In Section \ref{sec2}, we present the data reduction of the $\gamma$-ray observations and describe the SEDs’ observational characteristics. In Section \ref{sec3}, we provide our theoretical modeling of the broadband SEDs and their results. In Section \ref{sec4}, we conduct an in-depth analysis of the possible physical processes that may be responsible for their high-energy emission. Throughout this paper, we adopt a flat cosmological model with $\rm H_0 = 70.5\ \rm km\ s^{-1}\ Mpc^{-1}$, $\Omega_{\rm m} = 0.27$, and $\Omega_\Lambda = 0.73$.

\begin{deluxetable*}{cccccccccccccc}[t]
\centering
\tabletypesize{\tiny}
\tablenum{1}
\tablecaption{Observations and Environmental Parameters of the Blazar Sample \label{tab2}}
\tablehead{\colhead{Fermi name}&\colhead{Source name}&\colhead{R.A.}&\colhead{Dec.}&\colhead{$\rm \log \nu_{syn}$}& \colhead{$\rm \log \nu F_{syn}$}&  \colhead{$\rm \log \nu_{IC}$}   &\colhead{$\rm \log \nu F_{IC}$}&   \colhead{$\rm \log M_{BH}$}&\colhead{$\rm \log L_{disk}$} & \colhead{$R_{\rm BLR}$}& \colhead{$R_{\rm MT}$}&\colhead{$\rm L_{Edd}$}&\colhead{$\rm z$}
\\
\cline{5-13}
\colhead{}     & \colhead{}  & \colhead{}   &\colhead{}&\colhead{Hz}&  \colhead{$\rm erg \cdot cm^{-2}s^{-1}$}    &\colhead{Hz} & \colhead{$\rm erg \cdot cm^{-2}s^{-1}$} &\colhead{$\rm M_\odot$}& \colhead{$\rm erg \cdot s^{-1}$}&  \colhead{$\rm \times10^{17} \ (cm)$} & \colhead{$\rm \times10^{19} \ (cm)$} & \colhead{$\rm \times10^{47} \ (erg s^{-1})$} &
}
\startdata
J1510.1+5702 &GB 1508+5714 &227.54&57.04&12.07&-12.21&20.65&-11.72&8.56&46.80&7.94&1.99&0.46&4.314 \\
J1635.6-3628 &MG3 J163554+3629 &248.92&36.48&13.22&-12.88&21.60&-11.82&9.08&46.25&4.22&1.05&1.51&3.648 \\
J0539.6+1432 &TXS 0536+145 &84.91&14.54&12.44&-11.94&21.63&-11.33&10.01&46.14&3.72&0.93&12.89&2.710 \\
J0833.4-0458 &PMN J0833-0454 &128.37&-4.97&12.80&-12.62&21.12&-11.82&9.77&47.17&12.20&3.04&7.42&3.420 \\
J0337.8-1157 &PKS 0335-122 &54.47&-11.96&12.37&-12.53&21.64&-12.14&8.99&46.21&4.03&1.01&1.23&3.442 \\
J2320.8-0823 &PKS 2318-087 &350.22&-8.39&12.76&-12.99&20.65&-11.69&9.47&46.48&5.50&1.37&3.72&3.157 \\
J0539.9-2839 &PKS 0537-286 &84.99&-28.66&11.99&-11.99&20.76&-10.64&10.12&47.04&10.50&2.62&16.61&3.104 \\
J0805.4+6147 &TXS 0800+618 &121.36&61.79&12.20&-12.24&20.84&-10.80&9.47&46.69&7.00&1.75&3.72&3.033 \\
J1428.9+5406 &S4 1427+543 &217.23&54.11&12.38&-12.64&21.00&-11.87&9.33&46.19&3.94&0.98&2.69&3.012 \\
J0746.4+2546 &B2 0743+25 &116.60&25.77&12.26&-12.31&20.45&-10.60&9.37&46.61&6.38&1.60&2.95&2.994 \\
J1344.2-1723 &PMN J1344-1723 &206.06&-17.40&12.83&-11.53&23.02&-11.20&9.15&46.30&4.47&1.12&1.78&2.516 \\
J1127.4+5648 &S4 1124+57 &171.86&56.80&12.17&-12.64&20.95&-11.67&8.72&46.80&7.94&1.99&0.66&2.893 \\
J2313.9-4501 &PKS 2311-452 &348.49&-45.02&12.08&-12.50&20.48&-11.82&8.76&46.25&4.22&1.05&0.73&2.877 \\
J0440.3-4333 &PKS 0438-43 &70.09&-43.55&11.66&-12.11&20.50&-11.70&9.41&46.59&6.24&1.56&3.24&2.852 \\
J2015.4+6556 &S4 2015+65 &303.86&65.95&12.59&-12.23&20.75&-11.75&8.82&46.47&5.43&1.36&0.83&2.845 \\
J0836.5-2026 &PKS 0834-20 &129.13&-20.45&11.90&-11.90&20.85&-10.89&9.48&46.53&5.82&1.46&3.81&2.752 \\
J0224.9+1843 &TXS 0222+185 &36.23&18.72&12.56&-12.15&20.08&-10.78&9.66&46.73&7.33&1.83&5.76&2.690 \\
J0242.3+1102 &OD 166 &40.6&11.05&12.45&-11.89&21.18&-11.44&8.81&46.77&7.67&1.92&0.81&2.694 \\
J2339.6+0242 &RFC J2338+0251 &354.90&2.71&13.09&-12.53&21.38&-11.12&9.05&46.19&3.94&0.98&1.41&2.661 \\
J0910.6+2247 &TXS 0907+230 &137.67&22.80&12.62&-12.24&21.64&-11.34&8.30&45.93&2.92&0.73&0.25&2.677 \\
J1441.6-1522 &PMN J1441-1523 &220.41&-15.38&13.00&-12.43&22.28&-11.46&8.39&46.38&4.90&1.22&0.31&2.646 \\
J1054.2+3926 &GB6 B1051+3944 &163.56&39.43&12.60&-12.98&22.52&-12.38&8.59&45.78&2.45&0.61&0.49&2.635 \\
J1450.4+0910 &TXS 1448+093 &222.62&9.18&12.97&-12.13&21.60&-11.83&9.15&46.33&4.62&1.16&1.78&2.620 \\
J0226.5+0938 &TXS 0223+093 &36.63&9.64&13.05&-12.10&22.06&-11.57&10.09&46.61&6.38&1.60&15.50&2.605 \\
J0453.1-2806 &PKS 0451-28 &73.29&-28.11&12.50&-11.44&20.95&-10.53&9.21&46.88&8.71&2.18&2.04&2.564 \\
J0912.2+4127 &B3 0908+416B &138.06&41.46&12.77&-12.23&21.69&-11.58&9.42&45.85&2.66&0.67&3.31&2.568 \\
J1618.0+5139 &TXS 1616+517 &244.52&51.67&13.60&-12.84&21.27&-11.45&8.78&46.21&4.03&1.01&0.76&2.557 \\
J1625.7+4134 &4C+41.32 &246.45&41.57&12.03&-12.36&21.21&-11.48&7.85&45.94&2.95&0.74&0.09&2.550 \\
J1345.5+4453 &B3 1343+451 &206.39&44.88&12.73&-12.59&22.24&-10.84&9.06&46.06&3.39&0.85&1.45&2.542 \\
J2110.2-1021 &PKS 2107-105 &317.56&-10.36&12.35&-12.27&21.10&-11.54&9.04&46.99&9.89&2.47&1.38&2.500 \\
\enddata
\tablecomments{The observations and environmental parameters are presented in columns 5--14, which include the peak frequency and peak flux of synchrotron radiation and IC scattering, BH mass, accretion disk luminosity, radius of the BLR, inner radius of the MT, Eddington luminosity, and redshift, respectively \citep{2023ApJS..268....6C,2021ApJS..253...46P}. The BLR and MT are located at distances $R_{\rm BLR} \simeq 10^{17} \sqrt{L_{\rm disk}/10^{45}\rm erg \ s^{-1}}$ cm and $R_{\rm MT} \simeq 10^{18} \sqrt{L_{\rm disk}/10^{45}\rm erg \ s^{-1}}$ cm, respectively \citep{2008MNRAS.387.1669G}. $L_{\mathrm{Edd}}= \frac{4 \pi G M m_p c}{\sigma_T} \approx 1.26 \times 10^{38} \left( \frac{M}{M_\odot} \right) \ \mathrm{erg \ s^{-1}}$.
}
\end{deluxetable*}

\section{High-energy Emission of High-redshift Blazars\label{sec2}}

\subsection{Sample and Data Reduction \label{sec2.2}} 

Since 2008, the Fermi Large Area Telescope (Fermi-LAT) has provided a detailed view of the $\gamma$-ray sky, capturing images of the entire sky every 3 hr \citep{2009ApJ...697.1071A}.  The HE observations are based on the fourth catalog of Fermi-LAT AGNs (4FGL), which contains 3814 blazars, among which 792 are FSRQs, 1458 are BL~Lacertae objects, and 1493 are blazar candidates of uncertain type \citep{2020ApJS..247...33A,2022ApJS..260...53A}. The catalog reports the detection of 110 blazars with redshifts $z > 2$, and only 33 with redshifts $z > 2.5$. We exclude three blazars lacking a confirmed BH mass, synchrotron and IC scattering peak frequency or flux, and accretion disk luminosity. The remaining 30 blazars, which have redshifts $z > 2.5$, were selected as the sample. Detailed information about the sample is presented in Table \ref{tab2}.

The sample data, which span from 2008 August 4 to 2023 November 6 (MJD 54686--60262), were analyzed using Fermi {\tt ScienceTools} version 2.0.8 and the {\tt P8R3\_SOURCE\_V3} instrument response function. Data were focused on a $15^\circ$ radius region of interest (ROI) centered on each sample's $\gamma$-ray position, within the 100 MeV--500 GeV energy range. High-probability photon events ({\tt evclass = 128} and {\tt evtype = 3}) were selected with the {\tt gtselect} tool.

A $90^\circ$ zenith angle cut was applied to minimize contamination from the Earth's limb. Data were further filtered to include only intervals with good satellite operation and quality, using the {\tt gtmktime} filter expression \texttt{(DATA$\_$QUAL$>$0) \&\& (LAT$\_$CONFIG==1)}.

The binned data were prepared using the {\tt gtbin} tool, which allowed for binning the data in the spatial, spectral, and temporal dimensions. A live time cube, which is required for accurate exposure calculations, was created with the {\tt gtltcube} tool, accounting for variations in the effective area over the entire sky. A likelihood analysis was then performed with the {\tt gtlike} tool to determine the best-fit spectral parameters of each source.

A model file containing spectral parameters was created using the {\tt make4FGLxml.py} script based on the 4FGL catalog, covering the user-defined ROI plus an additional 10 degrees. In the likelihood fit, parameters for sources outside the ROI were fixed, while those within the ROI were allowed to vary. Galactic and extragalactic diffuse $\gamma$-ray emissions were modeled using {\tt gll\_iem\_v07.fits} and {\tt iso\_P8R3\_SOURCE\_V3\_v1.txt}, respectively. Note that the data used were not corrected for EBL absorption; observed values postabsorption were employed.

\subsection{High-energy Spectral Energy Distributions\label{sec2.2}}

Three spectral models were applied to analyze the $\gamma$-ray spectral shapes. The first is a simple power law (PL):

\begin{equation}
dN/dE\propto \left(\frac{E}{E_0}\right)^{-\alpha},
\end{equation}
where $E_0$ is the scale energy and $\alpha$ is the PL index.

The second form is a power law with an exponential cutoff (PLC):

\begin{equation}
dN/dE \propto\left(\frac{E}{E_0}\right)^{-\alpha}\exp\left(-\frac{E}{E_{\rm c}}\right),
\end{equation}
where $E_{\rm c}$ is the cutoff energy.

The third form is a log-parabola (LP):

\begin{equation}
dN/dE\propto \left(\frac{E}{E_0}\right)^{-\alpha-\beta\log(E/E_0)},
\end{equation}
where $\beta$ is the curvature index that describes the curvature around the peak. In our analysis, the value of $E_0$ is set to a fixed value given by {\tt FermiTools}.

All the best-fitting parameters and spectral normalization factors are presented in Table \ref{tab1}. To evaluate the curvature in the spectra, we computed the test statistic for curvature, $\rm TS_{curve} = 2(\log \mathcal{L}_{\rm PLC/LP} - \log \mathcal{L}_{\rm PL})$ \citep{2012ApJS..199...31N}, where $\mathcal{L}_{\rm PLC/LP}$ and $\mathcal{L}_{\rm PL}$ are the likelihood values estimated for the PLC (or LP) and PL models, respectively. A higher $\rm TS_{curve}$ value indicates a significant deviation from the PL model, favoring models that account for spectral curvature. Table \ref{tab1} lists the best-fitting model for each source, as determined by the selection criteria outlined above.

The photon indices are generally soft ($>2$), with the hardest and softest values being $2.03\pm0.10$ and $2.84\pm0.13$, respectively. The integrated $\gamma$-ray fluxes (above 0.1 GeV) derived from the best-fitting spectral models range from $\rm (0.49{-}113.92) \times 10^{-9} \ ph \ cm^{-2} \ s^{-1}$. The highest integrated flux is found for B3~1343+451, which has the highest-quality $\gamma$-ray data and is one of the most representative known high-redshift blazars \citep{2024ApJ...972..183W}, its SED is shown in Figure \ref{figx}, while the other SEDs are provided in Appendix \ref{A}.

Our results show that the $\gamma$-ray spectra of most sources exhibit significant curvature, deviating from the simple PL model. This curvature is well described by the LP and/or PLC models for the majority of the blazars in our sample. The exceptions are PKS~0438-43, TXS~1616+517, and 4C~+41.32, whose spectra are adequately fitted by the PL model.

\begin{figure}
\centering
\includegraphics[width=0.48\textwidth]{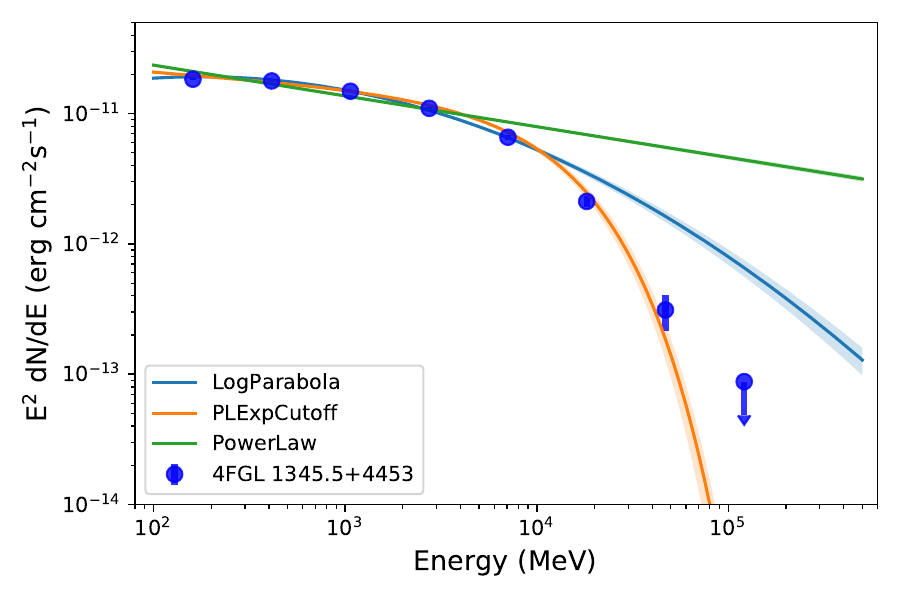}
\caption{$\gamma$-ray spectra of B3~1343+451. The data points are fitted with PL (green), PLC (orange), and LP (blue) models (see text).
\label{figx}}
\end{figure}

\begin{deluxetable*}{ccccccccc}[!ht]
\tabletypesize{\small}
\tablewidth{0pt}
\tablenum{2}
\tablecaption{Three Model Fits for the 30 High-redshift Blazars \label{tab1}}
\tablehead{\colhead{Fermi name}&\colhead{ $\rm F_{0.1-500GeV}$}&\colhead{$\alpha_{\rm PL}$}& \colhead{$ \rm \alpha_{\rm LP}$ }  &  \colhead{ $ \rm \alpha_{\rm PLC}$  }    &\colhead{$ \rm \beta$}&\colhead{$\rm E_{cut}$}& \colhead{$\rm best \ model$}\\
\cline{2-8}
\colhead{}&\colhead{$\rm 10^{-9} \ ph \ cm^{-2}s^{-1}$}& \colhead{ $\rm $}&\colhead{} & \colhead{   }       &\colhead{}&\colhead{GeV }&\colhead{}
}
\startdata
J1510.1+5702  & $5.21\pm0.81$&$2.62\pm0.08$&$2.48\pm0.11$&$2.11\pm 0.31$&$0.16\pm0.10$&$1.65\pm1.13$&LP \\
J1635.6-3628  & $8.76\pm1.84$&$2.75\pm0.13$&$2.75\pm0.13$&$2.72\pm 0.17$&$\bf (5.37\pm5928)\times10^{-8}$&$70.86\pm24.05$&LP \\
J0539.6+1432  & $3.26\pm2.57$ & $-$ &$1.09\pm0.18$&$1.16\pm0.89$&$0.70\pm0.10$&$2.07\pm1.67$&PLC \\
J0833.4-0458  & $0.49\pm0.13$ &$2.04\pm0.09$&$1.60\pm0.29$&$\bf 0.16\pm0.28$&$0.88\pm0.25$&$1.41\pm1.29$& LP \\
J0337.8-1157 &$4.60\pm0.95$&$2.54\pm0.09$&$2.38\pm0.16$&$1.54\pm0.43$&$0.44\pm0.16$&$0.85\pm0.45$&LP \\
J2320.8-0823   & $6.42\pm1.37$&$-$&$\bf 1.05\pm3.06$&$3.51\pm0.83$&$\bf 1.90\pm3.03$&$\bf 226.64\pm1619.96$&PLC \\
J0539.9-2839   & $47.62\pm1.24$ &$2.61\pm0.02$&$2.56\pm0.02$&$2.35\pm0.04$&$0.17\pm0.02$&$3.03\pm0.54$&LP  \\
J0805.4+6147&$12.53\pm1.08$&$2.79\pm0.06$&$2.64\pm0.09$&$2.49\pm0.16$&$0.14\pm0.06$&$2.76\pm1.63$&PLC \\
J1428.9+5406 &$7.78\pm1.19$ &$2.75\pm0.09$& $2.75\pm0.11$ &$2.75\pm0.10$&$\bf (7.67\pm29082)\times 10^{-9}$&$\bf 255.33\pm657.60$&PLC  \\
J0746.4+2546 & $18.25\pm1.35$&$2.92\pm0.06$& $2.90\pm0.08$ &$2.52\pm0.20$&$0.16\pm0.08$&$1.46\pm0.83$&LP \\
J1344.2-1723 &$11.87\pm0.98$ &$2.10\pm0.02$&$1.86\pm0.05$&$1.73\pm0.06$&$0.21\pm0.03$&$7.14\pm1.22$&LP \\
J1127.4+5648 & $9.16\pm0.91$ &$2.60\pm0.06$&$2.57\pm0.07$&$2.51\pm0.09$&$0.04\pm0.04$&$16.06\pm14.67$&PLC \\
J2313.9-4501 &$6.98\pm1.10$ &$2.84\pm0.13$&$2.84\pm0.14$&$2.83\pm0.13$&$\bf (5.31\pm 6431)\times 10^{-8}$&$309.92\pm30.07$&LP  \\
J0440.3-4333 &$71.52\pm1.48$ &$2.55\pm0.02$&$2.63\pm0.02$&$2.22\pm0.04$&$0.17\pm0.02$&$2.63\pm0.29$&PL  \\
J2015.4+6556 &$0.80\pm0.23$ &$2.10\pm0.16$&$2.23\pm0.38$&$-$&$1.19\pm0.33$&$-$&LP  \\
J0836.5-2026 &$8.24\pm1.47$ &$2.81\pm0.10$&$2.43\pm0.21$&$1.72\pm0.55$&$0.47\pm0.22$&$0.53\pm0.32$&PLC \\
J0224.9+1843 &$2.86\pm1.00$ &$-$&$1.79\pm0.36$&$1.60\pm1.01$&$0.56\pm0.29$&$\bf 0.94\pm1.47$&LP  \\
J0242.3+1102 &$11.55\pm1.30$ &$2.45\pm0.04$&$2.24\pm0.08$&$1.65\pm0.18$&$0.39\pm0.07$&$1.30\pm0.36$&LP  \\
J2339.6+0242 &$0.79\pm0.18$ &$2.09\pm0.10$&$1.08\pm0.28$&$-$&$0.87\pm0.18$&$0.97\pm0.40$&LP  \\
J0910.6+2247 & $5.69\pm1.13$&$2.32\pm0.07$&$2.22\pm0.10$&$2.08\pm0.14$&$0.10\pm0.06$&$11.89\pm6.99$&PLC  \\
J1441.6-1522  &$0.83\pm0.33$&$2.03\pm0.10$&$1.85\pm0.21$&$1.34\pm0.40$&$0.35\pm0.13$&$5.54\pm3.42$&LP  \\
J1054.2+3926 &$0.90\pm0.44$ &$2.10\pm0.16$&$\bf 1.83\pm3.38$&$0.75\pm0.45$&$\bf 0.66\pm1.68$&$1.94\pm2.86$&LP \\
J1450.4+0910 & $2.72\pm0.49$&$2.19\pm0.05$&$1.79\pm0.12$&$1.45\pm0.18$&$0.41\pm0.08$&$3.02\pm0.85$&LP  \\
J0226.5+0938 &$1.56\pm0.81$ &$2.12\pm0.10$&$2.03\pm0.17$&$1.87\pm0.21$&$0.14\pm0.09$&$21.70\pm19.37$&LP  \\
J0453.1-2806 &$53.06\pm1.26$ &$2.60\pm0.02$&$2.44\pm0.03$&$2.37\pm0.04$&$0.13\pm0.02$&$3.76\pm0.66$&PLC  \\
J0912.2+4127  &$15.76\pm1.06$&$2.48\pm0.04$&$2.47\pm0.04$&$2.40\pm0.06$&$0.07\pm0.03$&$19.53\pm13.20$&PLC  \\
J1618.0+5139 & $0.51\pm0.16$ &$2.43\pm0.15$&$0.53\pm0.32$&$-$&$1.14\pm0.28$&$0.68\pm0.31$&PL  \\
J1625.7+4134 & $8.03\pm0.79$&$2.39\pm0.05$&$1.82\pm0.13$&$1.49\pm0.21$&$0.44\pm0.09$&$1.81\pm0.55$&PL  \\
J1345.5+4453 & $113.92\pm1.15$ &$2.24\pm0.01$&$2.16\pm0.01$&$2.11\pm0.01$&$0.08\pm0.01$&$11.55\pm1.01$&PLC  \\
J2110.2-1021 & $3.41\pm1.28$ &$2.52\pm0.18$&$1.87\pm0.40$&$2.51\pm0.21$&$0.63\pm0.38$&$\bf 195.78\pm955.92$&PLC  \\
\enddata
\tablecomments{Column 2 is the $\gamma$-ray flux from the best-fit model. Columns 3--7 are the fitting parameters for each of the three models. Column 8 is the best-fit model for the observation period. Models that cannot be fitted are represented by "-." When the error of the fitted value is greater than the magnitude of the original data, it is considered an inaccurate parameter and is marked in \textbf{bold}.}
\end{deluxetable*}

\section{Broadband Spectral Energy Distribution Modeling\label{sec3}}

\subsection{Modeling the Broadband Emission\label{sec3.1}}

To construct broadband SEDs spanning from the IR to $\gamma$-ray bands, we collected multiwavelength data using the Space Science Data Center (SSDC) Sky Explorer.\footnote{\url{http://tools.ssdc.asi.it/SED/.}} Seven objects were discarded due to a lack of multiwavelength data, and ultimately, we selected 23 blazars with sufficient data coverage for our analysis. Hereafter, physical quantities denoted with a prime ($'$) are measured in the jet's comoving frame, while quantities without a prime are measured in the stationary AGN frame.

We assume that within the jet, there is a plasma blob with radius $R'$ moving along the jet axis with a bulk Lorentz factor $\Gamma$. Since the angle between the jet direction and the line of sight to Earth is close to zero, but not exactly zero, we have to consider the Doppler factor, $\delta_{\rm D} \approx \Gamma$. Inside the blob, there is an anisotropic magnetic field of strength $B'$. Relativistic electrons, $N_{\rm e}'(\gamma')$, are continuously injected, following a specific electron energy distribution (EED). Due to the clear deviation of the $\gamma$-ray spectra from a simple PL model, it is likely that the EED has some curvature, which can be modelled with an LP:

\begin{equation} N_{\rm e}'(\gamma') = N_{\rm pk}' \left( \frac{\gamma'}{\gamma_{\rm pk}'} \right)^{-s-r \log (\gamma' / \gamma_{\rm pk}')}, 
\end{equation}
where $N_{\rm pk}'$ is the normalization constant at the peak Lorentz factor $\gamma_{\rm pk}'$, and $s$ and $r$ are the spectral index and curvature parameter, respectively, which could be indicative of stochastic acceleration of electrons in the jet \citep[e.g.,][]{2006A&A...448..861M,2011ApJ...739...66T}. To maintain charge neutrality, we assume the presence of protons that are nearly stationary in the comoving frame. These "cold protons" contribute to the kinetic energy of the jet but do not contribute to its nonthermal radiation. The number density of cold protons is assumed to be 0.1 times that of the electrons.

Here, we assume an accretion efficiency of $\eta_{\rm acc} = 0.3$ for a rapidly rotating BH \citep{1974ApJ...191..507T}, and the accretion disk is modeled as a geometrically thin, optically thick Shakura--Sunyaev disk \citep{1973A&A....24..337S,2014ARA&A..52..529Y}. It extends from the inner radius $R_{\rm in,d} = 3 R_{\rm s}$ to the outer radius $R_{\rm out,d} = 10^3 R_{\rm s}$, where $R_{\rm s} \equiv 2 G M_{\rm BH} / c^2$ is the Schwarzschild radius and $G$ is the gravitational constant. For additional observational and environmental parameters, see Table \ref{tab2}.

The spectra of the BLR and MT are approximated as isotropic blackbodies peaking at energies corresponding to $\sim10.2$ eV and $3.93 k_{\rm B} T_{\rm MT}$, respectively, where $k_{\rm B}$ is the Boltzmann constant and $T_{\rm MT}$ is the characteristic temperature of the MT. The BLR photon energy is based on the strong Ly$\alpha$ line commonly observed in blazars. We assume $T_{\rm MT} = 10^3$ K for the MT radiation. The fractions of accretion disk luminosity reprocessed by the BLR and MT are adopted as $\tau_{\rm BLR} = 0.1$ and $\tau_{\rm MT} = 0.2$, respectively.

To determine the appropriate location of the radiation region, we consider potential external photon sources. We assume the dissipation region, $R_{\rm diss}$, is located near either the BLR or the MT, and we model the emission using the standard one-zone leptonic scenario proposed by \citet{1968PhRv..167.1159J,2002ApJ...575..667D,2011ApJ...739...66T}. For high-redshift blazars, the interaction between HE photons and the EBL can be significant. We account for this by converting the intrinsic $\gamma$-ray flux to observed flux using the EBL model from \citet{Finke2010ApJ}.

\movetabledown=68mm
\begin{rotatetable}
\begin{deluxetable*}{cccccccccccccccc}
\renewcommand{\arraystretch}{1.03}
\setlength{\tabcolsep}{3.8pt}
\centering
\tabletypesize{\tiny}
\tablenum{3}
\tablecaption{Parameters Obtained from the Modeling \label{tab3}}
\tablehead{\colhead{$\rm R_{diss}$}&\colhead{Fermi name}&\colhead{$N_{\rm pk}'$}&  \colhead{s}  &\colhead{r} &\colhead{$\delta_{\rm D}$}&   \colhead{$\gamma_{\rm pk}'$}&\colhead{$B'$} & \colhead{$R'$}& \colhead{$\rm P_e$}& \colhead{$\rm P_B$}& \colhead{$\rm P_p$}& \colhead{$\rm P_r$}& \colhead{$\rm P_{jet}$ }& \colhead{ $\rm CD$}& \colhead{ $\chi^2 / \chi^2_{\rm dof}$} \\
\cline{1-16}
\colhead{ } &\colhead{ } &\colhead{$\rm \times 10^{3}$ } &\colhead{ } &\colhead{ } &\colhead{ } &\colhead{$\rm \times10^{3}$ } &\colhead{ } &\colhead{$\rm \times10^{15}$} &\colhead{$\rm \times10^{44}$ } &\colhead{$\rm \times10^{44}$ } &\colhead{ $\rm \times10^{44}$} &\colhead{ $\rm \times10^{45}$} &\colhead{$\rm \times10^{45}$ }&\colhead{ } &\colhead{ } \\
\cline{3-14}
\colhead{ } &\colhead{ }&\colhead{$\rm (1/cm^{3})$}  &\colhead{ }&\colhead{ }&\colhead{ }&\colhead{ }&\colhead{$\rm (G)$}&\colhead{$\rm (cm)       $} &\colhead{$\rm (erg \ s^{-1})$}&\colhead{$\rm (erg \ s^{-1})$}&\colhead{$\rm (erg \ s^{-1})$}&\colhead{$\rm (erg \ s^{-1})$}&\colhead{$\rm (erg \ s^{-1})$}&\colhead{ }&\colhead{ }
}
\startdata
&J1510.1+5702 &$10.33^{+1.36}_{-1.19}      $&$7.53^{+0.74}_{-0.53}$&$1.99^{+0.25}_{-0.17}$&$15.02^{+0.80}_{-0.68}$&$9.38^{+1.32}_{-1.25} $&$2.04^{+0.20}_{-0.19}$&$4.15^{+0.40}_{-0.40}$&8.72  &0.60  &5.64  &1.55  &3.05 &14.45 &2.95/12.73 \\
&J1635.6+3628 &$1.81^{+0.24}_{-0.22}       $&$8.40^{+0.44}_{-0.40}$&$1.34^{+0.10}_{-0.09}$&$10.60^{+0.55}_{-0.48}$&$103.21^{+14.33}_{-13.81}$&$1.85^{+0.15}_{-0.13}$&$9.97^{+1.07}_{-1.04}$&4.22  &1.43  &2.85  &1.74  &2.59 & 2.96&7.41/4.75 \\
&J0539.6+1432 &$428.77^{+63.07}_{-51.22}   $&$2.77^{+0.14}_{-0.13}$&$1.14^{+0.04}_{-0.04}$&$26.03^{+1.24}_{-1.15}$&$0.16^{+0.01}_{-0.01} $&$4.79^{+0.26}_{-0.26}$&$1.32^{+0.13}_{-0.12}$&16.90 &1.01  &71.50 &1.72  &10.70 &16.74&159.2/6.80 \\
&J0539.9-2839 &$533.41^{+50.21}_{-49.78}   $&$7.45^{+0.29}_{-0.30}$&$1.44^{+0.07}_{-0.07}$&$5.37^{+0.44}_{-0.34} $&$101.02^{+12.07}_{-11.32}$&$2.48^{+0.26}_{-0.31}$&$1.13^{+0.07}_{-0.07}$&12.80 &0.008 &2.73  &150   &151 &1534.48 &8.06/7.89 \\
&J0805.4+6147 &$21.81^{+2.96}_{-2.86}      $&$7.68^{+1.17}_{-0.89}$&$6.11^{+0.63}_{-0.54}$&$8.54^{+0.39}_{-0.35} $&$2.63^{+0.39}_{-0.37} $&$1.79^{+0.14}_{-0.11}$&$3.61^{+0.35}_{-0.31}$&12.80  &0.11  &2.91  &10.10  &11.70 &112.80 &9.21/3.75 \\
&J1428.9+5406 &$80.28^{+10.16}_{-8.15}    $&$1.49^{+0.13}_{-0.14}$&$1.24^{+0.09}_{-0.09}$&$10.10^{+0.63}_{-0.62}$&$0.51^{+0.06}_{-0.06} $&$3.28^{+0.25}_{-0.21}$&$1.14^{+0.09}_{-0.09}$&4.19  &0.05  &1.50  &3.92  &4.50 &78.75 &2.67/3.79 \\
&J0746.4+2546 &$2.68^{+0.33}_{-0.40}  $&$6.16^{+0.35}_{-0.34}$&$0.69^{+0.05}_{-0.06}$&$31.80^{+1.39}_{-1.43}$&$4.40^{+0.72}_{-0.46} $&$6.24^{+0.88}_{-0.88}$&$34.3^{+2.39}_{-2.59}$&9.06  &1740.00  &451  &3.51  &223 &0.0052 &9.98/29.31 \\
&J1344.2-1723 &$24.70^{+5.00}_{-4.52} $&$8.70^{+0.92}_{-1.63}$&$1.92^{+0.25}_{-0.40}$&$45.01^{+4.16}_{-3.45}$&$24.50^{+6.25}_{-5.94}$&$8.01^{+0.91}_{-0.94}$&$0.29^{+0.04}_{-0.03}$&1.06  &0.41  &0.60  &0.11  &0.31 & 2.56&15.28/26.38 \\
&J1127.4+5648 &$282.03^{+51.78}_{-46.81}$&$3.48^{+0.14}_{-0.17}$&$0.18^{+0.03}_{-0.04}$&$6.93^{+0.48}_{-0.42} $&$127.78^{+24.02}_{-23.90}$&$0.70^{+0.05}_{-0.05}$&$24.81^{+3.49}_{-3.44}$&41.60  &0.53  &1170.00  &5.02 & 126.00 &78.64& 1.21/ 4.86 \\
&J2313.9-4501 &$3.22^{+0.52}_{-0.48}  $&$4.77^{+0.42}_{-0.37}$&$1.04^{+0.13}_{-0.14}$&$11.20^{+1.01}_{-0.74}$&$31.20^{+5.20}_{-4.72}$&$5.04^{+0.57}_{-0.55}$&$1.70^{+0.15}_{-0.15}$&0.74  &0.34  &0.17  &0.86  &0.99 &  2.16&2.3 /3.34 \\
&J0440.3-4333 &$1.60^{+0.18}_{-0.17}  $&$2.20^{+0.07}_{-0.06}$&$0.55^{+0.02}_{-0.02}$&$7.22^{+0.25}_{-0.24} $&$0.33^{+0.03}_{-0.03} $&$0.80^{+0.03}_{-0.03}$&$67.03^{+6.13}_{-5.44}$  &21.80  &5.61  &52.50  &25.00  &33.00 & 3.88 &0.1 /4.81 \\
$\rm R_{diss}=R_{BLR}$&J2015.4+6556 &$0.87^{+0.11}_{-0.11}$&$3.16^{+0.10}_{-0.10}$&$0.78^{+0.03}_{-0.03}$&$7.83^{+0.40}_{-0.39}$&$0.39^{+0.04}_{-0.04}$&$2.08^{+0.14}_{-0.13}$&$110.12^{+10.62}_{-9.22}$&16.50&119&90.40&12.90&35.50 &0.14&1.37 /2.68 \\
                      &J0836.5-2026 &$21.50^{+1.93}_{-1.95} $&$4.68^{+0.22}_{-0.19}$&$1.41^{+0.09}_{-0.08}$&$6.71^{+0.54}_{-0.42} $&$7.04^{+0.97}_{-0.86} $&$3.94^{+0.34}_{-0.33}$&$3.63^{+0.27}_{-0.26}$&5.06  &0.34  &1.79  &15.50  &16.20 & 14.83 &5.91/8.32 \\
&J0224.9+1843 &$2.52^{+0.48}_{-0.40}  $&$4.50^{+0.20}_{-0.17}$&$0.24^{+0.03}_{-0.03}$&$23.50^{+0,61}_{-1.52}$&$5.44^{+1.09}_{-1.03} $&$9.24^{+1.11}_{-0.88}$&$75.02^{+5.97}_{-5.12}$  &11.70  &9930  &1110  &2.44  &1110&0.0012 &0.49/7.47 \\
&J2339.6+0242 &$0.0053^{+0.001}_{-0.001}  $&$8.70^{+0.85}_{-0.12}$&$1.29^{+0.16}_{-0.23}$&$78.60^{+6.67}_{-5.95}$&$104.02^{+23.70}_{-25.44}$&$8.21^{+0.93}_{-0.87}$&$1.31^{+0.22}_{-0.18}$&0.007  &26.80  &0.008  &0.004  &2.68  &0.00027&1.95/15.46 \\
&J0910.6+2247 &$3.03^{+0.32}_{-0.30}  $&$7.95^{+0.42}_{-0.38}$&$3.15^{+0.24}_{-0.05}$&$11.02^{+0.57}_{-0.50}$&$7.29^{+0.78}_{-0.74} $&$2.02^{+0.15}_{-0.14}$&$4.26^{+0.30}_{-0.31}$&3.48  &0.33  &0.94  &0.97  &1.45 &10.42 &15.19/12.57 \\
&J1054.2+3926 &$19.80^{+3.04}_{-2.87} $&$8.27^{+0.53}_{-0.47}$&$1.72^{+0.15}_{-0.14}$&$11.20^{+0.87}_{-0.81}$&$30.40^{+3.89}_{-4.00}$&$2.66^{+0.27}_{-0.23}$&$2.46^{+0.27}_{-0.26}$&3.74  &0.20  &2.12  &0.71  &1.32 &18.68 &2.64/3.24 \\
&J0226.5+0938 &$0.0036^{+0.0006}_{-0.0007} $&$7.75^{+0.83}_{-0.63}$&$3.03^{+0.35}_{-0.34}$&$5.55^{+0.60}_{-0.51} $&$4.79^{+0.86}_{-0.70} $&$2.17^{+0.26}_{-0.21}$&$206.43^{+29.03}_{-28.33}$ &1.57  &227.00  &0.66  &7.59  &30.50  &0.0069&15.65/18.27 \\
&J0453.1-2806 &$11.60^{+0.98}_{-1.01} $&$6.11^{+0.23}_{-0.25}$&$1.00^{+0.06}_{-0.06}$&$10.90^{+0.49}_{-0.44}$&$56.80^{+5.84}_{-5.79}$&$2.83^{+0.09}_{-0.08}$&$6.32^{+0.44}_{-0.38}$&11.60  &1.42  &7.76  &11.10  &13.20 &8.19 &3.01/7.79 \\
&J0912.2+4127 &$2.39^{+0.37}_{-0.36}  $&$4.86^{+0.12}_{-0.12}$&$0.41^{+0.02}_{-0.02}$&$20.40^{+1.29}_{-1.13}$&$40.60^{+5.15}_{-4.20}$&$3.73^{+0.25}_{-0.23}$&$23.4^{+1.96}_{-1.90}$&2.74  &119.00  &77.00  &1.14  &21.00  &0.023&9.11/4.18 \\
&J1618.0+5139 &$9.35^{+1.40}_{-1.26}  $&$7.28^{+0.31}_{-0.32}$&$1.30^{+0.09}_{-0.09}$&$11.50^{+0.65}_{-0.66}$&$55.40^{+6.26}_{-6.81}$&$2.30^{+0.18}_{-0.15}$&$2.94^{+0.23}_{-0.25}$&2.70  &0.23  &1.51  &0.53  &0.97 & 11.96&0.2/4.98 \\
&J1625.7+4134 &$298.43^{+66.31}_{-55.13}   $&$7.03^{+1.04}_{-0.78}$&$4.54^{+0.57}_{-0.46}$&$32.60^{+3.07}_{-2.44}$&$1.45^{+0.26}_{-0.21} $&$3.09^{+0.35}_{-0.29}$&$0.20^{+0.02}_{-0.02}$&3.40  &0.02  &1.77  &0.07  &0.59 & 225.07&5.49/3.69 \\
&J1345.5+4453 &$23.50^{+2.37}_{-2.50} $&$8.02^{+0.24}_{-0.24}$&$1.50^{+0.07}_{-0.06}$&$36.40^{+2.64}_{-1.84}$&$38.01^{+3.43}_{-3.17} $&$1.49^{+0.11}_{-0.10}$&$0.88^{+0.06}_{-0.06}$&4.74  &0.09  &3.44  &0.76  &1.58  &55.07 &23.27/10.98 \\
\cline{2-14}
&Average &$7.75\times 10^4 $&$6.04$&$1.65$&$18.87$&$3.29\times 10^4$&$3.51$&$2.55 \times 10^{16}$&$8.75 \times 10^{44}  $&$5.29 \times 10^{46} $&$1.33 \times 10^{46} $&$1.12 \times 10^{46} $ &$7.82 \times 10^{46} $ &  &  \\
\cline{1-16}
\colhead{ } &\colhead{ } &\colhead{$\rm \times 10^{3}$ } &\colhead{ } &\colhead{ } &\colhead{ } &\colhead{ } &\colhead{ } &\colhead{$\rm \times10^{16}$} &\colhead{$\rm \times10^{45}$ } &\colhead{$\rm \times10^{44}$ } &\colhead{$\rm \times10^{46}$ } &\colhead{$\rm \times10^{44}$ } &\colhead{$\rm \times10^{46}$ } &\colhead{ }& \\
\cline{1-16}
&J1510.1+5702 &$0.11^{+0.03}_{-0.02}  $&$1.54^{+0.12}_{-0.13}$&$0.42^{+0.04}_{-0.04}$&$40.40^{+2.82}_{-2.80}$&$3.03^{+0.76}_{-0.63} $&$0.20^{+0.02}_{-0.02}$&$28.30^{+3.51}_{-2.99}$&6.91  &203.79  &20.91  &5.09  &23.69 &0.34 &1.86/14.94 \\
&J1635.6+3628 &$1.06^{+0.16}_{-0.16}  $&$1.06^{+0.13}_{-0.13}$&$0.60^{+0.04}_{-0.03}$&$31.50^{+2.21}_{-2.41}$&$9.67^{+1.44}_{-1.28} $&$0.18^{+0.02}_{-0.02}$&$7.61^{+0.90}_{-0.84}$&11.12  &6.59  &8.62  &4.56  &9.84  &16.87&2.31/4.92 \\
&J0539.6+1432 &$0.09^{+0.01}_{-0.01}  $&$1.89^{+0.07}_{-0.07}$&$0.46^{+0.02}_{-0.02}$&$37.30^{+2.52}_{-1.88}$&$6.54^{+0.90}_{-0.72} $&$0.31^{+0.02}_{-0.02}$&$41.7^{+3.55}_{-3.44}$&9.26  &870.94  &31.50  &6.95  &41.21 &0.11 &13.07/6.44 \\
&J0539.9-2839 &$1.35^{+0.18}_{-0.16}  $&$1.12^{+0.10}_{-0.10}$&$0.75^{+0.02}_{-0.02}$&$41.10^{+1.55}_{-1.40}$&$7.87^{+1.19}_{-1.08} $&$0.11^{+0.007}_{-0.007}$&$14.9^{+1.03}_{-0.94}$&58.25  &18.26  &71.66  &31.92  &77.99 &31.90 &5.62/7.34 \\
&J0805.4+6147 &$0.017^{+0.002}_{-0.002}  $&$1.82^{+0.09}_{-0.11}$&$0.76^{+0.03}_{-0.03}$&$75.80^{+2.91}_{-4.39}$&$7.69^{+1.15}_{-1.04} $&$0.56^{+0.04}_{-0.04}$&$25.62^{+2.32}_{-1.58}$&2.94  &4408.95  &8.80  &3.53  &53.21 &0.0067 &7.75/4.09 \\
&J1428.9+5406 &$16.22^{+1.57}_{-1.44} $&$2.67^{+0.10}_{-0.11}$&$1.06^{+0.07}_{-0.06}$&$55.60^{+0.37}_{-0.32}$&$2042.01^{+238.32}_{-227.19} $&$2.83^{+0.27}_{-0.28}$&$0.43^{+0.03}_{-0.03}$&0.39  &0.17  &0.013  &131.25  &1.37 &23.65 &2.21/3.78 \\
&J0746.4+2546 &$0.0015^{+0.0002}_{-0.0002} $&$1.32^{+0.08}_{-0.09}$&$0.89^{+0.02}_{-0.02}$&$72.30^{+3.67}_{-2.50}$&$5.87^{+0.60}_{-0.62} $&$0.42^{+0.02}_{-0.02}$&$65.13^{+4.23}_{-4.47}$&2.11  &14296.31  &4.70  &7.02  &147.95 &0.0015 &2.88/27.80 \\
&J1344.2-1723 &$2.37^{+0.53}_{-0.49}  $&$1.36^{+0.27}_{-0.23}$&$1.70^{+0.41}_{-0.29}$&$103.03^{+8.63}_{-8.77}$&$313.01^{+74.44}_{-56.31}$&$0.75^{+0.10}_{-0.09}$&$0.12^{+0.02}_{-0.02}$&0.88  &0.29  &0.050  &0.25  &0.14 &29.73 &12.61/16.07 \\
&J1127.4+5648 &$10123^{+1834.22}_{-1711.09} $&$4.23^{+0.39}_{-0.36}$&$0.46^{+0.08}_{-0.07}$&$85.80^{+6.42}_{-6.92}$&$3101.77^{+644.21}_{-576.89} $&$0.33^{+0.03}_{-0.03}$&$0.10^{+0.01}_{-0.01}$&38.21  &0.03  &109.52  &0.42  &113.34  &12528.67&0.61/5.31 \\
&J2313.9-4501 &$5.50^{+0.81}_{-0.81}  $&$2.44^{+0.09}_{-0.09}$&$0.51^{+0.06}_{-0.05}$&$30.40^{+2.47}_{-2.22}$&$2760.21^{+458.03}_{-413.41} $&$1.24^{+0.15}_{-0.13}$&$0.14^{+0.01}_{-0.01}$&0.25  &0.10  &0.014  &1.07  &0.05 &24.11 &1.76/3.27 \\
&J0440.3-4333 &$0.39^{+0.04}_{-0.03}  $&$4.43^{+0.14}_{-0.14}$&$1.14^{+0.05}_{-0.05}$&$6.82^{+0.29}_{-0.29} $&$27917.36^{+3041.78}_{-3030.07}$&$0.19^{+0.01}_{-0.01}$&$5.45^{+0.35}_{-0.33}$&5.86  &0.18  &0.075  &454.71  &5.21 &334.16 &2.55/5.62 \\
$\rm R_{diss}=R_{MT}$&J2015.4+6556 &$0.44^{+0.07}_{-0.06}$&$1.55^{+0.08}_{-0.08}$&$0.48^{+0.02}_{-0.02}$&$49.40^{+2.74}_{-2.73}$&$3.14^{+0.33}_{-0.32}$&$0.51^{+0.04}_{-0.04}$&$10.90^{+0.10}_{-0.10}$&5.52&283.66&18.11&2.05&21.51 &0.19 &0.12/2.58 \\
&J0836.5-2026 &$89.10^{+8.86}_{-8.55} $&$2.10^{+0.13}_{-0.14}$&$1.28^{+0.08}_{-0.08}$&$29.50^{+1.82}_{-1.65}$&$788.03^{+91.23}_{-90.03}$&$0.84^{+0.09}_{-0.07}$&$0.093^{+0.007}_{-0.006}$&2.36  &0.02  &0.095  &6.95  &0.40 &1172.97 &4.16/8.36 \\
&J0224.9+1843 &$3.92^{+0.46}_{-0.43}  $&$5.15^{+0.23}_{-0.20}$&$1.40^{+0.07}_{-0.07}$&$9.70^{+0.38}_{-0.35} $&$13404.97^{+1881.05}_{-1533.78}$&$0.17^{+0.01}_{-0.01}$&$2.79^{+0.20}_{-0.18}$&14.55  &0.076  &0.40  &170.92  &3.57 &1910.79 &1.03/7.98 \\
&J2339.6+0242 &$0.27^{+0.09}_{-0.09}  $&$1.57^{+0.30}_{-0.29}$&$0.31^{+0.08}_{-0.10}$&$42.30^{+6.01}_{-5.62}$&$3.34^{+1.48}_{-1.07} $&$0.22^{+0.03}_{-0.03}$&$5.53^{+1.43}_{-1.17}$&0.86  &9.83  &2.11  &0.30  &2.29 & 0.87&4.98/15.66 \\
&J0910.6+2247 &$3.43^{+0.37}_{-0.37}  $&$2.82^{+0.32}_{-0.26}$&$2.89^{+0.09}_{-0.07}$&$36.51^{+2.11}_{-2.11}$&$1750.33^{+194.24}_{-170.39} $&$0.23^{+0.02}_{-0.01}$&$0.28^{+0.02}_{-0.02}$&2.86  &0.02  &0.05  &1.39  &0.35 &1425.59 &6.19/12.29 \\
&J1054.2+3926 &$0.42^{+0.06}_{-0.06}  $&$3.12^{+0.17}_{-0.15}$&$0.99^{+0.06}_{-0.06}$&$19.91^{+1.63}_{-1.56} $&$1722.31^{+220.11}_{-224.18} $&$0.16^{+0.02}_{-0.02}$&$3.15^{+0.34}_{-0.36}$&3.29  &0.39  &0.23  &1.92  &0.59 &84.14 &0.81/3.00 \\
&J0226.5+0938 &$0.0007^{+0.00002}_{-0.00001}$&$1.09^{+0.17}_{-0.18}$&$1.08^{+0.06}_{-0.05}$&$71.70^{+4.70}_{-4.89}$&$10.40^{+1.58}_{-1.36}$&$0.73^{+0.08}_{-0.06}$&$30.9^{+4.04}_{-3.71}$&0.43  &9774.60  &0.48  &1.41  &98.29 &0.00044 &12.34/19.13 \\
&J0453.1-2806 &$2303.94^{+220.35}_{-206.44}  $&$1.81^{+0.06}_{-0.06}$&$0.49^{+0.04}_{-0.03}$&$81.80^{+2.80}_{-2.50}$&$4.75^{+0.53}_{-0.58} $&$0.65^{+0.23}_{-0.18}$&$0.41^{+0.03}_{-0.03}$&101.21  &1.82  &371.63  &2.86  &381.75  &556.72&2.61/5.92 \\
&J0912.2+4127 &$7.52^{+1.08}_{-1.03}  $&$1.36^{+0.09}_{-0.10}$&$0.42^{+0.02}_{-0.02}$&$38.82^{+2.12}_{-2.08}$&$4.98^{+0.60}_{-0.60} $&$0.20^{+0.01}_{-0.01}$&$3.50^{+0.29}_{-0.29}$&11.72  &2.71  &19.62  &2.36  &20.85 &43.28 &1.64/3.70 \\
&J1618.0+5139 &$0.18^{+0.02}_{-0.03}  $&$1.91^{+0.12}_{-0.12}$&$0.29^{+0.02}_{-0.02}$&$39.11^{+3.07}_{-2.69}$&$2.90^{+0.43}_{-0.40} $&$0.29^{+0.03}_{-0.03}$&$15.5^{+1.79}_{-1.68}$&2.13  &118.12  &9.20  &0.71  &10.60 &0.18 &0.32/4.76 \\
&J1625.7+4134 &$0.004^{+0.0009}_{-0.0008} $&$1.25^{+0.13}_{-0.14}$&$0.66^{+0.04}_{-0.04}$&$45.87^{+3.82}_{-3.43}$&$6.57^{+1.23}_{-0.97} $&$0.23^{+0.02}_{-0.02}$&$45.03^{+6.11}_{-5.53}$&1.39  &830.89  &2.20  &2.67  &10.68 &0.017 &89.27/4.73 \\
&J1345.5+4453 &$0.016^{+0.002}_{-0.002}  $&$1.23^{+0.05}_{-0.05}$&$0.44^{+0.01}_{-0.01}$&$54.50^{+2.89}_{-3.01}$&$6.02^{+0.64}_{-0.56} $&$0.089^{+0.005}_{-0.005}$&$21.12^{+1.68}_{-1.61}$&2.58  &39.07  &3.07  &7.40  &3.79 & 0.66&1.01/10.30 \\
\cline{2-14}
&Average &$5.45\times 10^5 $&$2.12$&$0.85$&$47.78$&$2.34\times 10^3$&$0.50$&$1.43\times 10^{17}$&$1.24 \times 10^{46}  $&$1.34 \times 10^{47} $&$2.97 \times 10^{47} $&$3.69 \times 10^{45}$ &$4.47 \times 10^{47} $ &  &  \\
\enddata
\tablecomments{\parbox{1.4\textwidth}{Column 3 denotes the electron number density. Columns 4--5 denote the spectral index and curvature of the EED, respectively. Column 6 denotes the Doppler factor. Column 7 denotes the peak Lorentz factor. Column 8 denotes the magnetic field strength. Column 9 denotes the radius of the emitting region. Columns 10--13 are, respectively, the power of the jet in the form of relativistic electrons ($P_{\rm e}$), magnetic field ($P_{\rm B}$), and cold protons ($P_{\rm P}$) with $n_{\rm e}/n_{\rm p}=1$ and radiation field ($P_{\rm r}$), where $P_{\rm r}$ is the sum of all radiative components. Column 14 is the total power ($P_{\rm jet}=P_{\rm e}+P_{\rm B}+P_{\rm r}+P_{\rm p}$). Column 15 is the ratio of energy density, i.e., Compton dominant $U_e/U_B$. Column 16 denotes the $\chi^2$ derived from the $\gamma$-ray band and the reduced $\chi^2_{\rm dof} =\rm \chi^2/dof$ of the fit to all bands.}}
\end{deluxetable*}
\end{rotatetable}
\clearpage

\begin{figure*}[t]
\centering
\includegraphics[width=1\textwidth]{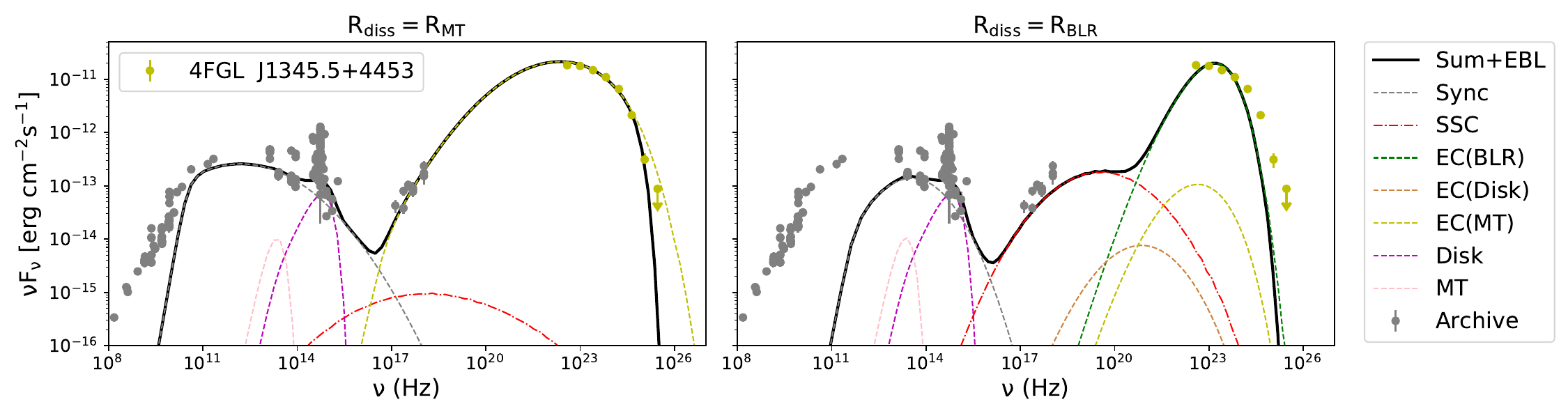}
\caption{The left and right panels represent the best fits with $R_{\rm diss}=R_{\rm MT}$ and $R_{\rm diss}=R_{\rm BLR}$, respectively. The grey points from the radio to X-ray bands represent the historical data obtained from the SSDC Sky Explorer. Separate synchrotron, MT, accretion disk, SSC, EC-disk, EC-BLR, and EC-MT components are shown. The black solid line in all plots represents the sum of all components, which have been corrected for EBL absorption considering the model of \citet{Finke2010ApJ}
\label{figg}}
\end{figure*}

\begin{deluxetable}{ccc}
\tabletypesize{\normalsize}
\tablenum{4}
\tablecaption{Normality Test Results (Shapiro--Wilk) \label{tab4}}
\tablehead{\colhead{Parameter}&\colhead{Shapiro--Wilk Statistic}& \colhead{$p$-value}
}
\startdata
$\rm B'$&	0.9527	&0.3328 \\
$\rm \log P_e$	&0.9771&	0.8518 \\
$\rm \log P_B$	&0.9368&	0.1530 \\
$\rm \log P_r$	&0.9153	&0.0530 \\
$\rm \log P_{jet}$	&0.9702	&0.6935 \\
$\rm \log \nu_{syn}$	&0.9786	&0.8817 \\
$\rm \log \nu_{IC}$	&0.9494	&0.2843 \\
$\rm \log M_{BH}$	&0.9759	&0.8265 \\
$\rm \log L_{disk}$	&0.9649	&0.5695 \\
\enddata
\tablecomments{A Shapiro--Wilk statistic close to 1 and $p> 0.05$ indicates that the parameter set follows a normal distribution.}
\end{deluxetable}

\begin{figure*}[ht]
\centering
\includegraphics[width=0.32\textwidth]{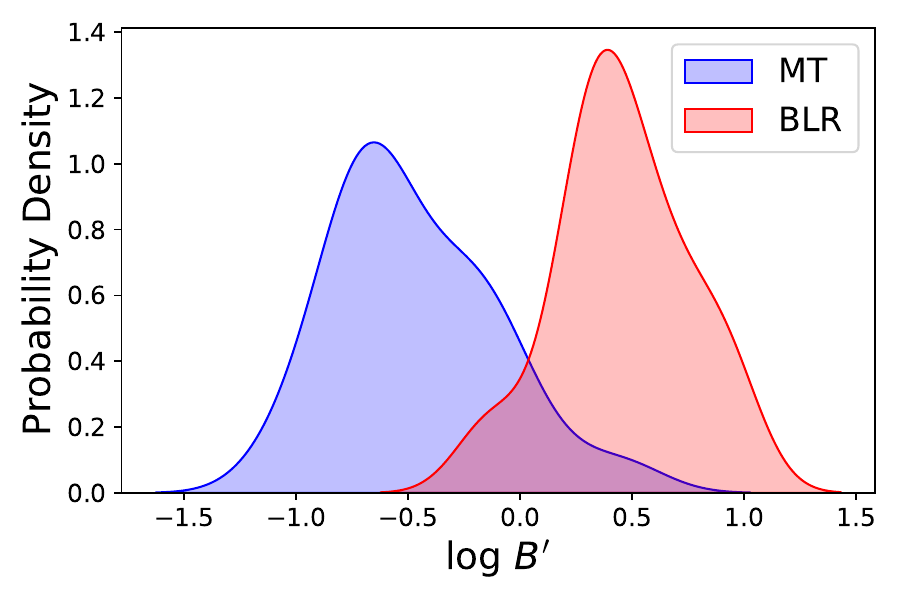}
\includegraphics[width=0.32\textwidth]{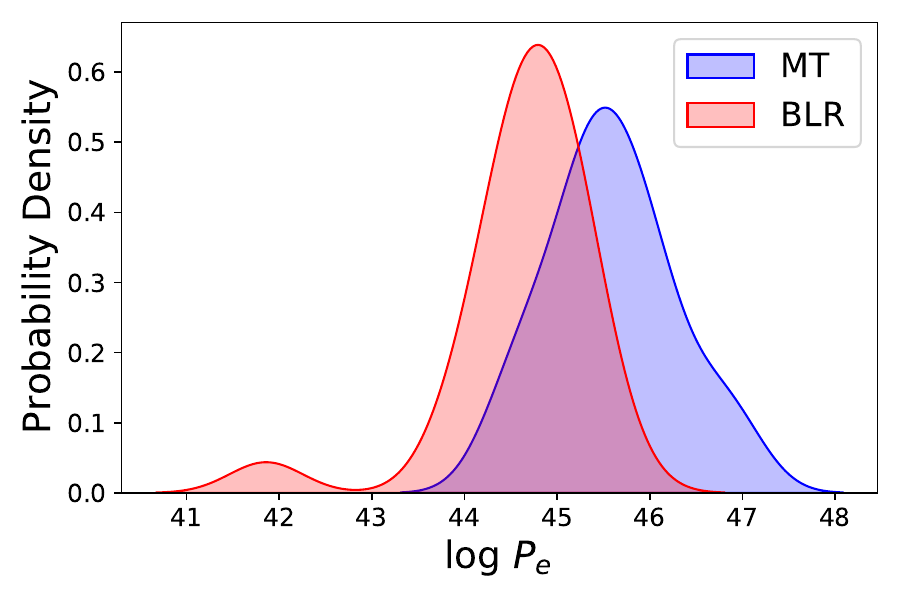}
\includegraphics[width=0.32\textwidth]{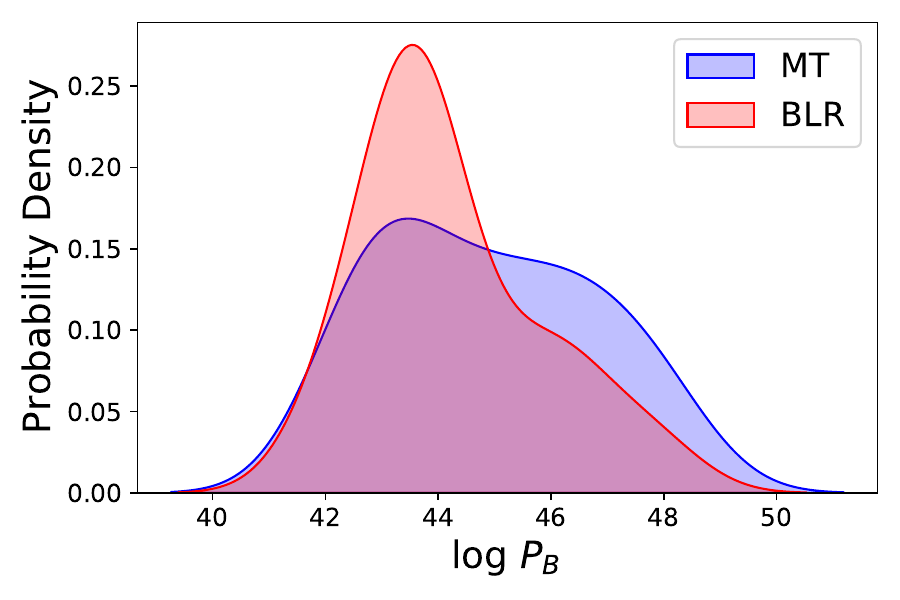}
\includegraphics[width=0.32\textwidth]{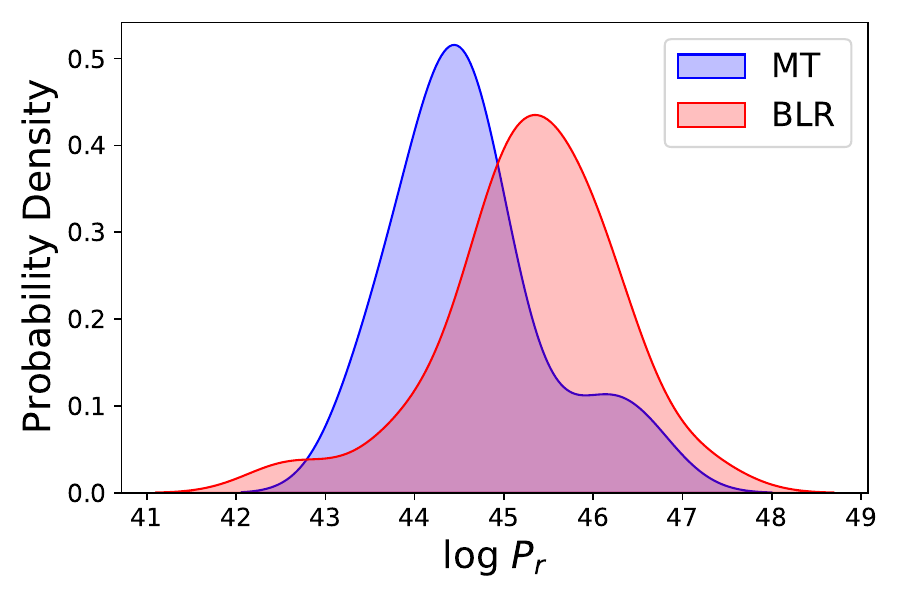}
\includegraphics[width=0.32\textwidth]{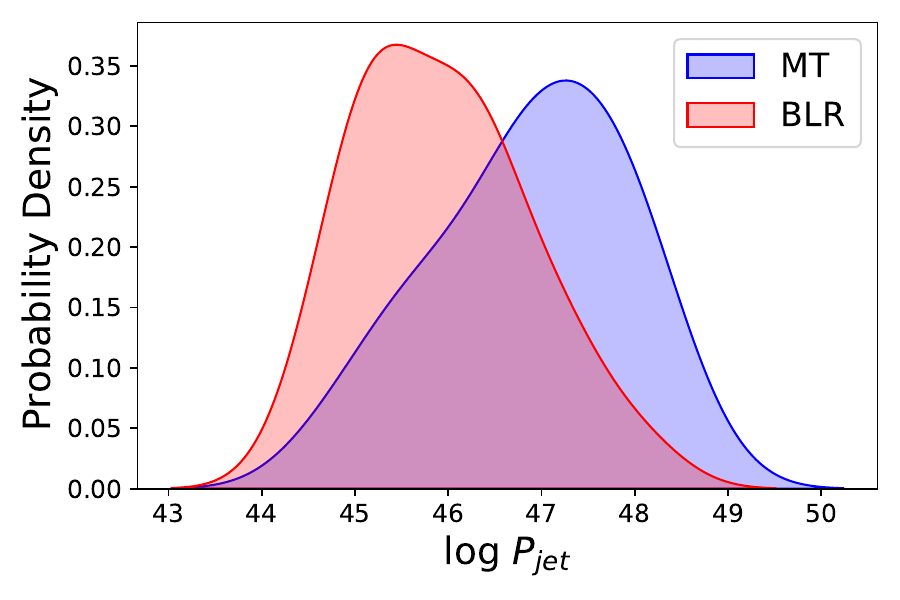}
\includegraphics[width=0.32\textwidth]{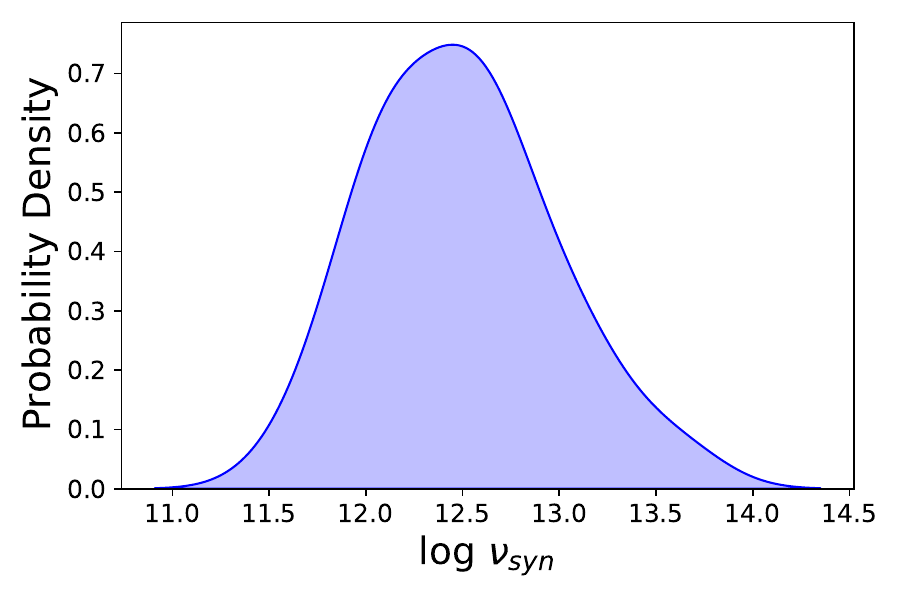}
\includegraphics[width=0.32\textwidth]{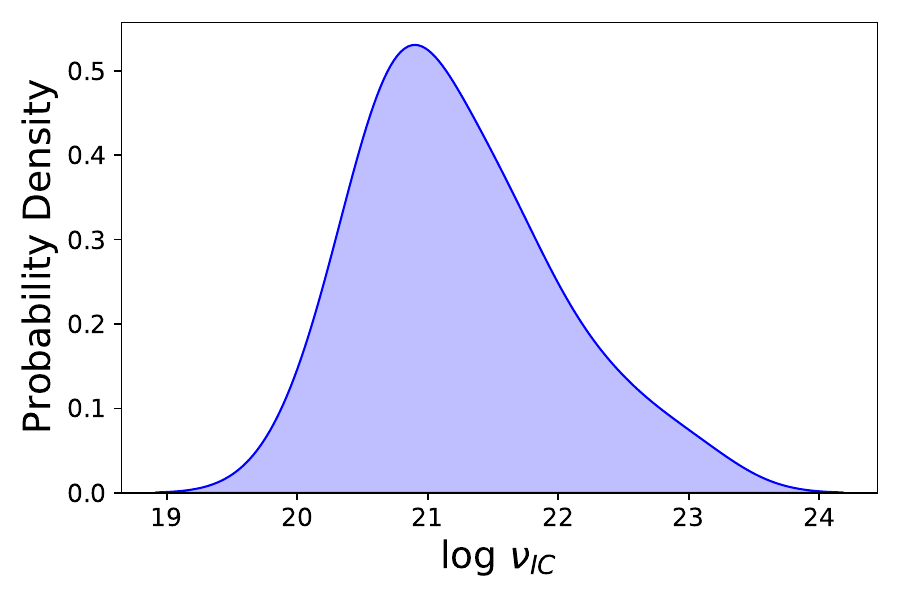}
\includegraphics[width=0.32\textwidth]{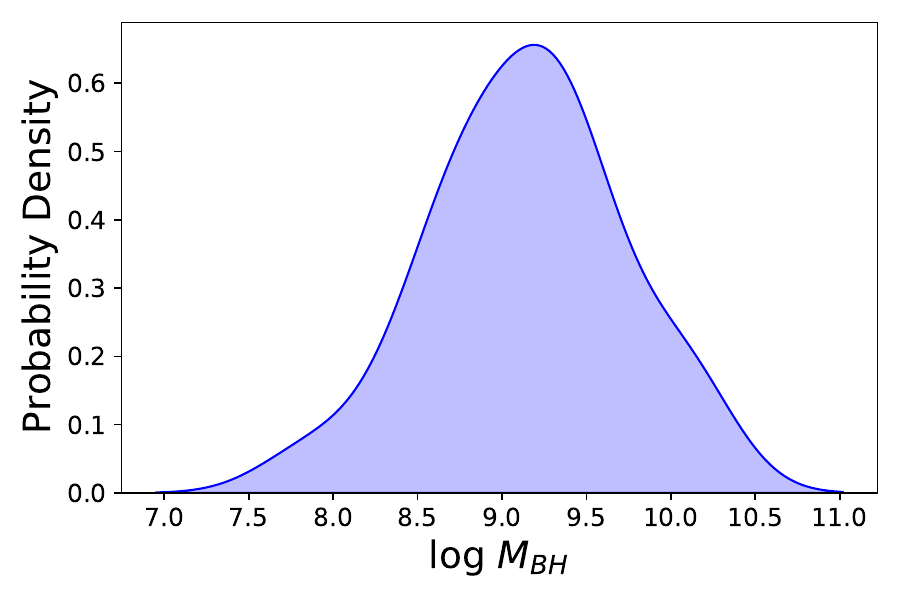}
\includegraphics[width=0.32\textwidth]{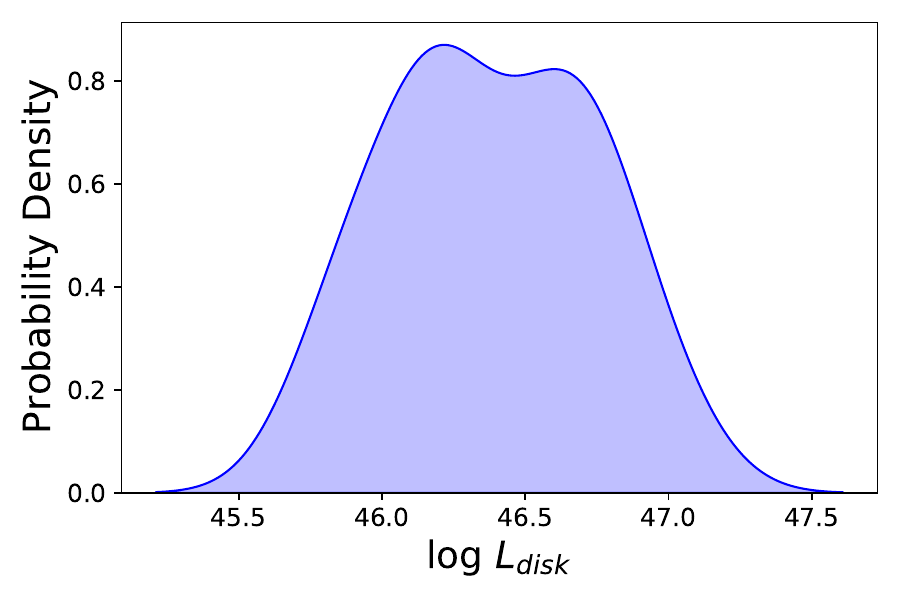}
\caption{The distributions of the physical parameters. The red line is $R_{\rm diss}=R_{\rm BLR}$ and the blue line is $R_{\rm diss}=R_{\rm MT}$.
\label{fig1}}
\end{figure*}

\subsection{Spectral Energy Distribution Modeling Results\label{sec3.2}}

We fit the SEDs over the energy range from IR to $\gamma$-rays ($\rm 10^{12}{-}10^{28}$ Hz). The LE peak is assumed to be dominated by synchrotron radiation, while the HE peak is primarily due to IC scattering. The seed photons for IC scattering originate from two sources: (1) reprocessed emission from the BLR in the optical--UV range, and (2) reprocessed IR emission from the MT \citep{2009ApJ...704...38S}. The radio band, which is thought to originate from a larger region outside the jet, is excluded from the fit. Since our study focuses on high-redshift blazars, we also account for the effects of EBL absorption.

The dissipation region, $R_{\rm diss}$, is constrained to be near either $R_{\rm BLR}$ or $R_{\rm MT}$. Seven free parameters are included in the fit: $N_{\rm pk}'$, $s$, $r$, $\delta_{\rm D}$, $\gamma_{\rm pk}'$, $B'$, and $R'$. The SEDs are derived from theoretical modeling, calculated using the publicly available \texttt{JetSet} package \citep{2006A&A...448..861M,2009A&A...501..879T,2011ApJ...739...66T}. We consider data above $\rm 10^{12}$ Hz, and the model is optimized using the \texttt{emcee} optimizer \citep{2013PASP..125..306F}. 

We collect the peak frequency and peak flux of the synchrotron and IC bumps from \citet{2022ApJS..260...53A,2023ApJS..268....6C}, while the BH masses and disk luminosities are taken from \citet{2022ApJ...925...40X}. Based on the environmental parameters listed in Table \ref{tab2}, we obtained the model parameters, power parameters, and fit curves for the 23 blazars with good data coverage. The best-fit values and their $1\sigma$ errors are summarized in Table \ref{tab3}. As an example, the typical SED of B3~1343+451 is shown in Figure \ref{figg}, while the complete fitting results are provided in Appendix \ref{B}. Our results represent the general state during the period of observation. It can be observed that for blazars with redshift $>2.5$, the effect of EBL absorption only becomes significant at energy levels $>10^{25}$ Hz.

In the BLR case, the X-ray emission is mostly attributed to the SSC process, except for sources B2~0743+25, TXS~0222+185, and B3~0908+416B. In the MT case, the X-ray band is dominated by either SSC or EC processes. Table \ref{tab3} lists the chi-square values for the $\gamma$-ray band fits. The lower chi-square values for the MT scenario compared to the BLR suggest that the radiation region is more likely located near the MT, where seed photons are provided by IR photons, consistent with previous studies of high-redshift blazars \citep{2020MNRAS.498.2594S,2024ApJ...972..183W}.

Following \citet{2008MNRAS.385..283C}, we evaluated the power carried by the magnetic field ($P_{\rm B}$), emitting electrons/positrons ($P_{\rm e}$), radiation ($P_{\rm r}$), and cold protons ($P_{\rm p}$) in the jet. The cold proton number density is assumed to be 10\% of the electron number density. 

No matter where the radiation region is, we find that the jet kinetic power $ P_{\text{kin}} = P_{\rm e} + P_{\rm B} + P_{\rm p}$ is higher than $P_{\rm r} $ or $L_{\rm disk} $, ensuring the continuous and stable ejection of jet. This is consistent with the conclusions of \citet{2017ApJ...851...33P, 2014Natur.515..376G, 2023ApJS..268....6C}. We also find that $P_{\rm e}$ is lower than $ P_{\rm B}$, which is likely due to high-redshift blazars exist in a photon-rich environment, allowing the electrons within the blob to cool more efficiently via the IC process \citep{2023ApJS..268...23F}.

The Compton dominance (CD) is used to characterize the proportion of IC scattering relative to synchrotron radiation. As shown in Table \ref{tab3}, we find that for high-redshift blazars, it is challenging to distinguish between particle-dominated or Poynting flux-dominated jets directly solely through parameter fitting, consistent with the conclusions of \citet{2023ApJS..268...23F}. While equipartition might exist between the magnetic field and nonthermal electron energy density, this balance is disrupted when considering the protons in the jet, whose content/proportion remains uncertain \citep{2020MNRAS.496.5518S}. Future observations, particularly those investigating polarization properties and variability timescales, could provide additional constraints on the jet content and structure.

\begin{figure}[h]
\centering
\includegraphics[width=0.48\textwidth]{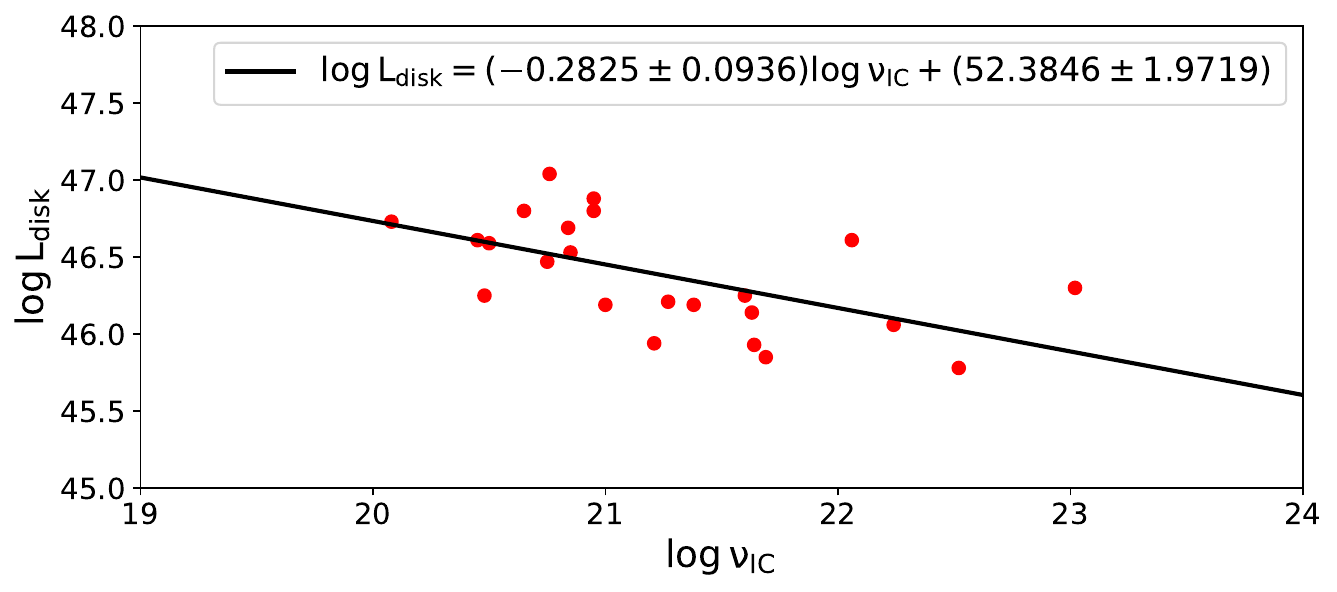}
\includegraphics[width=0.48\textwidth]{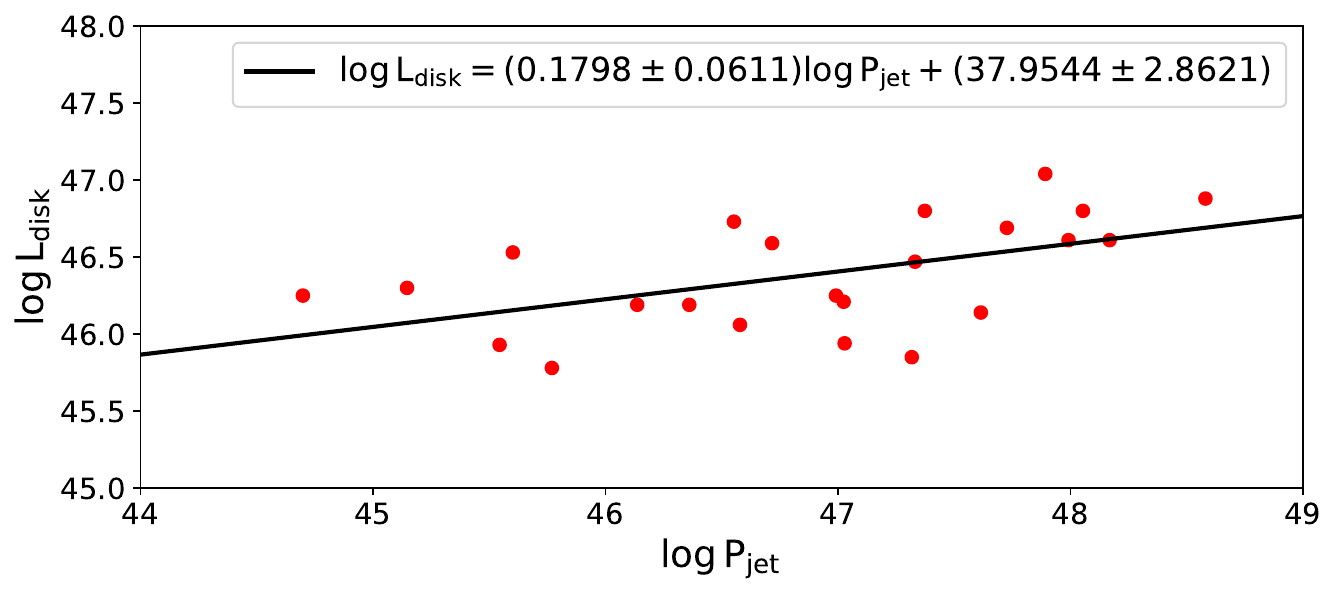}
\includegraphics[width=0.48\textwidth]{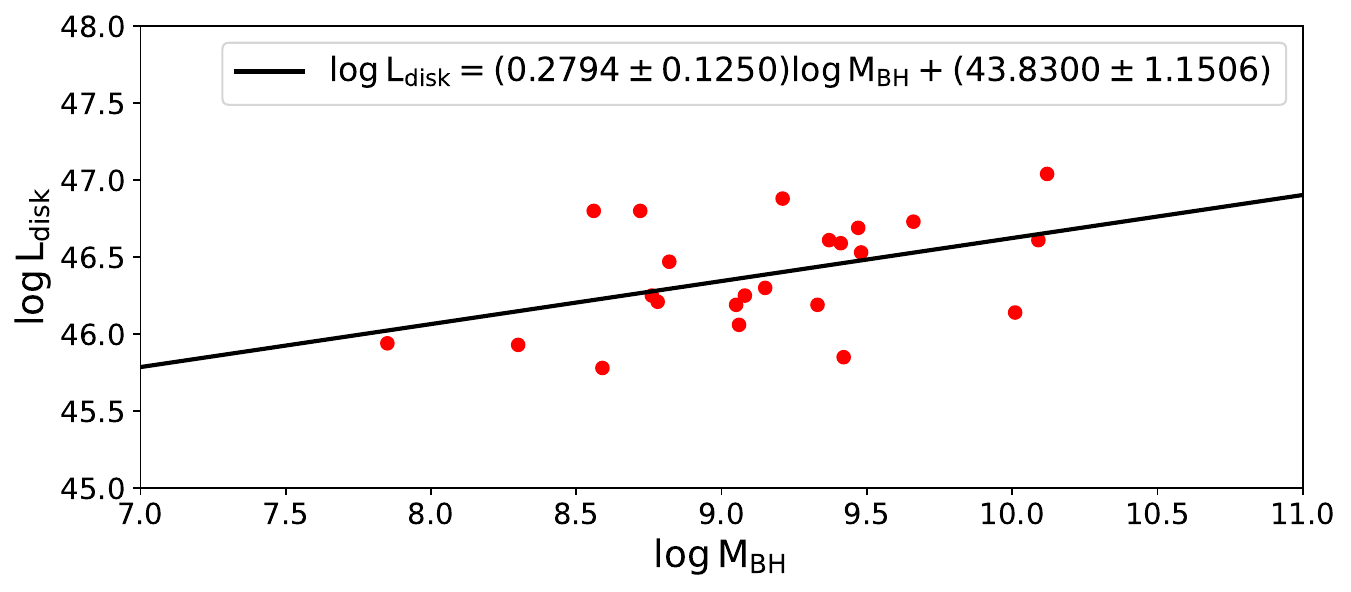}
\caption{Linear relationships between the accretion disk luminosity and IC peak frequency ($r = -0.5774$, $p = 0.0039$), jet power ($r = 0.5245$, $p = 0.0102$), and BH mass ($r = 0.4418$, $p = 0.0348$) are shown.
\label{fig2}}
\end{figure}

\subsection{Correlation among the Parameters}

We used kernel density estimation to determine the probability density function of each parameter, as shown in Figure \ref{fig1}. The figure indicates that the parameters cover similar ranges across the sample. To evaluate the parameter distributions further, we applied the Shapiro--Wilk normality test. The corresponding Shapiro--Wilk statistics and chance probability (i.e., $p$-values) are presented in Table \ref{tab4}, which confirm that all of the parameters in the table follow a normal distribution.

\begin{figure}[h]
\centering
\includegraphics[width=0.48\textwidth]{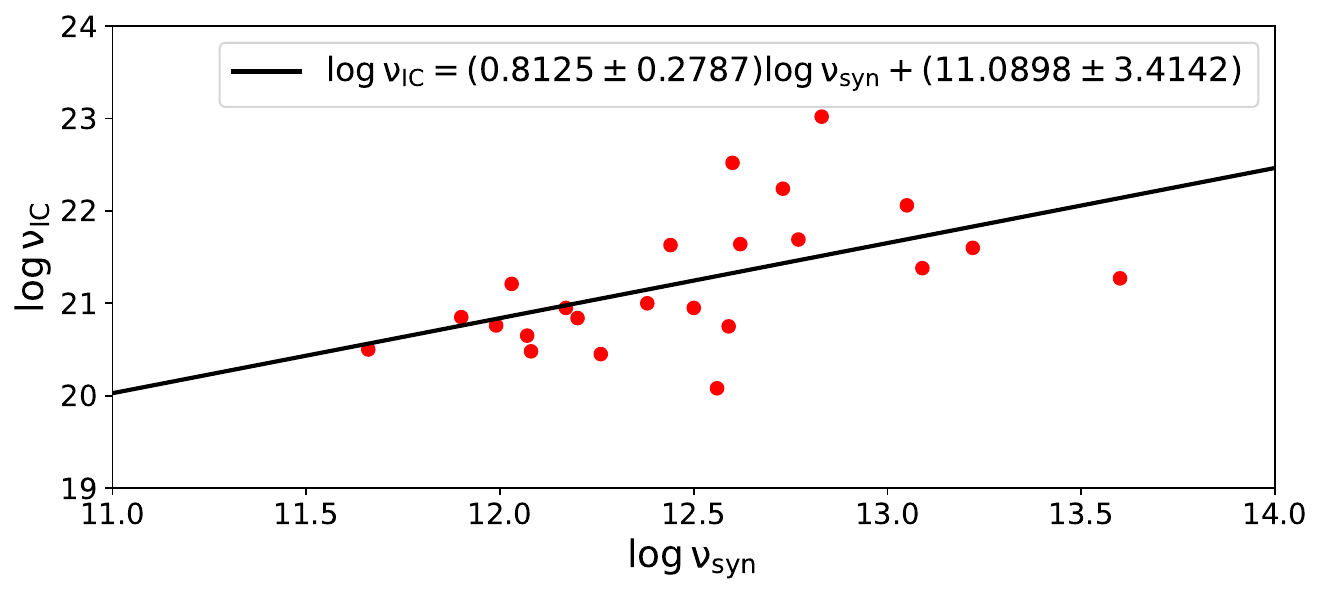}
\includegraphics[width=0.48\textwidth]{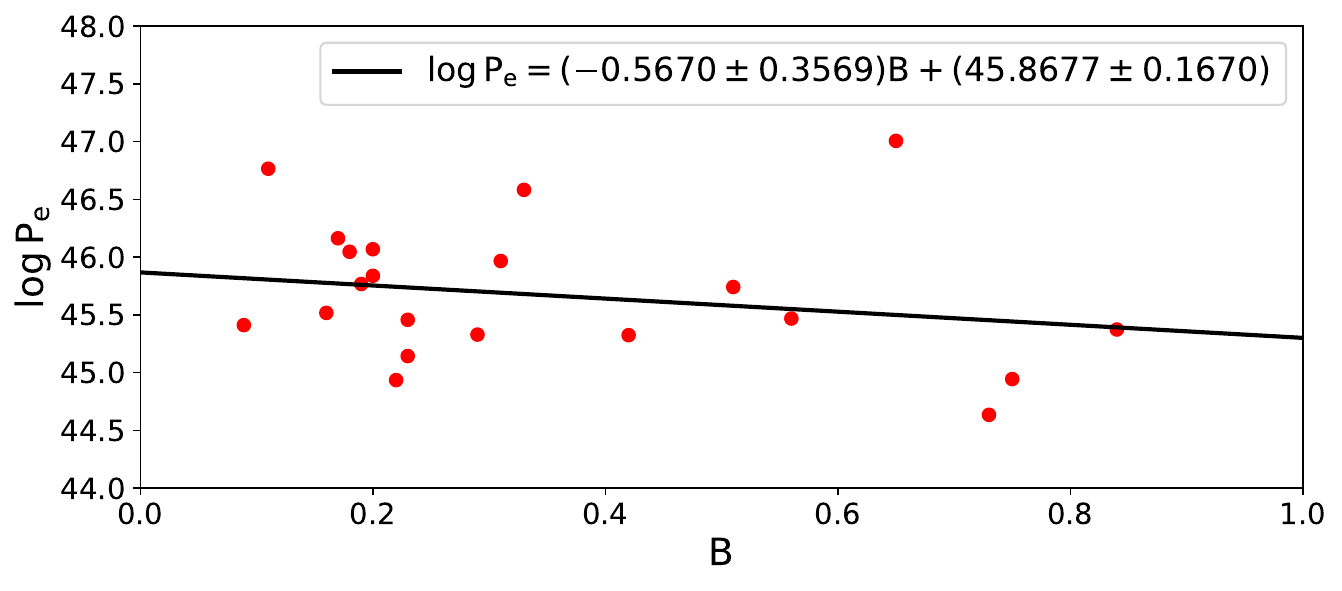}
\caption{Linear relationships between the synchrotron peak frequency and IC peak frequency ($r = 0.5244$, $p = 0.0102$) and magnetic field and electron power ($r = -0.4903$, $p = 0.0175$) are shown.
\label{fig3}}
\end{figure}

Understanding correlations between parameters is crucial for interpreting the physical processes in blazars. Above, we found that these parameters all passed the normality test. Next, we proceeded to compute the Pearson correlation coefficient \footnote{When using the Pearson correlation coefficient to measure a linear relationship between two variables, it is recommended that the variables approximately follow a normal distribution.} ($r$) to analyze potentially correlated parameter pairs. Figures \ref{fig2} and \ref{fig3} present the correlation fitting curves among selected parameters, which quantify the strength of their linear relationships and the statistical significance of the results. 

The correlation coefficients of the parameter pairs are between 0.4 and 0.7, indicating moderate correlation---there is a clear relationship between the variables, but the data points show considerable scatter, and the chance probability is $p<0.05$---meaning the empirical relations can be obtained by linear regression analysis.

\section{Discussion and Conclusion\label{sec4}}

\subsection{High-energy Radiation Properties}

The $\gamma$-ray fluxes we obtained range from $(0.49{-}113.92) \times 10^{-9}\ \mathrm{ph\ cm^{-2}\ s^{-1}}$, which is comparable to that of their low-redshift counterparts \citep{2024ApJS..275...41Z}. The fact that such distant blazars can still receive such high fluxes may be that early AGNs have a higher external photon field density, which leads to increased IC scattered luminosities. Additionally, selection effects can also provide an explanation, as high-redshift AGNs must be powerful enough to be detected. Among them, the object with the highest flux is B3~1343+451, and the lowest is PMN~J0833-0454.

The synchrotron peak frequencies are in the range of $(0.46{-}39.81) \times 10^{12}\ \mathrm{Hz}$, which is typical for low-synchrotron-peaked blazars. The accretion disk luminosities are between $(0.69{-}10.96) \times 10^{46}\ \mathrm{erg\ s^{-1}}$, with an average value of $3.29 \times 10^{46}\ \mathrm{erg\ s^{-1}}$; these values are characteristic of powerful blazars. We find that the jet powers are larger than their disk luminosities, which is consistent with result of \citet{2014MNRAS.441.3375X,2010MNRAS.402..497G}. The BH masses in our sample are generally $> 10^{8.5} \ M_{\rm \odot}$, because blazars with larger disk luminosities have larger BH masses \citep{2010MNRAS.402..497G, 2013MNRAS.431.1840P}. The disk luminosities of these sources do not exceed their Eddington luminosities, however, where the ratio $L_{\rm disk}/L_{\rm Edd} \sim 0.01{-}0.37$. This suggests that high-redshift blazars are likely in a sub-Eddington accretion state, where the typical standard thin disk, i.e., the Shakura--Sunyaev disk, can sufficiently explain their accretion structure \citep{1973A&A....24..337S, 2014ARA&A..52..529Y,2024ApJ...976...78X}.

We find that, in general, the closer the radiation region is to the central BH, the stronger its magnetic field. For high-redshift blazars, we find that the energy density of external photons from the MT---mainly IR photons from the MT---is approximately $(10^{-3}{-}10^{-2})\ \mathrm{erg\ cm^{-3}}$, while the energy density of the external photon field near the BLR---mainly UV photons from the BLR---is $(10^{-1}{-}10^{0})\ \mathrm{erg\ cm^{-3}}$. This is related to the stronger UV photon field in the BLR and the weaker IR photon field in the MT region. Additionally, these energy densities of high-redshift FSRQs are generally larger than their low-redshift counterparts \citep{2024ApJS..271...27Z}.

When the radiation region is near the MT, the EED spectral index is $\sim2$, consistent with shock acceleration theories. Near the BLR, the EED spectral index is typically $>4$, which is also within the expected range from standard particle acceleration theories \citep{2001MNRAS.328..393A,2002A&A...394.1141O,2005PhRvL..94k1102K,2005ApJ...621..313V}. Moreover, since the kinetic power in both radiation region positions is greater than the radiation power, the jets can exist stably, making it impossible to distinguish the radiation region position directly based solely on the fitted parameters. However, since the chi-square ($\chi^2$) values obtained when the radiation region is near the MT are generally smaller, we infer that the radiation region may be near the MT, which is assumed in the subsequent analysis.

In the MT case, except for S4~1427+543, PKS~2311-452, PKS~0438-43, PKS~0834-20, and TXS~0222+185, the soft photons in the FSRQ $\gamma$-ray emission can be entirely explained by IR photons from the MT. The total jet power ranges from $(0.05{-}381.75) \times 10^{46}\ \mathrm{erg\ s^{-1}}$, with an average value of $4.47 \times 10^{47}\ \mathrm{erg\ s^{-1}}$, significantly larger than the jet power of low-redshift AGNs \citep{2023ApJS..268....6C}. Whether this result is due to selection effects or other factors, the high-redshift blazars considered here likely have larger kinetic and jet powers than their low-redshift counterparts. The disk luminosities and jet powers are nearly of the same order of magnitude, and the average ratio $\langle L_{\rm disk}/L_{\rm jet} \rangle \sim 2.93$, which is consistent with other studies \citep{1991Natur.349..138R,2003ApJ...593..667M,2010MNRAS.402..497G,2024MNRAS.532.3729L}, but differs from those obtained by \citet{2020MNRAS.498.2594S,2014MNRAS.441.3375X}, possibly due to their omission of proton content in the jets.

\subsection{Disk--Jet Connection Condition}

By analyzing the average parameters and comparing them with their low-redshift counterparts, we can obtain the average properties of high-redshift blazars. Correlation analysis can further uncover the potential physical mechanisms. To explore the possible close connection between jets and accretion, we performed a correlation analysis on several parameters with a normal distribution.

There is a moderate negative correlation between the IC peak frequency ($\nu_{\rm IC}$) and the disk luminosity ($\log L_{\rm disk}$), 
indicating that blazars with higher accretion disk luminosities tend to have lower IC peak frequencies, possibly reflecting evolutionary effects or differences in the physical conditions of the radiation region \citep{2011MNRAS.411..901G}. High-luminosity accretion disks provide stronger external photon fields, leading to more efficient cooling of HE electrons, which reduces the IC peak frequency and increases the IC scattering flux \citep{Fossati1998MNRAS,Prandini2022Galax}.

The positive correlation between jet power ($P_{\rm jet}$) and disk luminosity ($\log L_{\rm disk}$), 
highlights the relationship between the accretion process and the jet launching mechanism, further supporting the accretion--jet connection theory \citep{2003ApJ...593..667M, 2011MNRAS.416..216V,Ghisellini2013MNRAS,2024MNRAS.527.2672S}, with disk luminosity possibly dominating over jet power \citep{2014Natur.515..376G,2022PhRvD.106f3001R}. Here, for high-redshift blazars, we have also found evidence supporting the accretion--jet connection.

Several well-established correlations further validate the validity of our results. First, a positive correlation between BH mass ($M_{\rm BH}$) and disk luminosity ($L_{\rm disk}$) is observed, as expected by \citet{2010MNRAS.402..497G, 2013MNRAS.431.1840P}, since more massive BHs can accrete more material, leading to higher luminosities. Next, a positive correlation exists between the synchrotron peak frequency ($\nu_{\rm syn}$) and IC peak frequency ($\nu_{\rm IC}$). When $\nu_{\rm syn}$ increases, corresponding to a larger $\Gamma$, IC-scattered photons are boosted to higher energies. This is also consistent with radiation models where emission at both peaks originates from the same population of electrons \citep{2020OAst...29..168I}. Third, the negative correlation between $B'$ and $\log P_{\rm e}$ arises because the magnetic field strength is related to synchrotron power. Stronger magnetic fields enhance synchrotron cooling, thereby affecting the EED and electron power.

Our analysis of high-redshift blazars provides valuable insights into the radiation properties of their jets, which exhibit higher $\gamma$-ray luminosities and softer spectral indices than their lower-redshift counterparts. By constructing and analyzing the average SEDs of a sample of high-redshift blazars ($z>2.5$), we reduced the uncertainties caused by lower-quality data typical of individual high-redshift blazars. 

In our analysis, we found that the majority of the $\gamma$-ray seed photons are likely IR photons from the MT, and hence the radiation region is more likely located near the MT. Compared to their low-redshift counterparts, high-redshift blazars exhibit higher $\gamma$-ray luminosities, jet power, kinetic power, energy densities, BH mass, and accretion disk luminosities.

All objects have an accretion disk luminosity lower than their Eddington luminosity, indicating sub-Eddington accretion. The accretion structure can be explained by the Shakura--Sunyaev disk model. The EED spectral index is $\sim2$, and the electron acceleration mechanism is consistent with diffusive shock acceleration of nonrelativistic electrons. The $\nu_{\rm IC}{-}L_{\rm disk}$ correlation supports the idea that early AGNs have a higher external photon field density. The $P_{\rm jet}{-}L_{\rm disk}$ we observed correlation further supports the accretion--jet connection.

\begin{acknowledgments}
This work was partly supported by the National Science Foundation of China (grant nos. 12263007 and 12233006) and the High-level Talent Support Program of Yunnan Province.
\end{acknowledgments}

\clearpage

\appendix

\section{$\gamma$-ray band Spectral Energy Distributions \label{A} }

In this appendix, the $\gamma$-ray band SEDs of the objects in our high-redshift blazar sample are presented.

\begin{figure*}[h]
\centering
\includegraphics[width=0.32\textwidth]{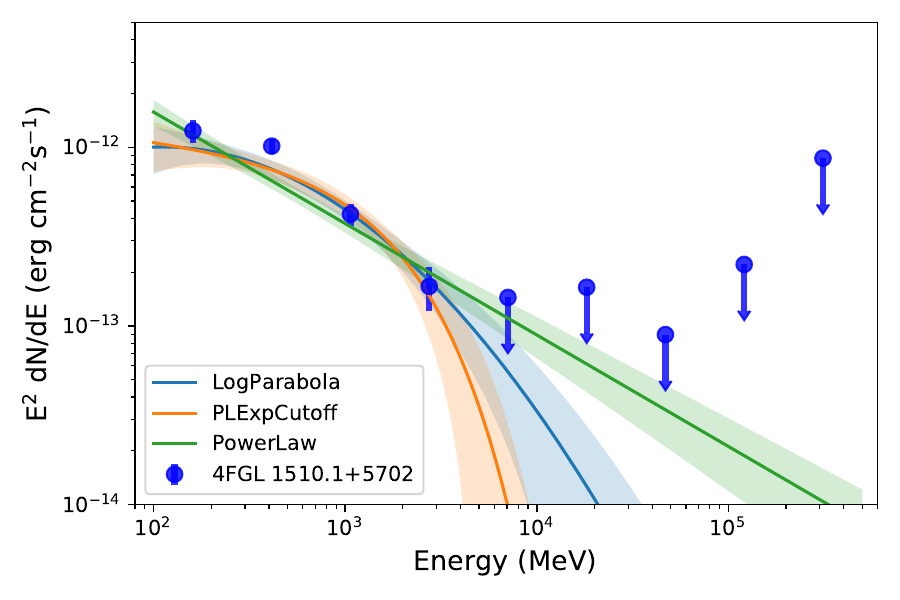}
\includegraphics[width=0.32\textwidth]{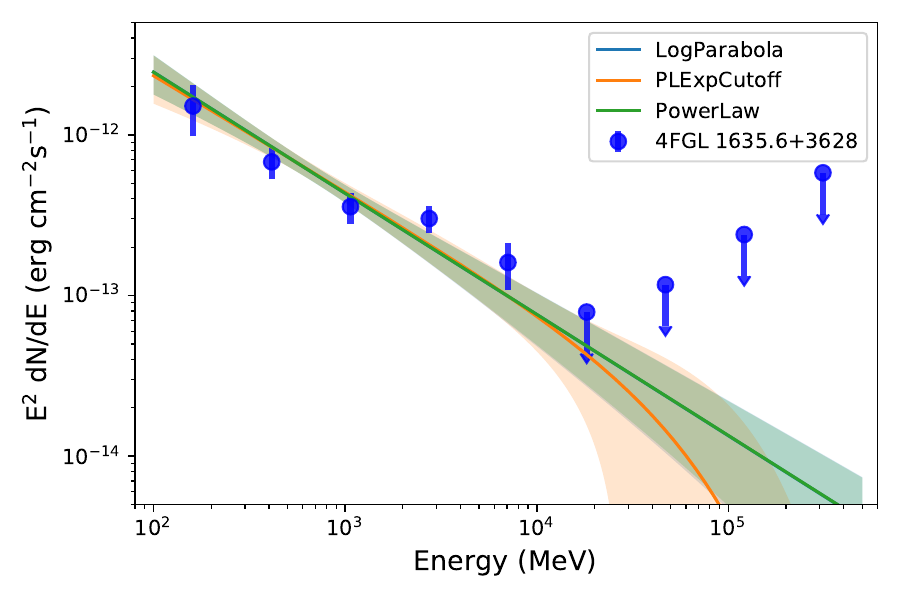}
\includegraphics[width=0.32\textwidth]{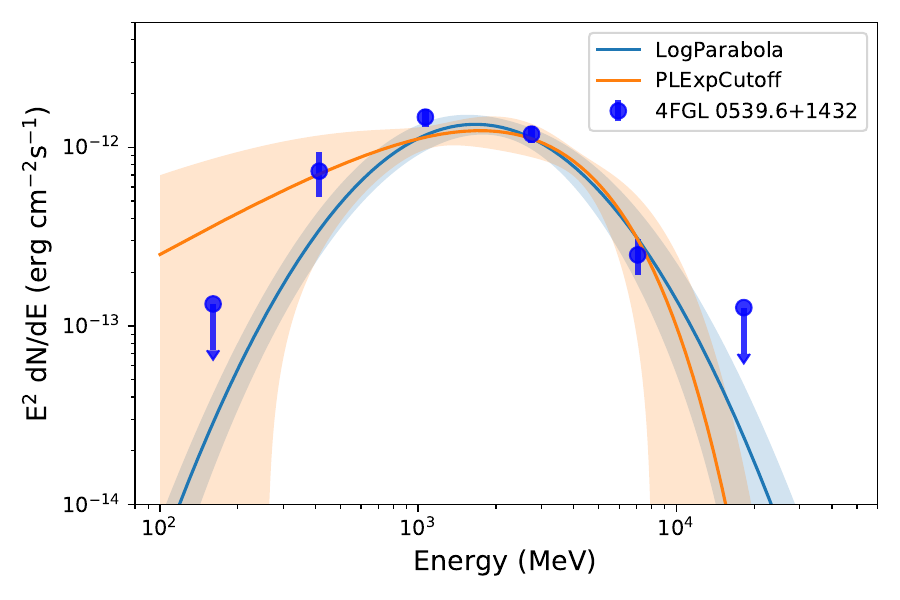}
\includegraphics[width=0.32\textwidth]{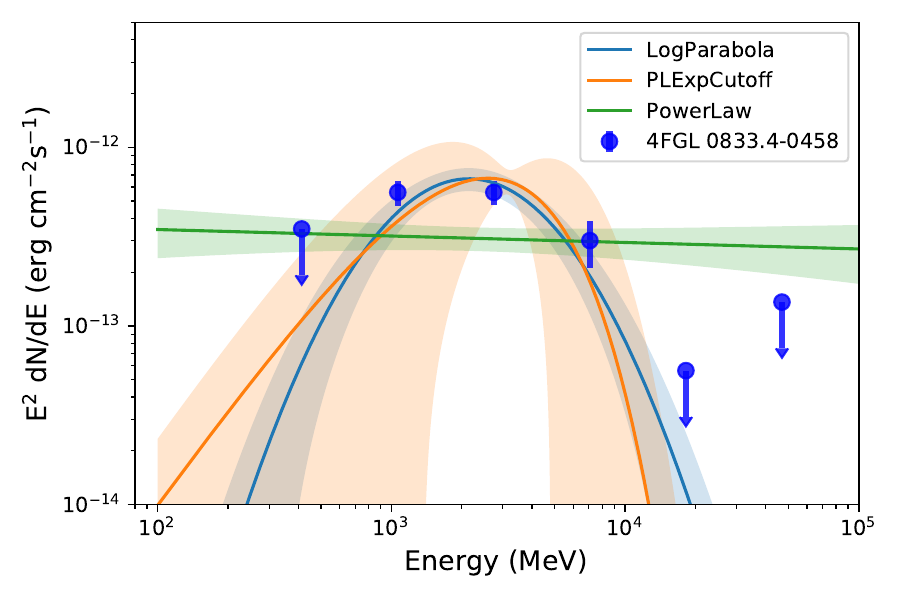}
\includegraphics[width=0.32\textwidth]{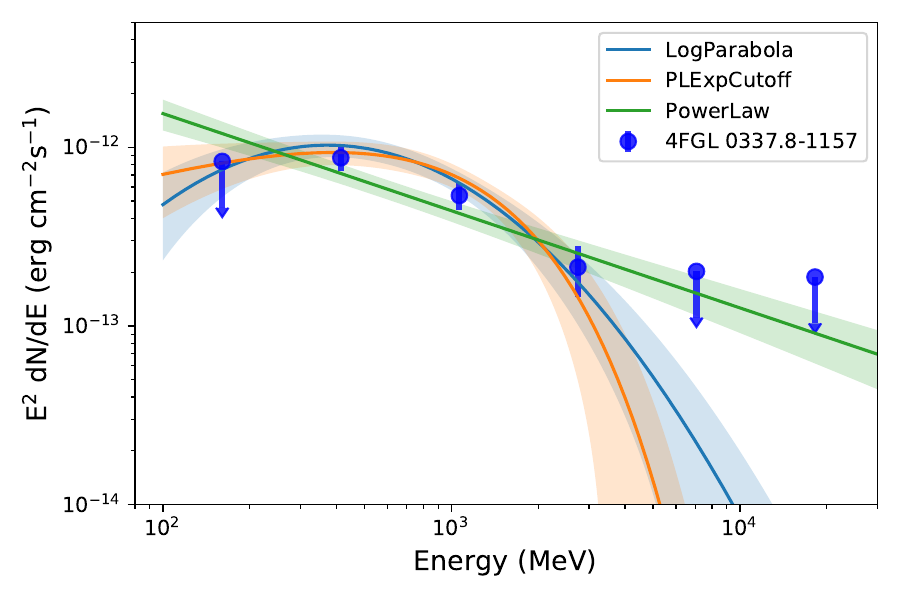}
\includegraphics[width=0.32\textwidth]{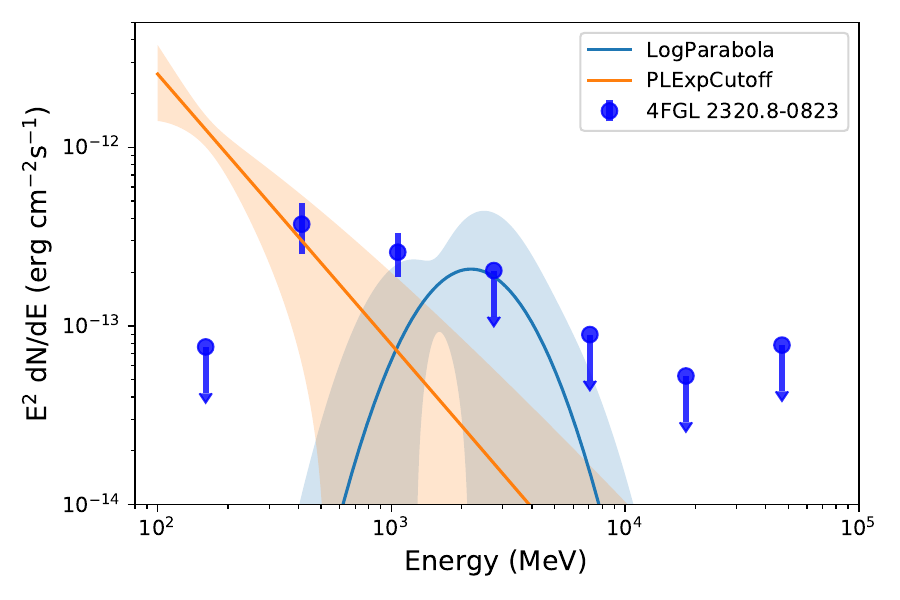}
\includegraphics[width=0.32\textwidth]{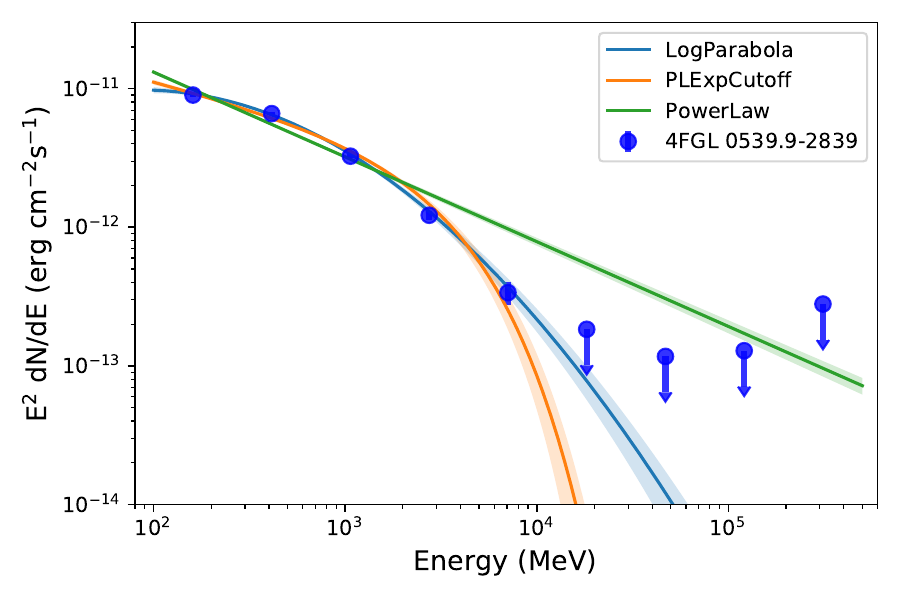}
\includegraphics[width=0.32\textwidth]{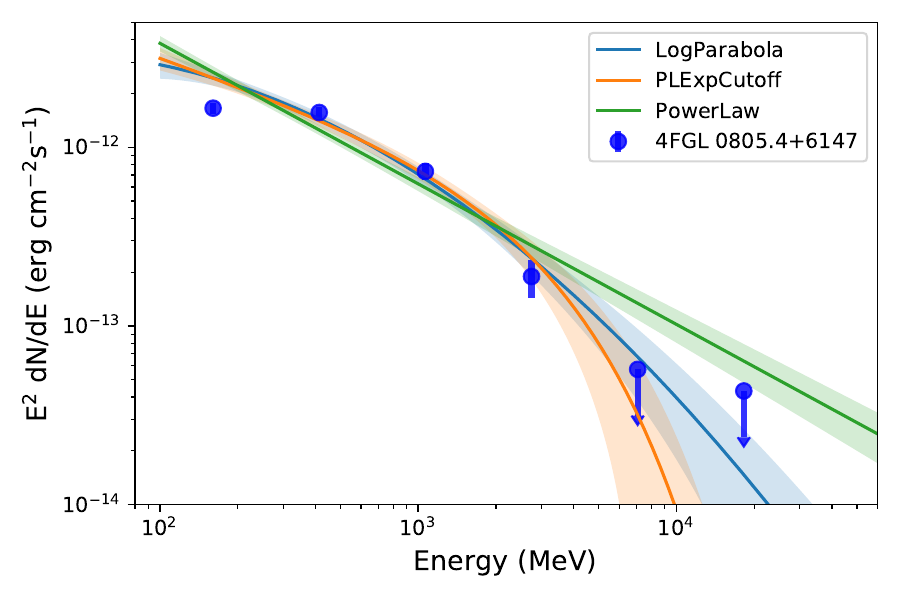}
\includegraphics[width=0.32\textwidth]{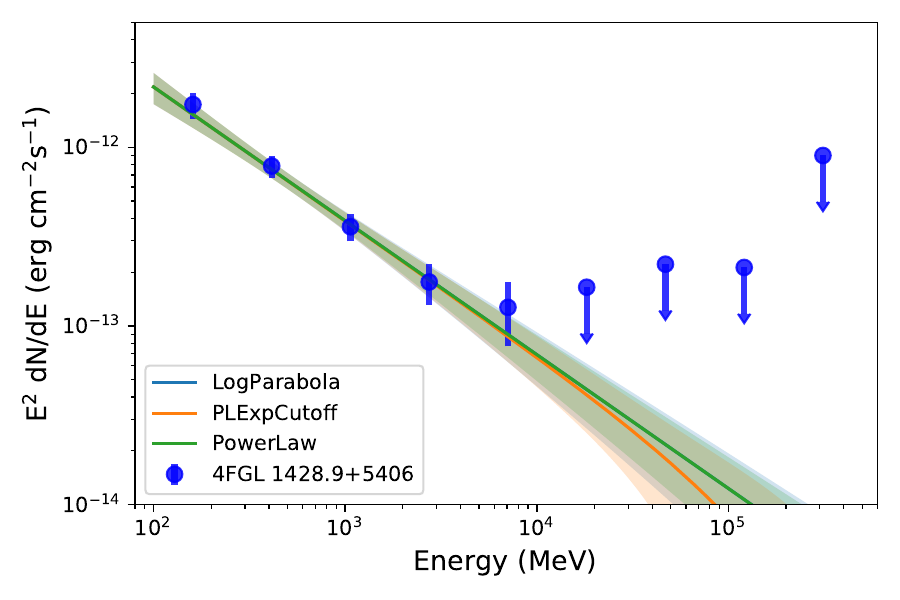}
\includegraphics[width=0.32\textwidth]{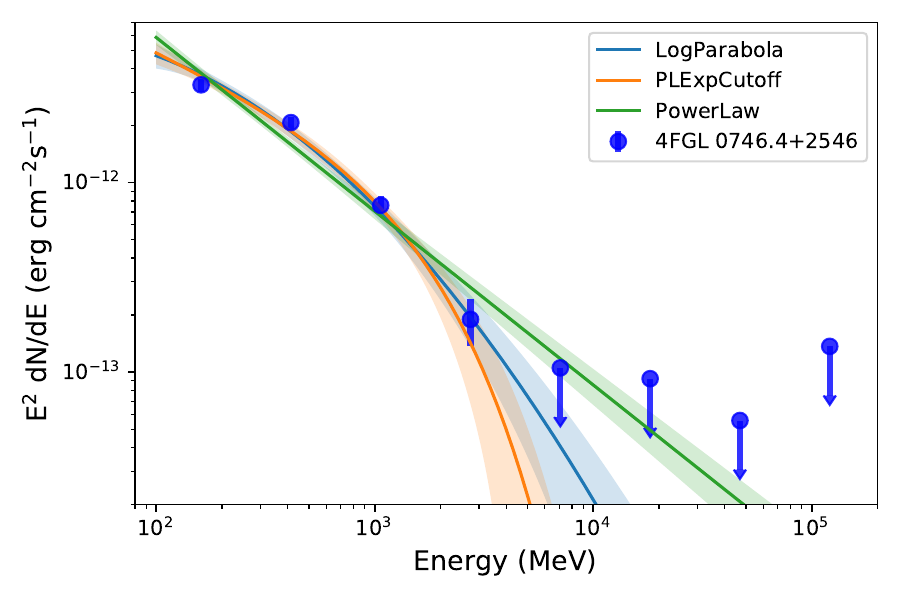}
\includegraphics[width=0.32\textwidth]{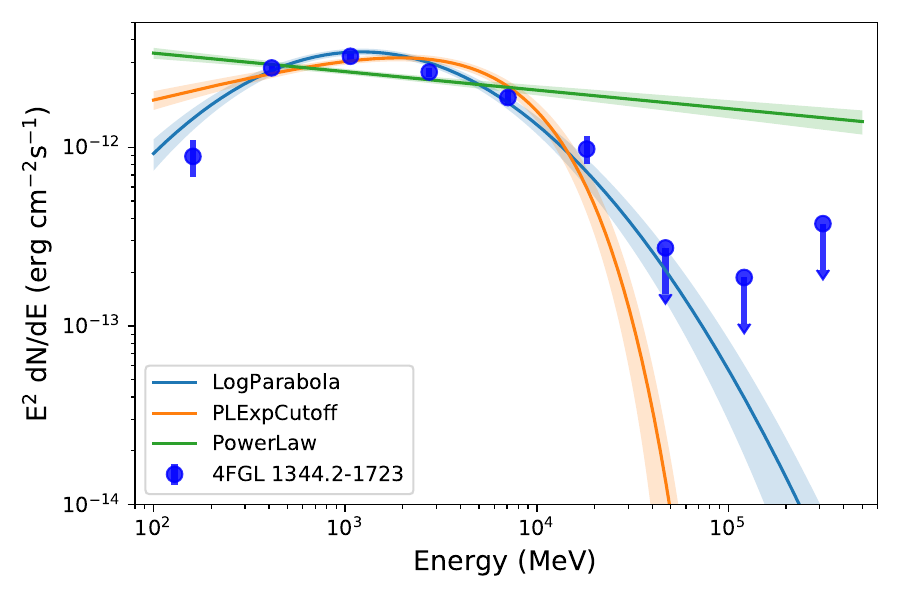}
\includegraphics[width=0.32\textwidth]{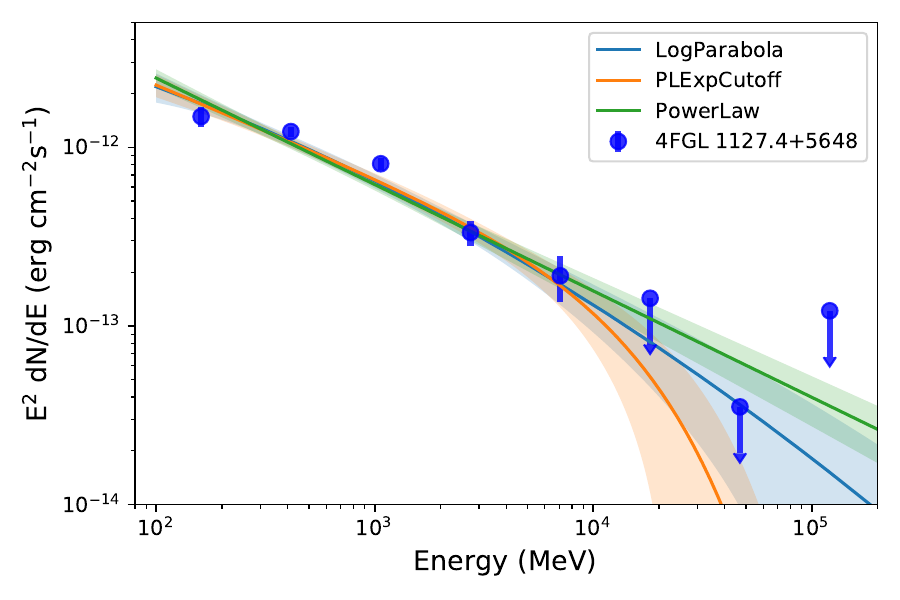}
\caption{$\gamma$-ray spectra of high-redshift blazars. The symbols and lines are the same as shown in Figure \ref{figx}. \label{figxx}}
\end{figure*}

\begin{figure*}
\centering
\includegraphics[width=0.32\textwidth]{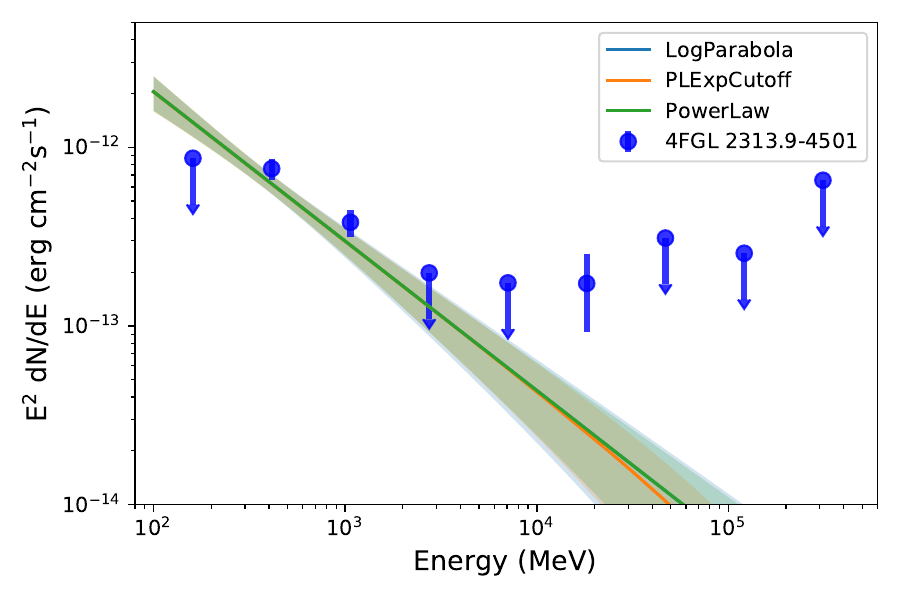}
\includegraphics[width=0.32\textwidth]{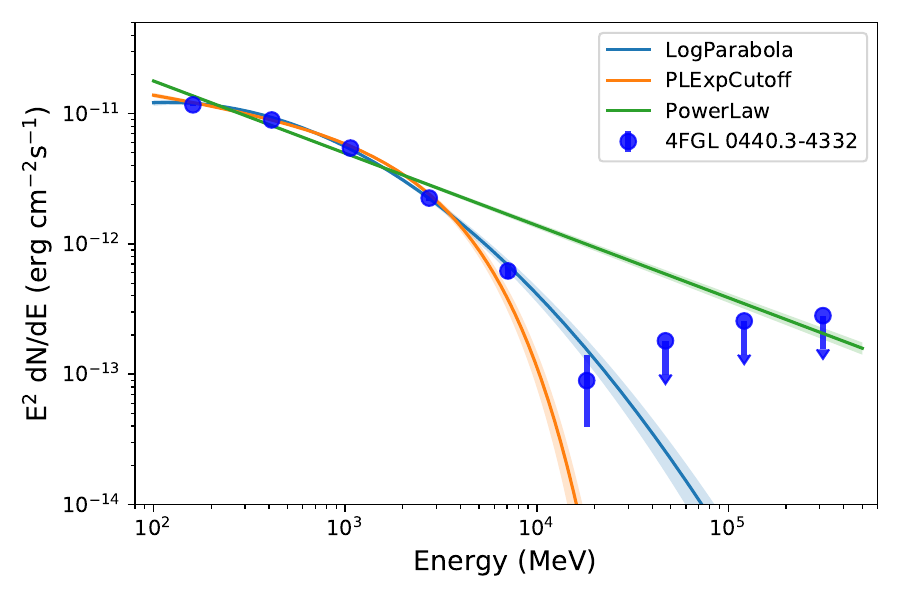}
\includegraphics[width=0.32\textwidth]{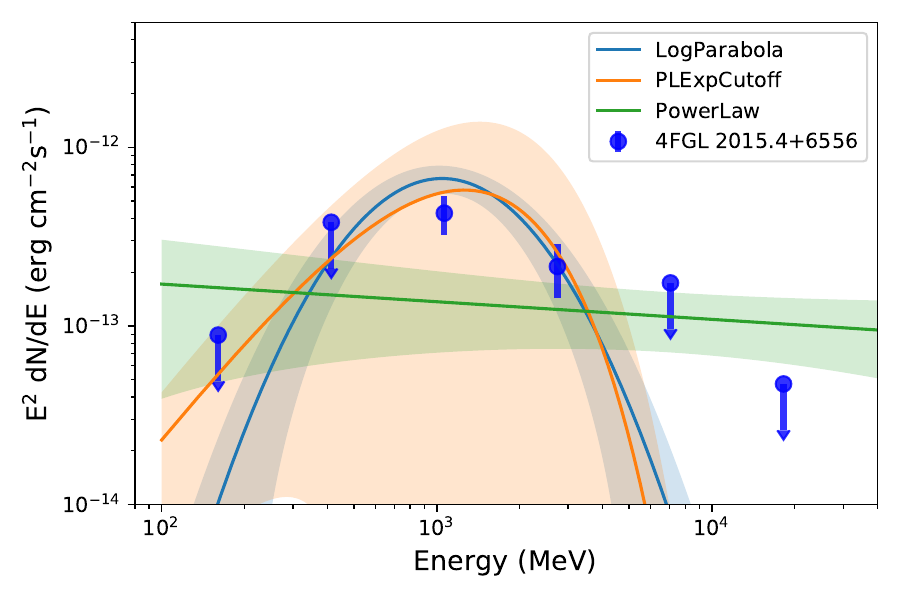}
\includegraphics[width=0.32\textwidth]{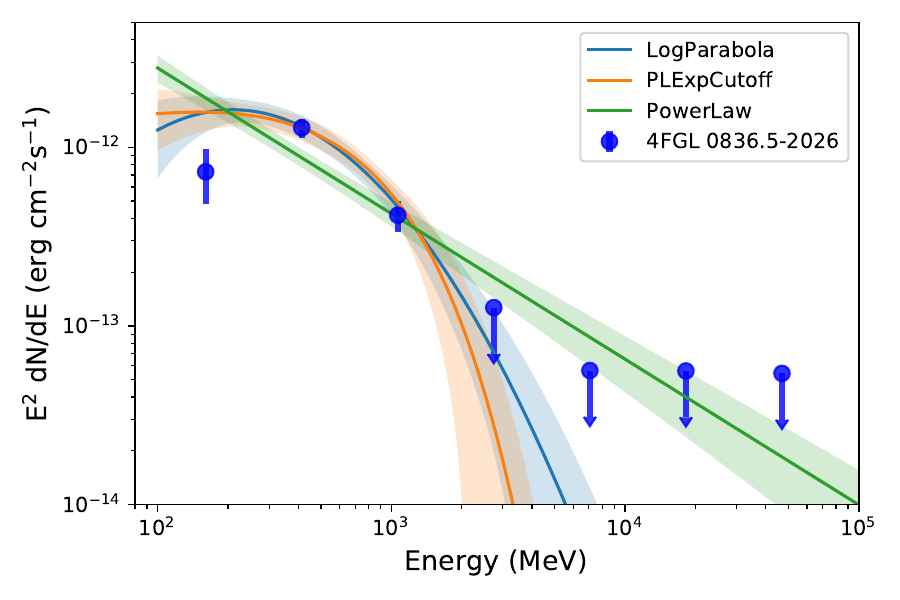}
\includegraphics[width=0.32\textwidth]{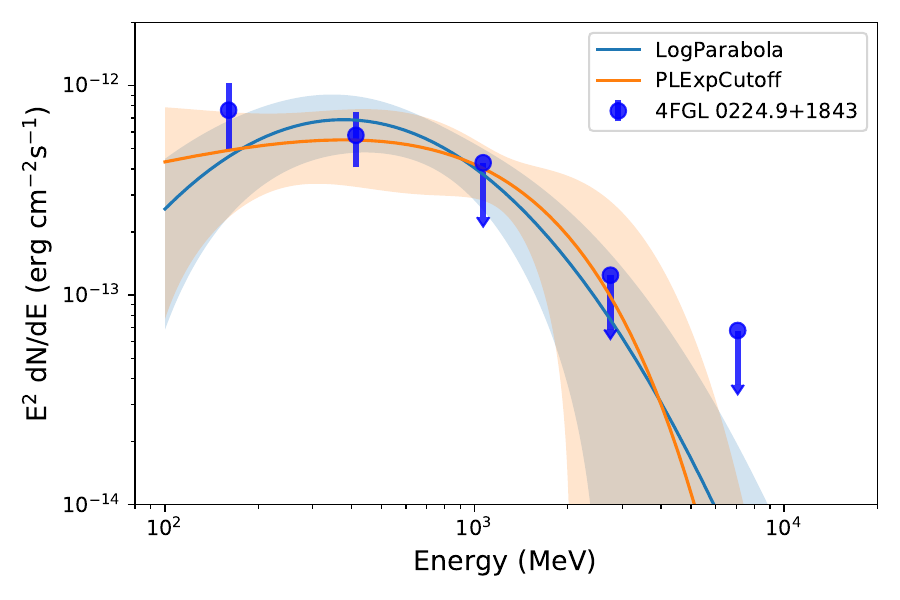}
\includegraphics[width=0.32\textwidth]{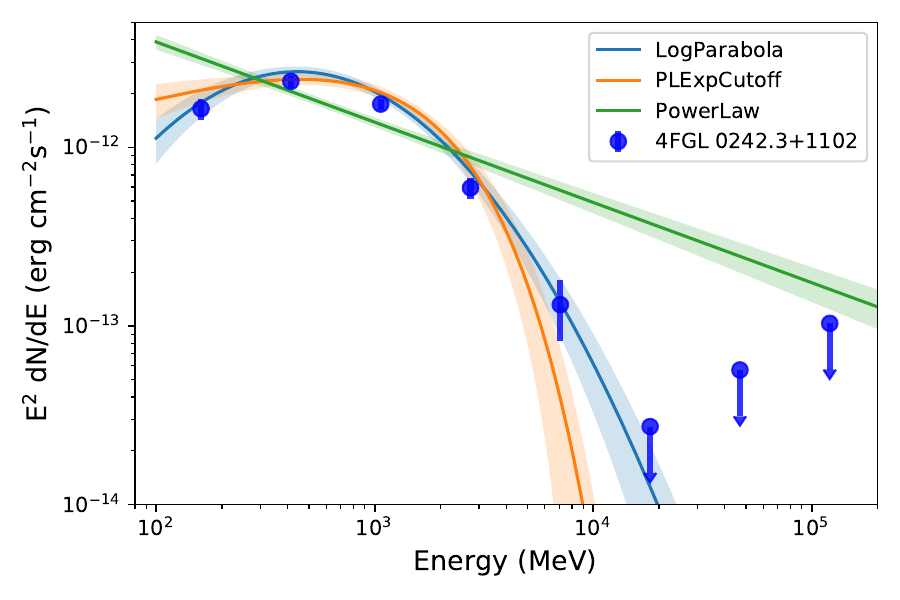}
\includegraphics[width=0.32\textwidth]{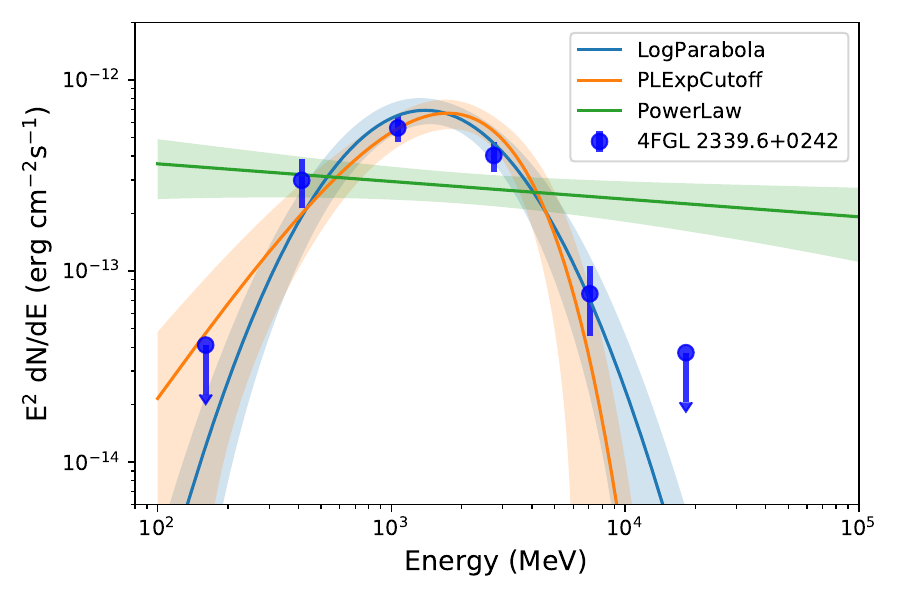}
\includegraphics[width=0.32\textwidth]{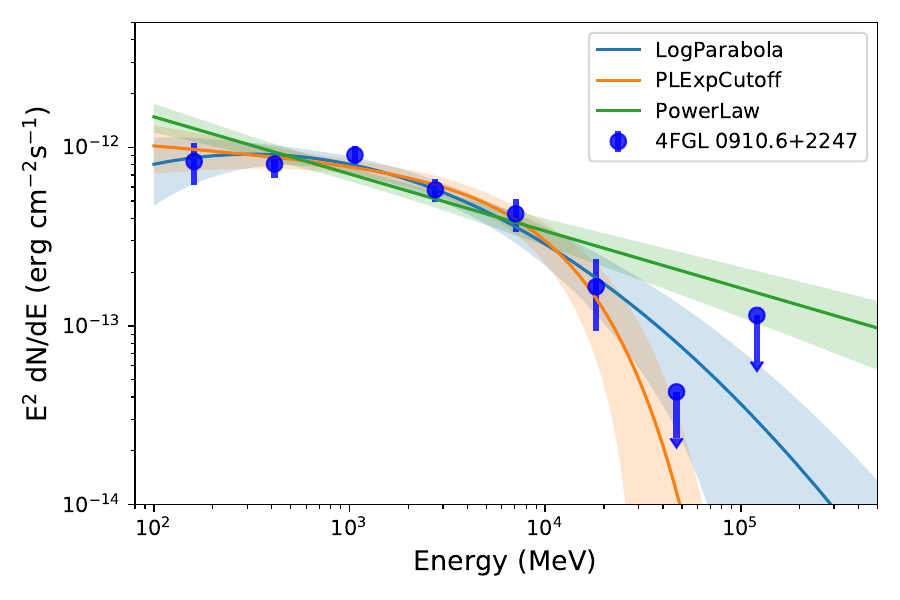}
\includegraphics[width=0.32\textwidth]{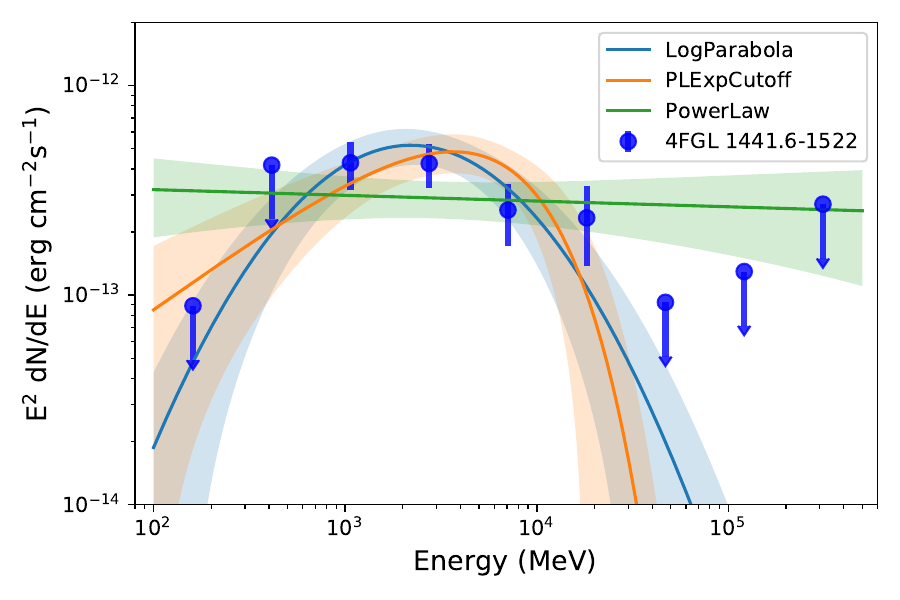}
\includegraphics[width=0.32\textwidth]{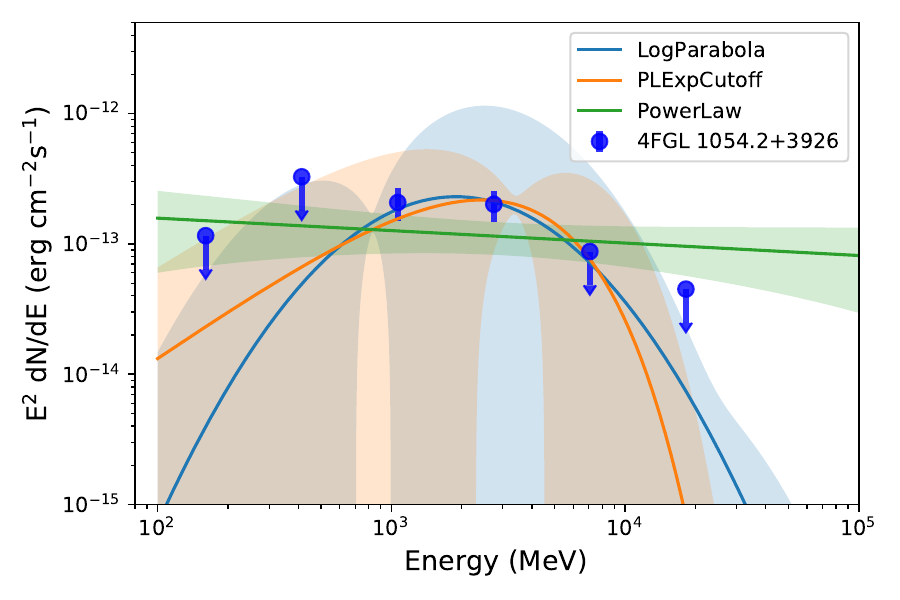}
\includegraphics[width=0.32\textwidth]{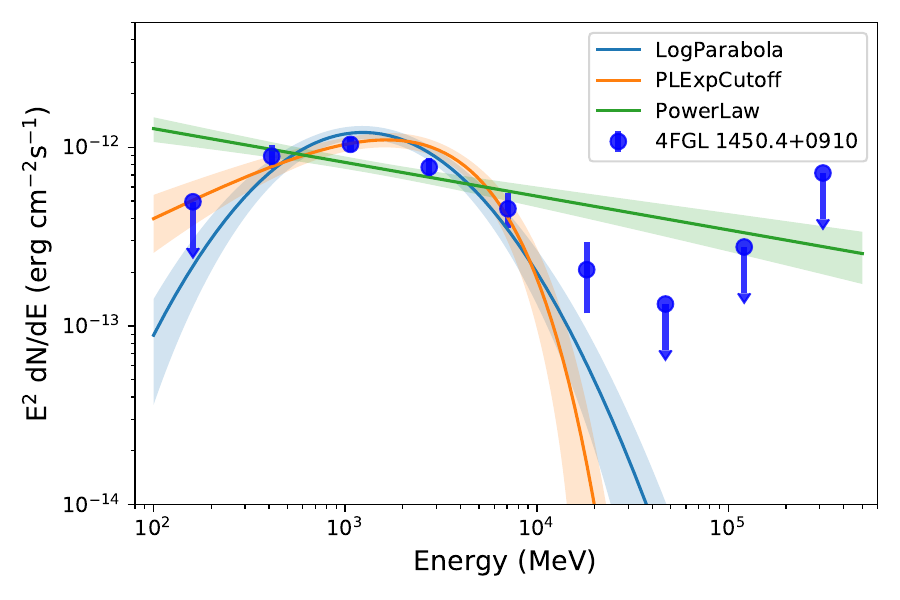}
\includegraphics[width=0.32\textwidth]{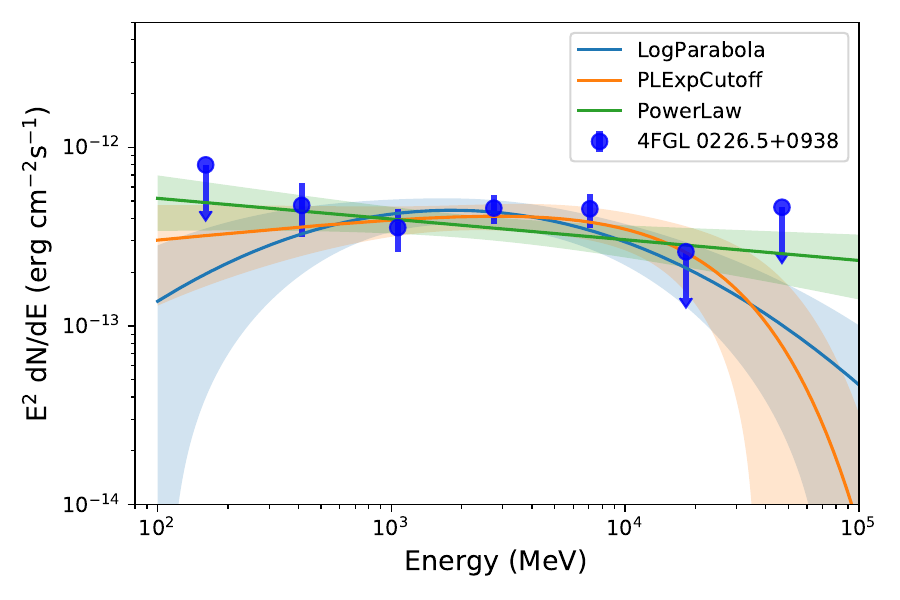}
\includegraphics[width=0.32\textwidth]{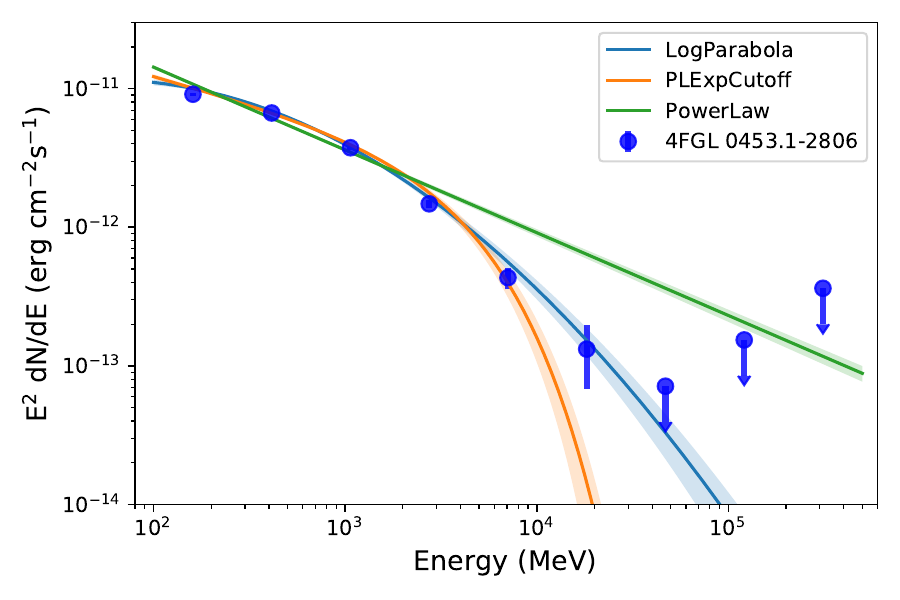}
\includegraphics[width=0.32\textwidth]{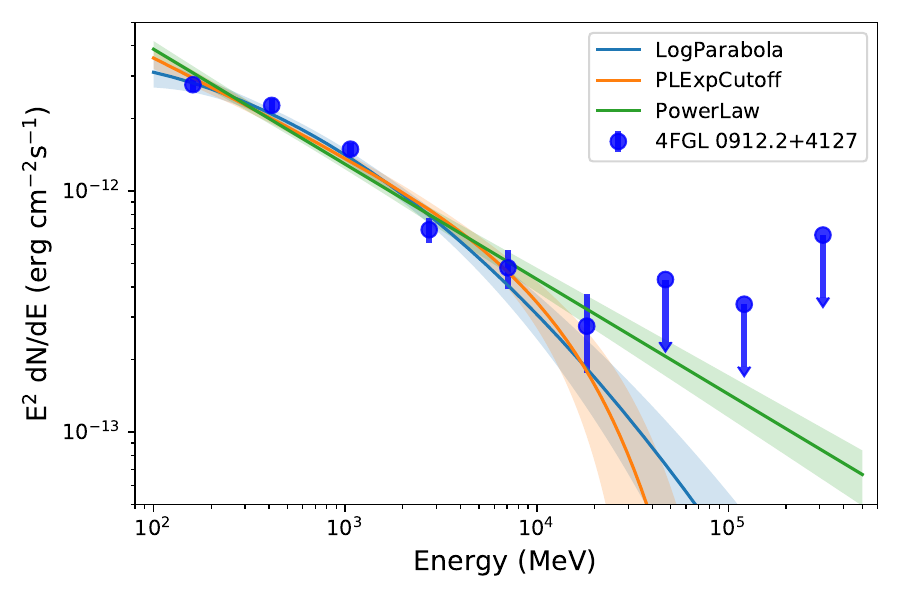}
\includegraphics[width=0.32\textwidth]{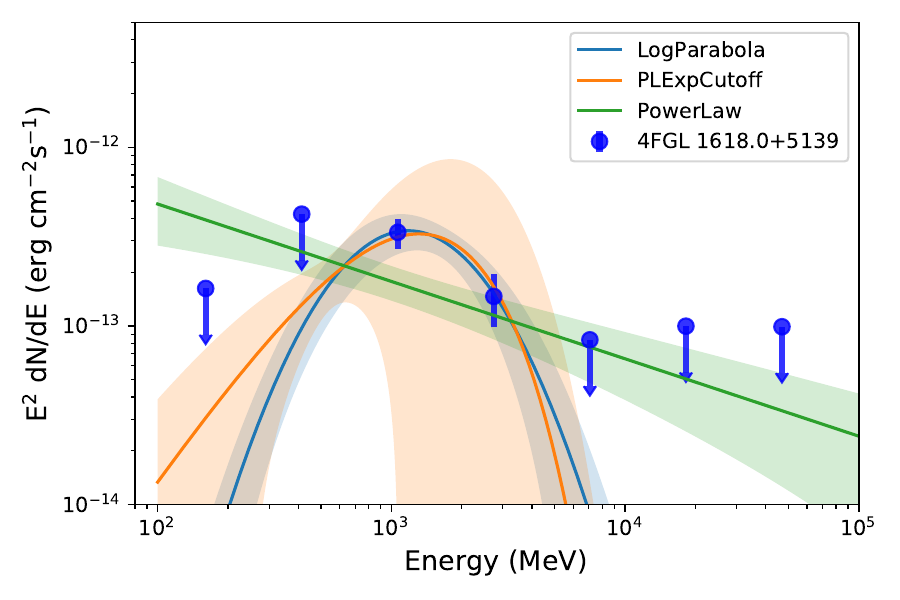}
\includegraphics[width=0.32\textwidth]{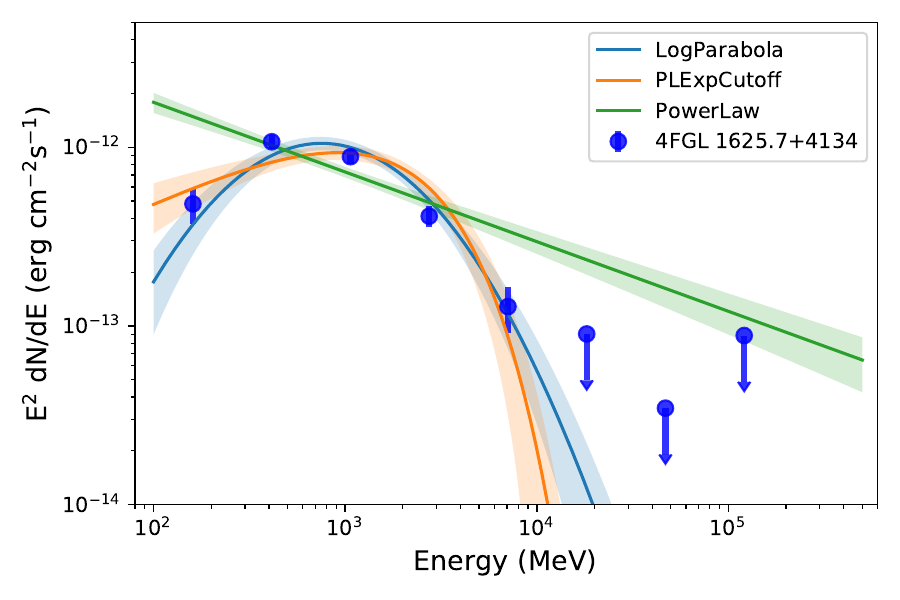}
\includegraphics[width=0.32\textwidth]{34.pdf}
\includegraphics[width=0.32\textwidth]{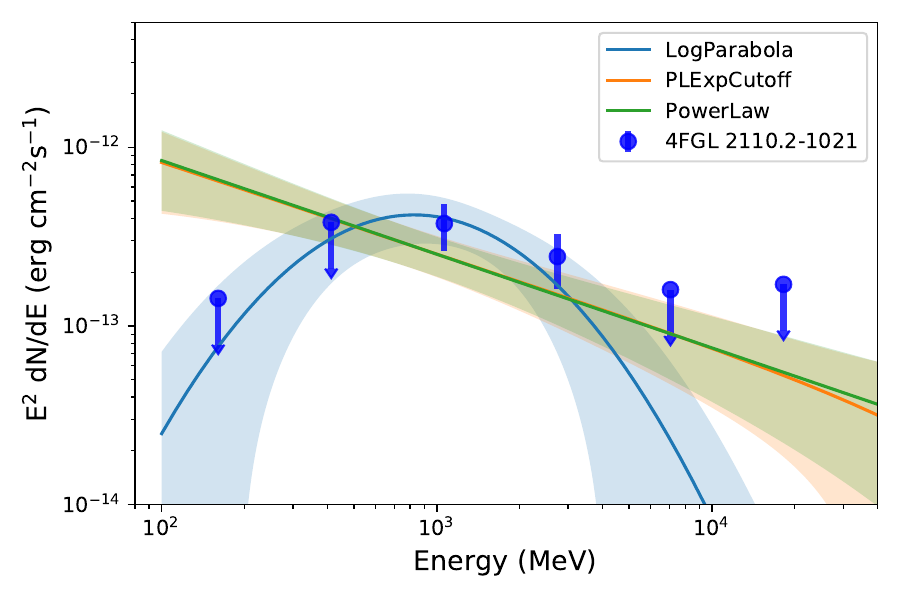}
\caption{$\gamma$-ray spectra of high-redshift blazars. The symbols and lines are the same as shown in Figure \ref{figx}.
\label{figxxx}}
\end{figure*}

\clearpage
\section{Multiwavelength Spectral Energy Distribution Fitting Diagrams\label{B}}

In this appendix, the results of modelling the high-redshift blazars in terms of theoretical SEDs are presented.

\begin{figure*}[h]
\centering
\includegraphics[width=1\textwidth]{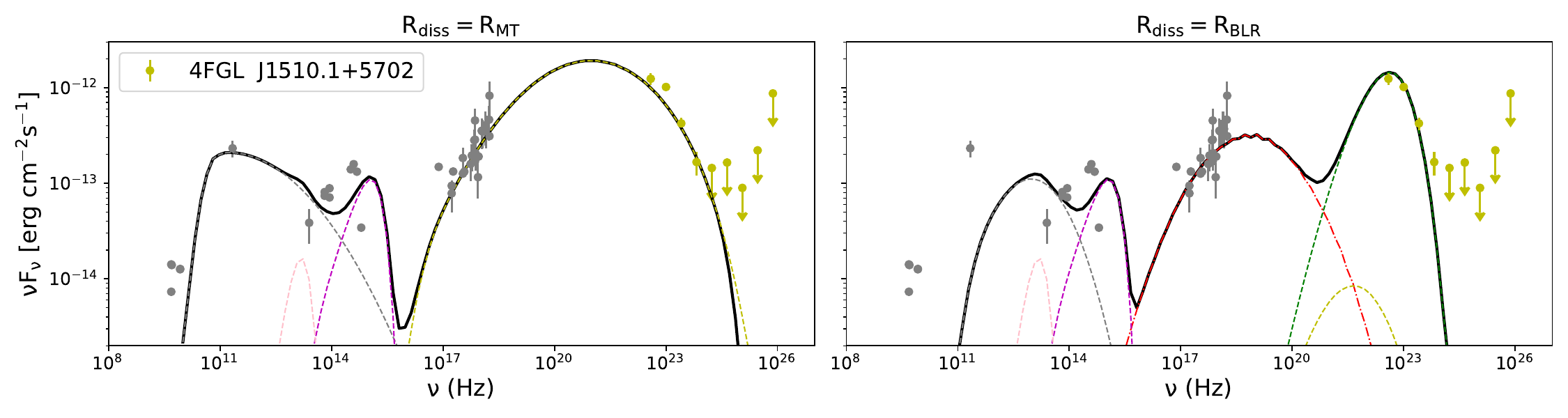}
\includegraphics[width=1\textwidth]{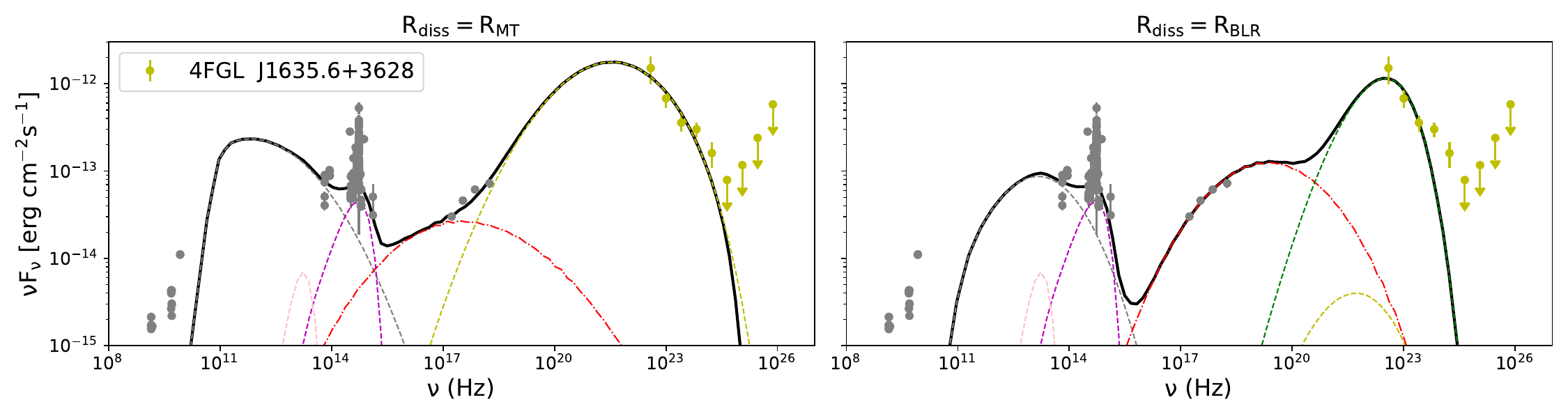}
\includegraphics[width=1\textwidth]{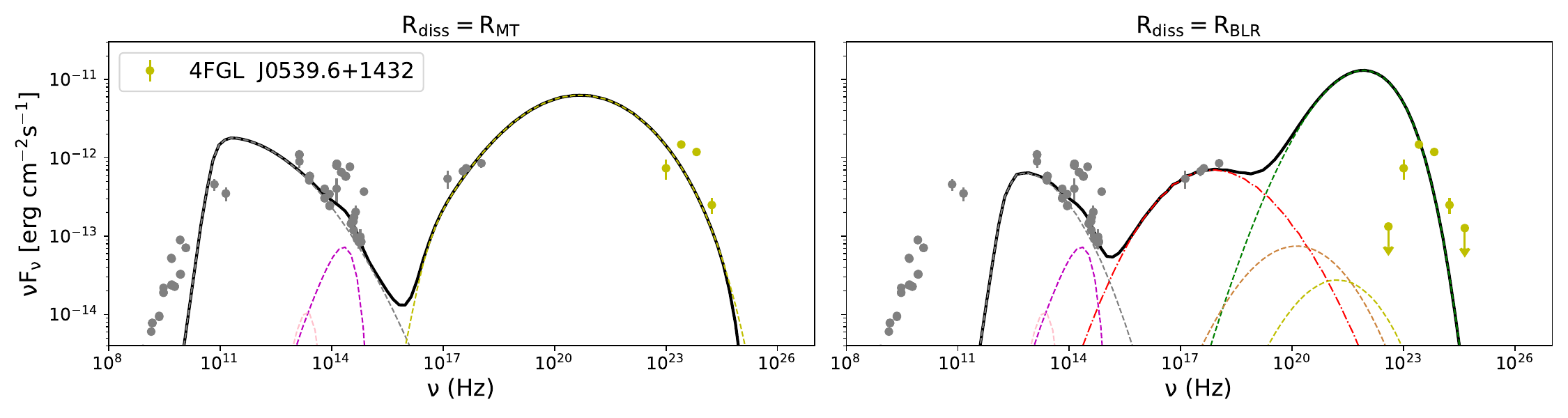}
\caption{Theoretical modeling of the SEDs of 23 high-redshift blazars. The symbols and lines are the same as shown in Figure \ref{figg}.
\label{fig6}}
\end{figure*}

\begin{figure*}[t]
\centering
\includegraphics[width=1\textwidth]{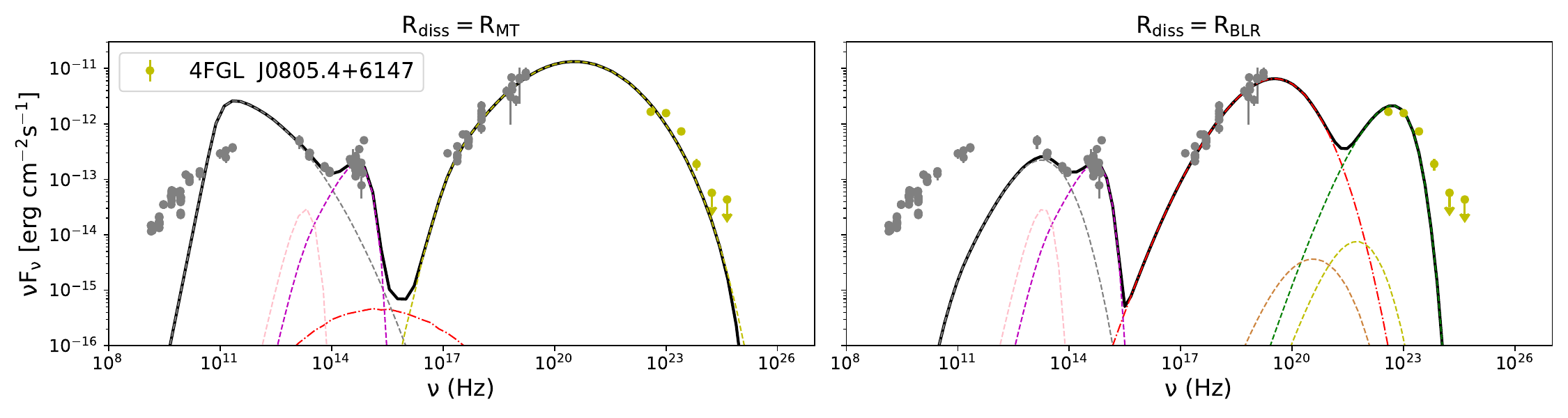}
\includegraphics[width=1\textwidth]{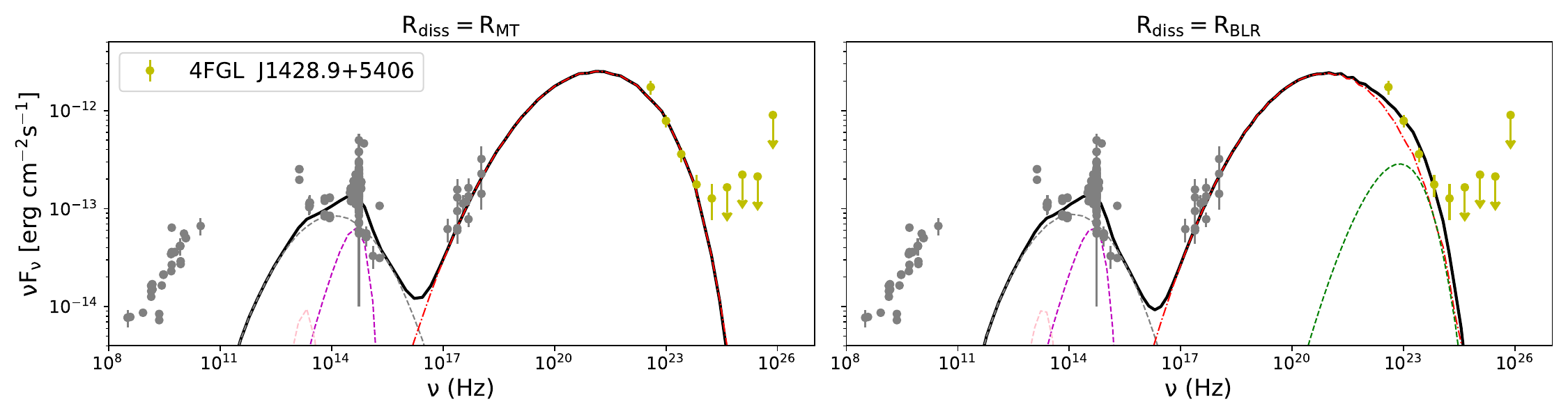}
\includegraphics[width=1\textwidth]{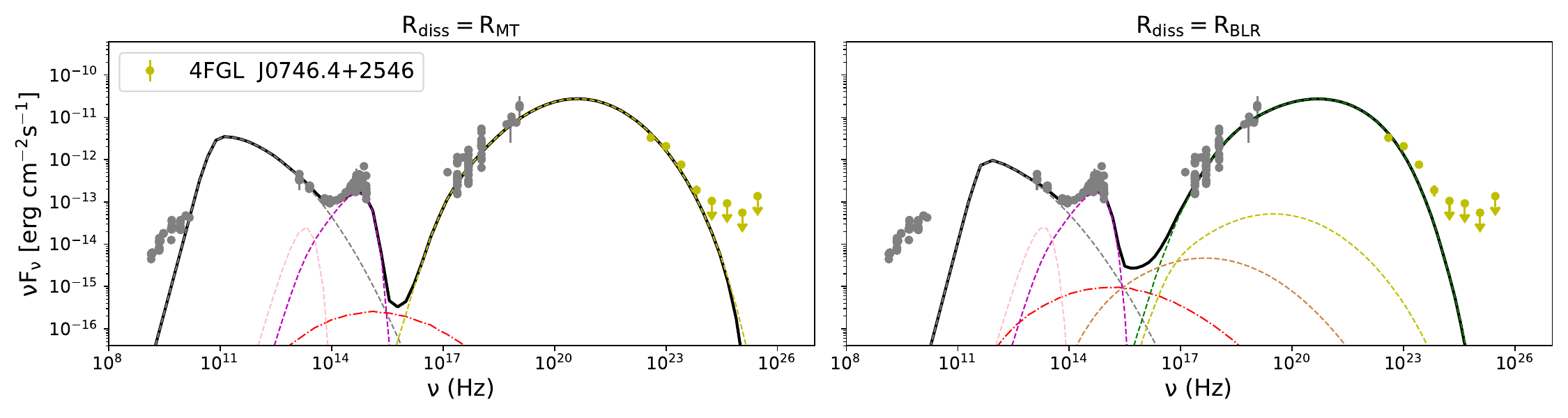}
\includegraphics[width=1\textwidth]{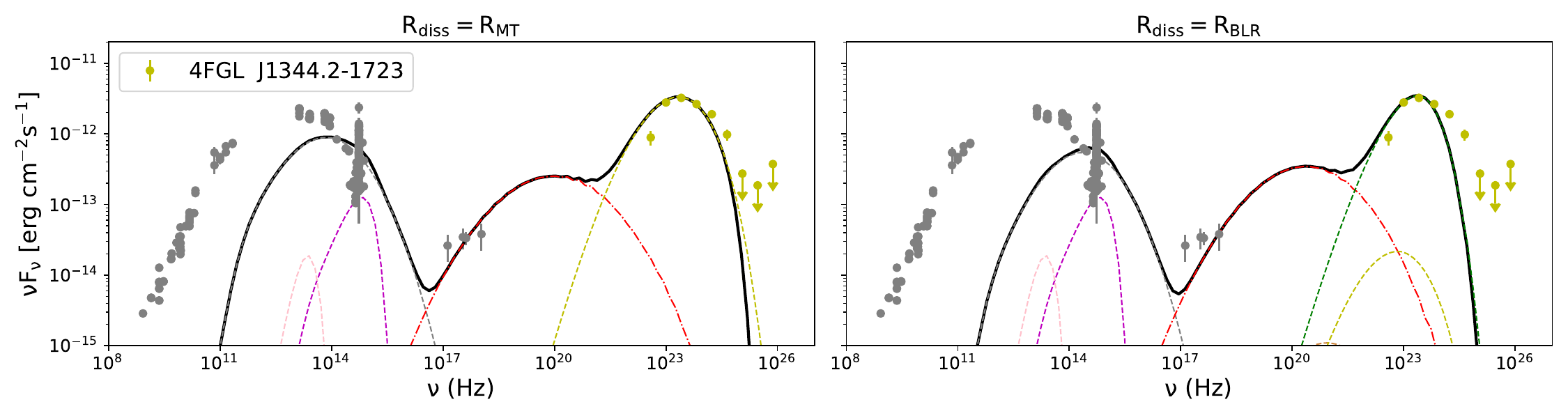}
\includegraphics[width=1\textwidth]{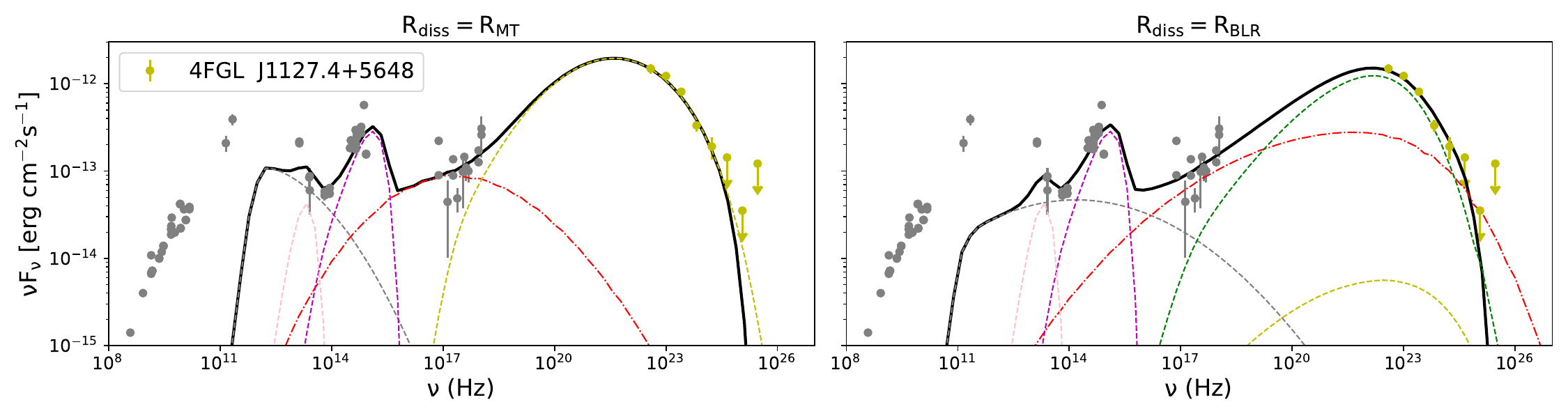}
\caption{The symbols and lines are the same as shown in Figure \ref{figg}. 
\label{fig7}}
\end{figure*}

\begin{figure*}[t]
\centering
\includegraphics[width=1\textwidth]{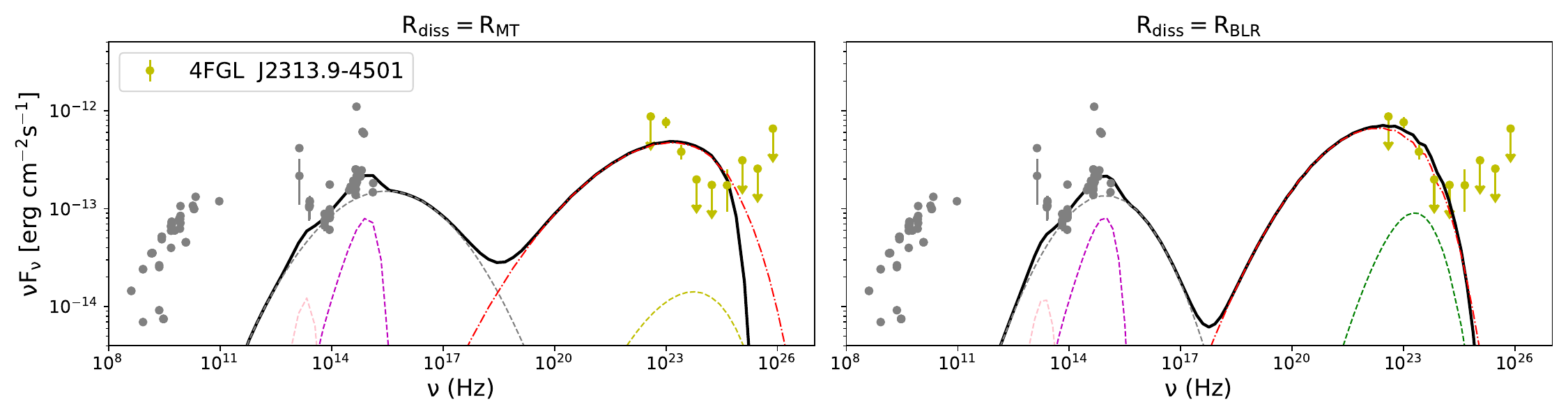}
\includegraphics[width=1\textwidth]{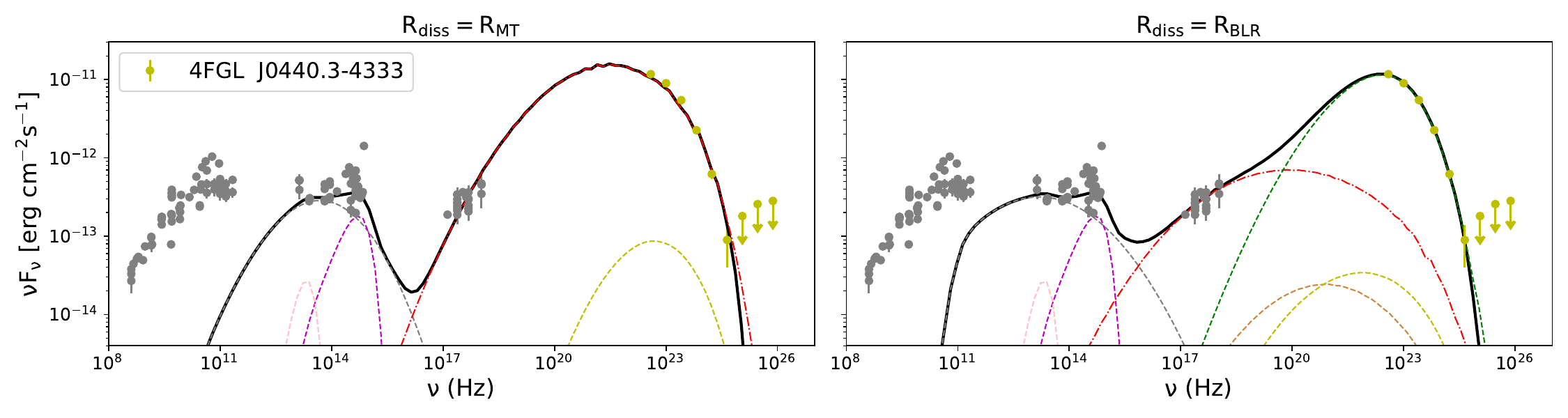}
\includegraphics[width=1\textwidth]{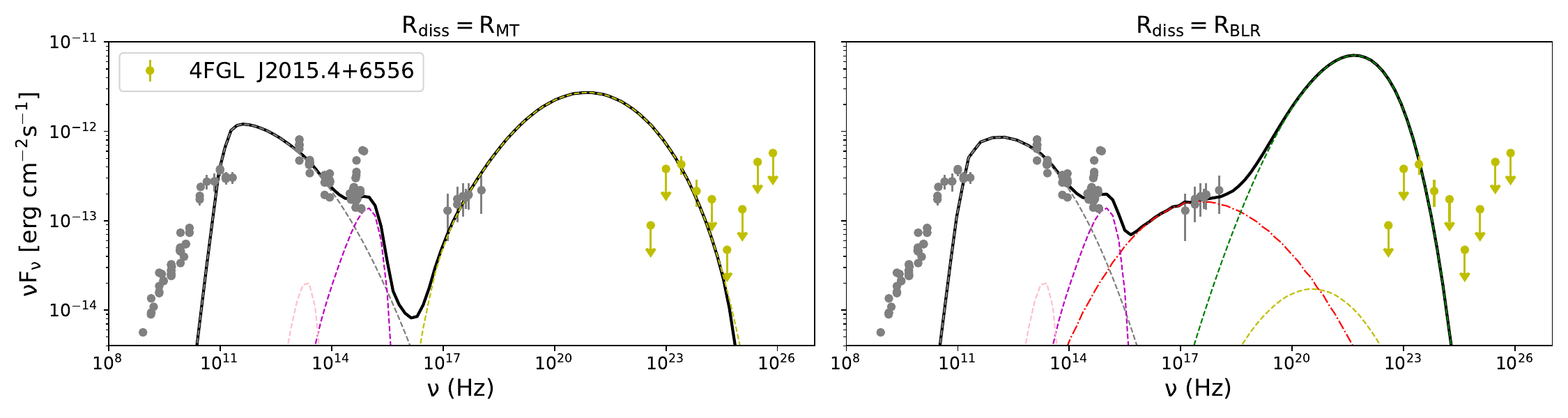}
\includegraphics[width=1\textwidth]{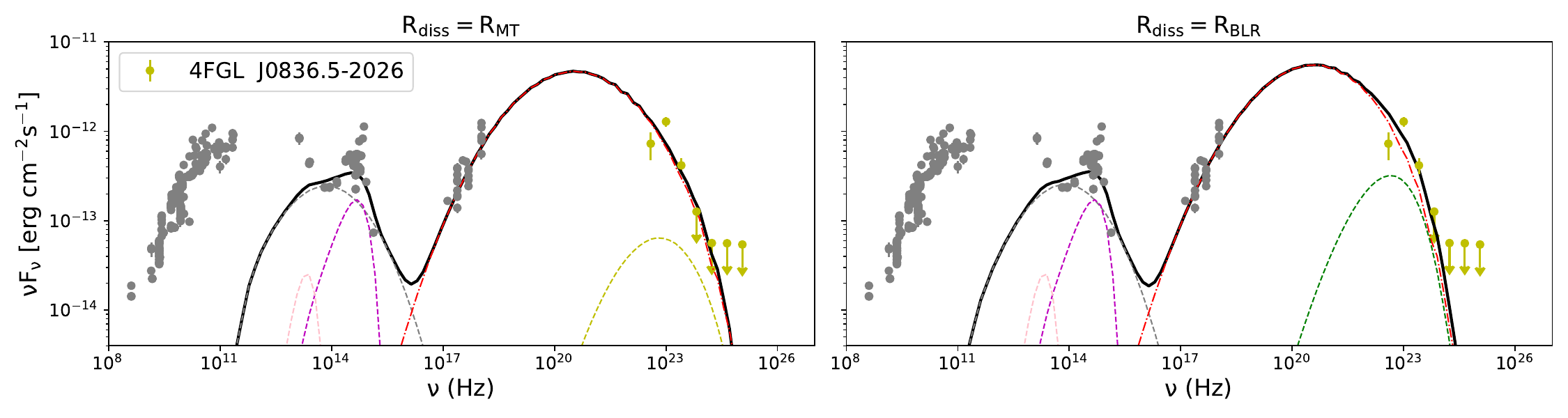}
\includegraphics[width=1\textwidth]{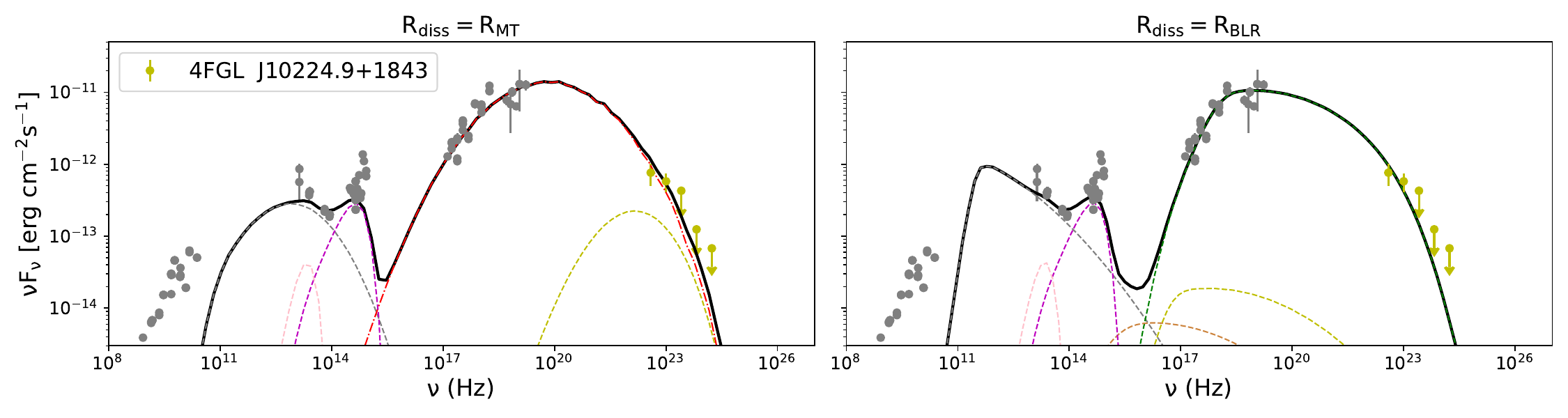}
\caption{The symbols and lines are the same as shown in Figure \ref{figg}. 
\label{fig8}}
\end{figure*}

\begin{figure*}[t]
\centering
\includegraphics[width=1\textwidth]{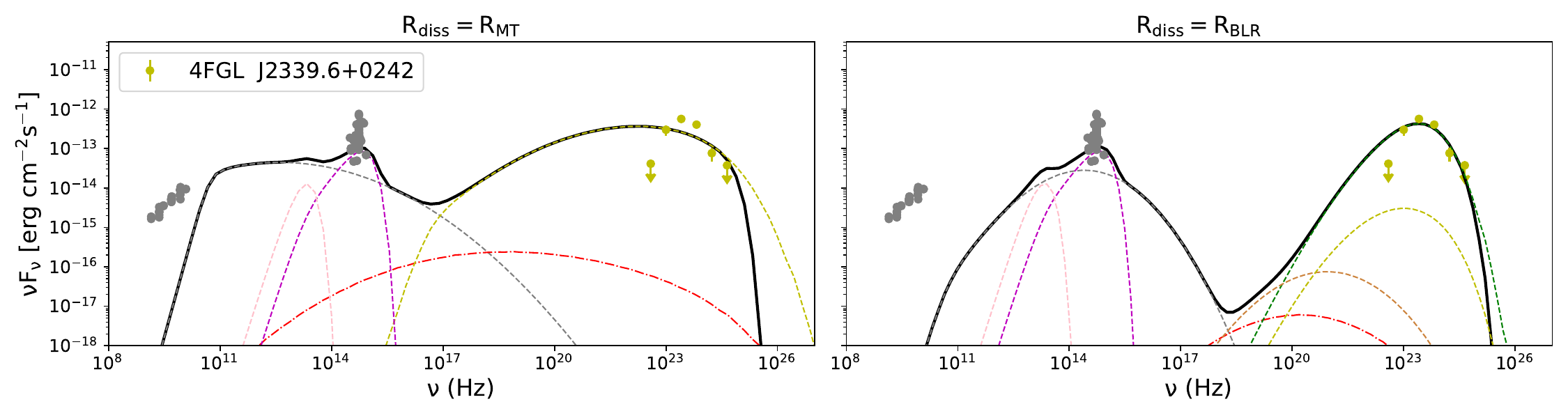}
\includegraphics[width=1\textwidth]{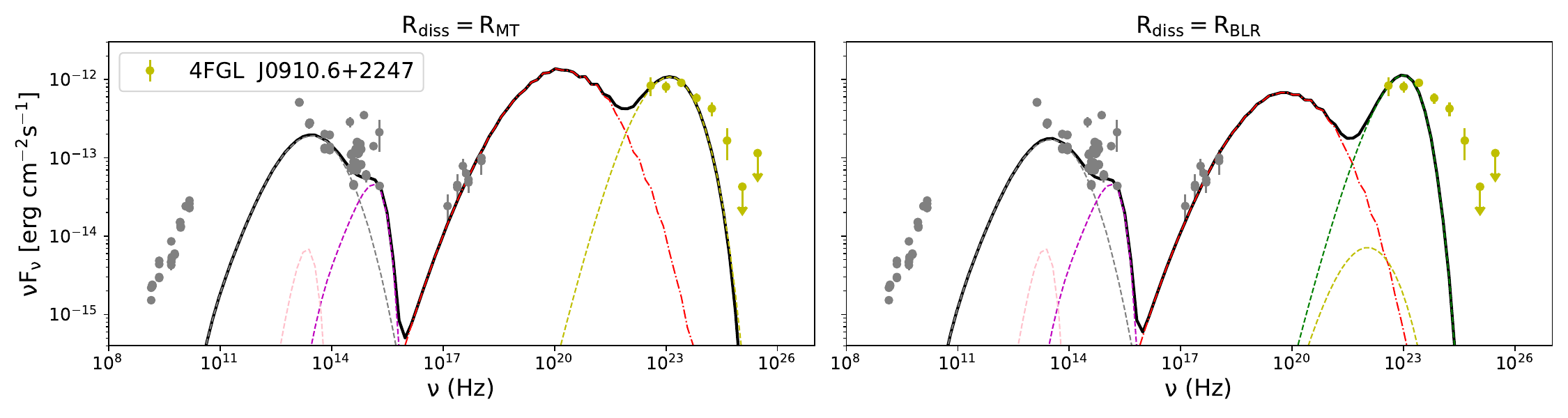}
\includegraphics[width=1\textwidth]{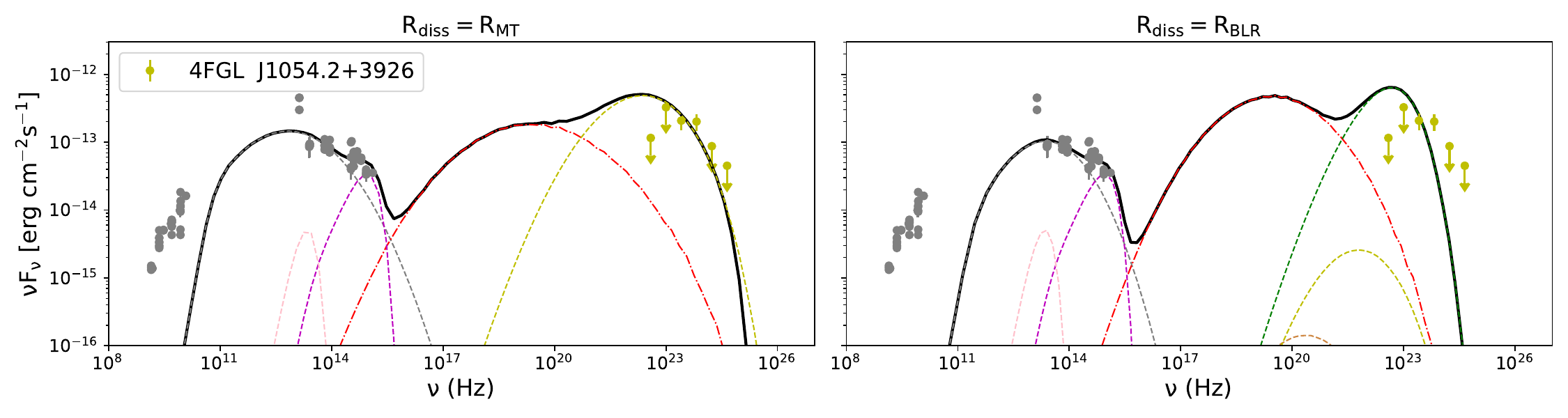}
\includegraphics[width=1\textwidth]{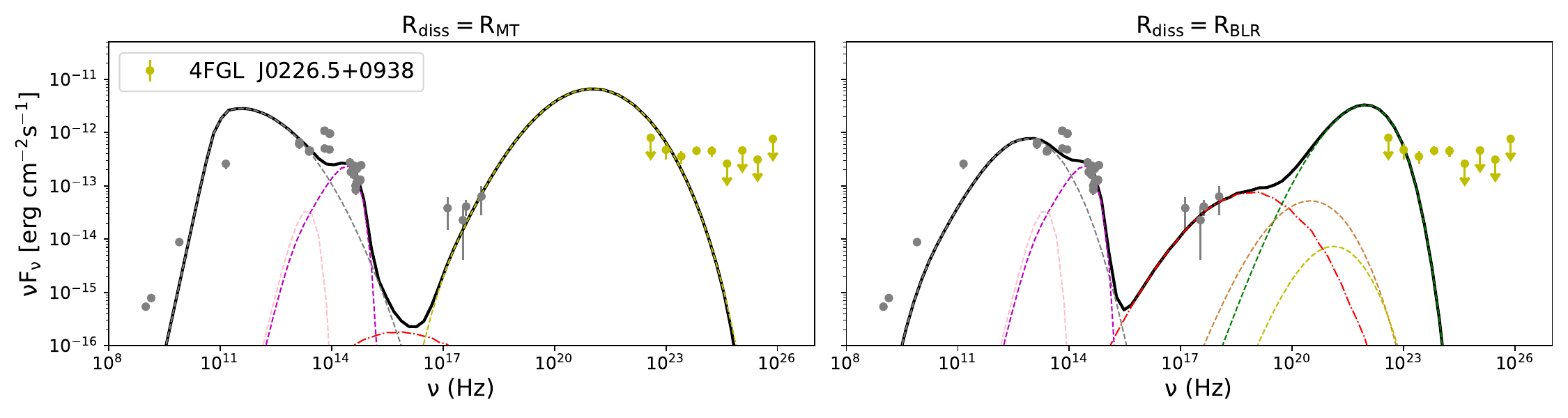}
\includegraphics[width=1\textwidth]{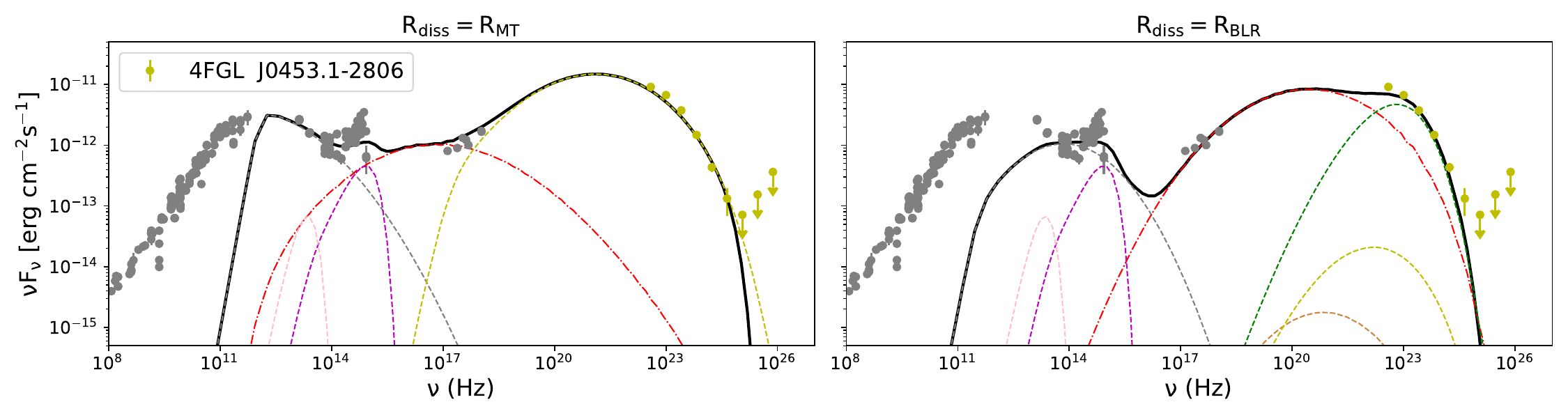}
\caption{The symbols and lines are the same as shown in Figure \ref{figg}. 
\label{fig9}}
\end{figure*}

\begin{figure*}[t]
\centering
\includegraphics[width=1\textwidth]{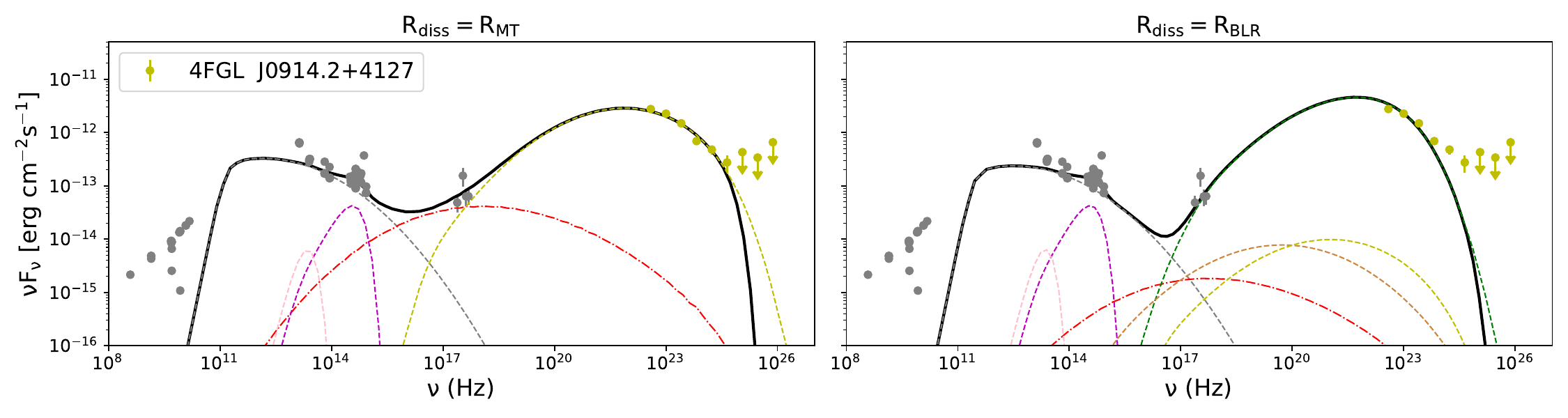}
\includegraphics[width=1\textwidth]{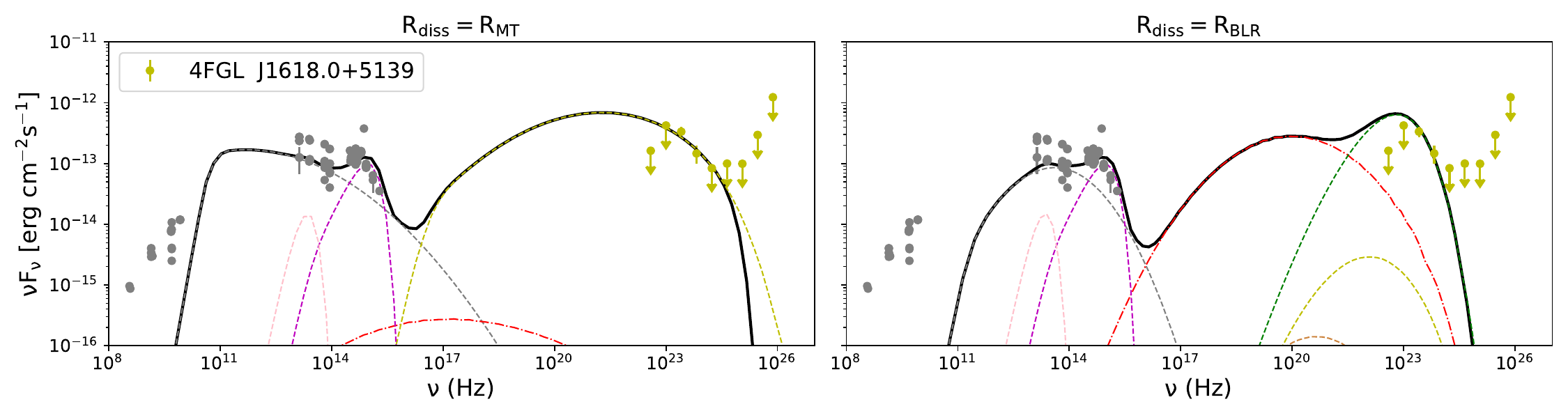}
\includegraphics[width=1\textwidth]{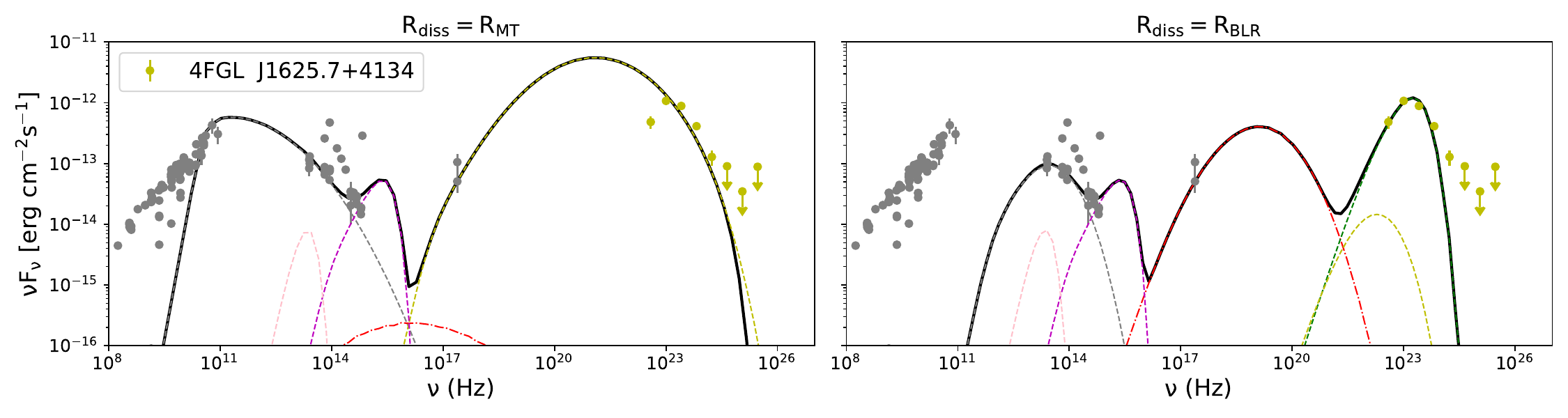}
\includegraphics[width=1\textwidth]{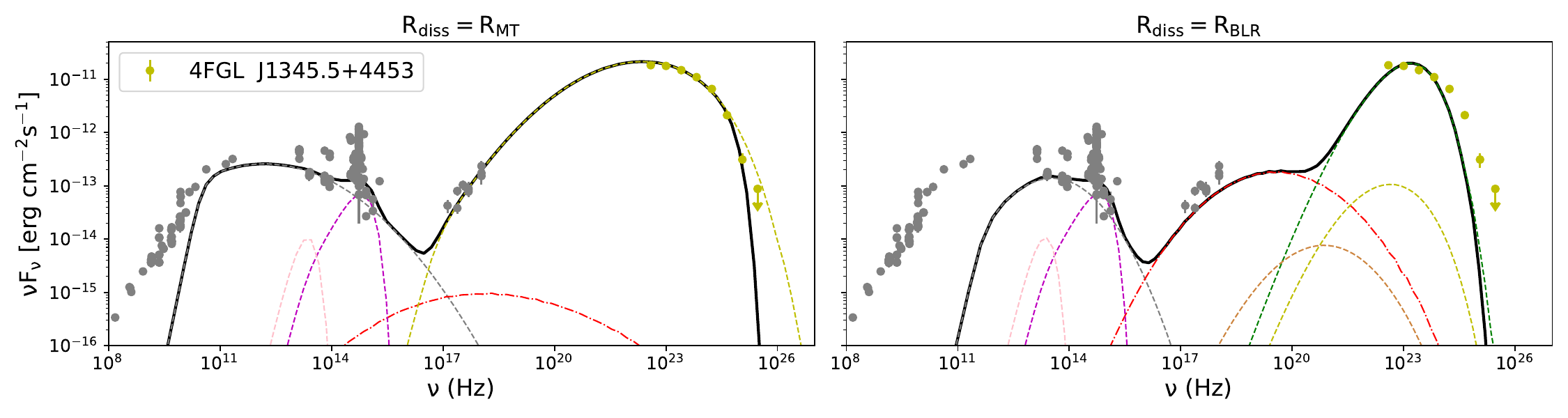}
\includegraphics[width=1\textwidth]{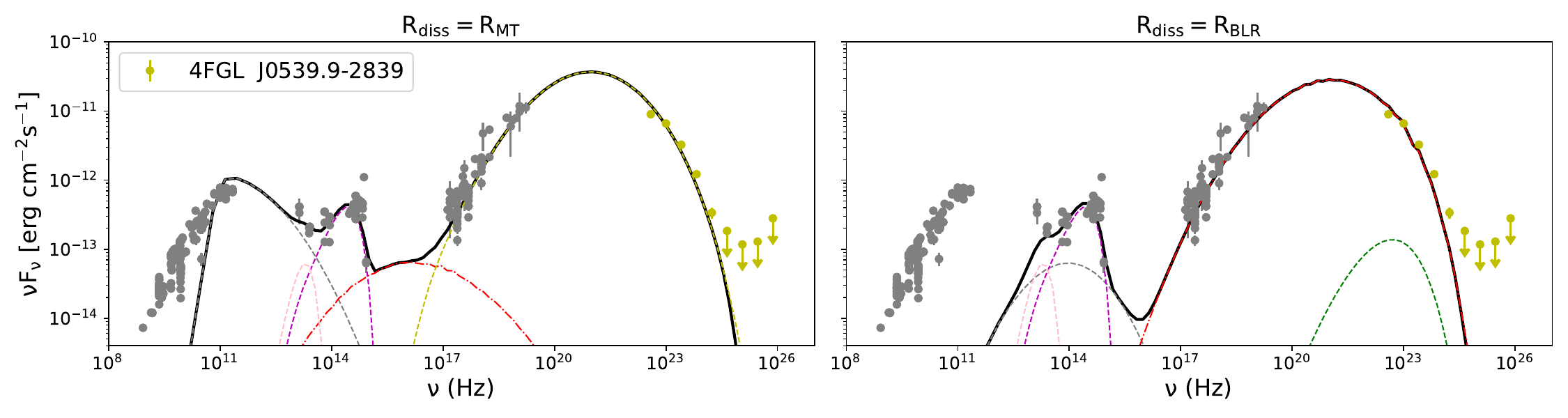}
\caption{The symbols and lines are the same as shown in Figure \ref{figg}. 
\label{fig10}}
\end{figure*}
\clearpage

\bibliography{sample631}{}

\begin{thebibliography}{}
\expandafter\ifx\csname natexlab\endcsname\relax\def\natexlab#1{#1}\fi
\providecommand{\url}[1]{\href{#1}{#1}}
\providecommand{\dodoi}[1]{doi:~\href{http://doi.org/#1}{\nolinkurl{#1}}}
\providecommand{\doeprint}[1]{\href{http://ascl.net/#1}{\nolinkurl{http://ascl.net/#1}}}
\providecommand{\doarXiv}[1]{\href{https://arxiv.org/abs/#1}{\nolinkurl{https://arxiv.org/abs/#1}}}

\bibitem[{{Abdollahi} {et~al.}(2020){Abdollahi}, {Acero}, {Ackermann},
  {Ajello}, {Atwood}, {Axelsson}, {Baldini}, {Ballet}, {Barbiellini},
  {Bastieri}, {Becerra Gonzalez}, {Bellazzini}, {Berretta}, {Bissaldi},
  {Blandford}, {Bloom}, {Bonino}, {Bottacini}, {Brandt}, {Bregeon}, {Bruel},
  {Buehler}, {Burnett}, {Buson}, {Cameron}, {Caputo}, {Caraveo}, {Casandjian},
  {Castro}, {Cavazzuti}, {Charles}, {Chaty}, {Chen}, {Cheung}, {Chiaro},
  {Ciprini}, {Cohen-Tanugi}, {Cominsky}, {Coronado-Bl{\'a}zquez}, {Costantin},
  {Cuoco}, {Cutini}, {D'Ammando}, {DeKlotz}, {de la Torre Luque}, {de Palma},
  {Desai}, {Digel}, {Di Lalla}, {Di Mauro}, {Di Venere}, {Dom{\'\i}nguez},
  {Dumora}, {Fana Dirirsa}, {Fegan}, {Ferrara}, {Franckowiak}, {Fukazawa},
  {Funk}, {Fusco}, {Gargano}, {Gasparrini}, {Giglietto}, {Giommi}, {Giordano},
  {Giroletti}, {Glanzman}, {Green}, {Grenier}, {Griffin}, {Grondin}, {Grove},
  {Guiriec}, {Harding}, {Hayashi}, {Hays}, {Hewitt}, {Horan},
  {J{\'o}hannesson}, {Johnson}, {Kamae}, {Kerr}, {Kocevski}, {Kovac'evic'},
  {Kuss}, {Landriu}, {Larsson}, {Latronico}, {Lemoine-Goumard}, {Li},
  {Liodakis}, {Longo}, {Loparco}, {Lott}, {Lovellette}, {Lubrano}, {Madejski},
  {Maldera}, {Malyshev}, {Manfreda}, {Marchesini}, {Marcotulli},
  {Mart{\'\i}-Devesa}, {Martin}, {Massaro}, {Mazziotta}, {McEnery}, {Mereu},
  {Meyer}, {Michelson}, {Mirabal}, {Mizuno}, {Monzani}, {Morselli},
  {Moskalenko}, {Negro}, {Nuss}, {Ojha}, {Omodei}, {Orienti}, {Orlando},
  {Ormes}, {Palatiello}, {Paliya}, {Paneque}, {Pei}, {Pe{\~n}a-Herazo},
  {Perkins}, {Persic}, {Pesce-Rollins}, {Petrosian}, {Petrov}, {Piron}, {Poon},
  {Porter}, {Principe}, {Rain{\`o}}, {Rando}, {Razzano}, {Razzaque}, {Reimer},
  {Reimer}, {Remy}, {Reposeur}, {Romani}, {Saz Parkinson}, {Schinzel},
  {Serini}, {Sgr{\`o}}, {Siskind}, {Smith}, {Spandre}, {Spinelli}, {Strong},
  {Suson}, {Tajima}, {Takahashi}, {Tak}, {Thayer}, {Thompson}, {Tibaldo},
  {Torres}, {Torresi}, {Valverde}, {Van Klaveren}, {van Zyl}, {Wood},
  {Yassine}, \& {Zaharijas}}]{2020ApJS..247...33A}
{Abdollahi}, S., {Acero}, F., {Ackermann}, M., {et~al.} 2020, \apjs, 247, 33,
  \dodoi{10.3847/1538-4365/ab6bcb}

\bibitem[{{Abdollahi} {et~al.}(2022){Abdollahi}, {Acero}, {Baldini}, {Ballet},
  {Bastieri}, {Bellazzini}, {Berenji}, {Berretta}, {Bissaldi}, {Blandford},
  {Bloom}, {Bonino}, {Brill}, {Britto}, {Bruel}, {Burnett}, {Buson}, {Cameron},
  {Caputo}, {Caraveo}, {Castro}, {Chaty}, {Cheung}, {Chiaro}, {Cibrario},
  {Ciprini}, {Coronado-Bl{\'a}zquez}, {Crnogorcevic}, {Cutini}, {D'Ammando},
  {De Gaetano}, {Digel}, {Di Lalla}, {Dirirsa}, {Di Venere}, {Dom{\'\i}nguez},
  {Fallah Ramazani}, {Fegan}, {Ferrara}, {Fiori}, {Fleischhack}, {Franckowiak},
  {Fukazawa}, {Funk}, {Fusco}, {Galanti}, {Gammaldi}, {Gargano}, {Garrappa},
  {Gasparrini}, {Giacchino}, {Giglietto}, {Giordano}, {Giroletti}, {Glanzman},
  {Green}, {Grenier}, {Grondin}, {Guillemot}, {Guiriec}, {Gustafsson},
  {Harding}, {Hays}, {Hewitt}, {Horan}, {Hou}, {J{\'o}hannesson}, {Karwin},
  {Kayanoki}, {Kerr}, {Kuss}, {Landriu}, {Larsson}, {Latronico},
  {Lemoine-Goumard}, {Li}, {Liodakis}, {Longo}, {Loparco}, {Lott}, {Lubrano},
  {Maldera}, {Malyshev}, {Manfreda}, {Mart{\'\i}-Devesa}, {Mazziotta}, {Mereu},
  {Meyer}, {Michelson}, {Mirabal}, {Mitthumsiri}, {Mizuno}, {Moiseev},
  {Monzani}, {Morselli}, {Moskalenko}, {Negro}, {Nuss}, {Omodei}, {Orienti},
  {Orlando}, {Paneque}, {Pei}, {Perkins}, {Persic}, {Pesce-Rollins},
  {Petrosian}, {Pillera}, {Poon}, {Porter}, {Principe}, {Rain{\`o}}, {Rando},
  {Rani}, {Razzano}, {Razzaque}, {Reimer}, {Reimer}, {Reposeur},
  {S{\'a}nchez-Conde}, {Saz Parkinson}, {Scotton}, {Serini}, {Sgr{\`o}},
  {Siskind}, {Smith}, {Spandre}, {Spinelli}, {Sueoka}, {Suson}, {Tajima},
  {Tak}, {Thayer}, {Thompson}, {Torres}, {Troja}, {Valverde}, {Wood}, \&
  {Zaharijas}}]{2022ApJS..260...53A}
{Abdollahi}, S., {Acero}, F., {Baldini}, L., {et~al.} 2022, \apjs, 260, 53,
  \dodoi{10.3847/1538-4365/ac6751}

\bibitem[{{Achterberg} {et~al.}(2001){Achterberg}, {Gallant}, {Kirk}, \&
  {Guthmann}}]{2001MNRAS.328..393A}
{Achterberg}, A., {Gallant}, Y.~A., {Kirk}, J.~G., \& {Guthmann}, A.~W. 2001,
  \mnras, 328, 393, \dodoi{10.1046/j.1365-8711.2001.04851.x}

\bibitem[{{Ackermann} {et~al.}(2012){Ackermann}, {Ajello}, {Allafort},
  {Schady}, {Baldini}, {Ballet}, {Barbiellini}, {Bastieri}, {Bellazzini},
  {Blandford}, {Bloom}, {Borgland}, {Bottacini}, {Bouvier}, {Bregeon},
  {Brigida}, {Bruel}, {Buehler}, {Buson}, {Caliandro}, {Cameron}, {Caraveo},
  {Cavazzuti}, {Cecchi}, {Charles}, {Chaves}, {Chekhtman}, {Cheung}, {Chiang},
  {Chiaro}, {Ciprini}, {Claus}, {Cohen-Tanugi}, {Conrad}, {Cutini},
  {D'Ammando}, {de Palma}, {Dermer}, {Digel}, {do Couto e Silva},
  {Dom{\'\i}nguez}, {Drell}, {Drlica-Wagner}, {Favuzzi}, {Fegan}, {Focke},
  {Franckowiak}, {Fukazawa}, {Funk}, {Fusco}, {Gargano}, {Gasparrini},
  {Gehrels}, {Germani}, {Giglietto}, {Giordano}, {Giroletti}, {Glanzman},
  {Godfrey}, {Grenier}, {Grove}, {Guiriec}, {Gustafsson}, {Hadasch},
  {Hayashida}, {Hays}, {Jackson}, {Jogler}, {Kataoka}, {Kn{\"o}dlseder},
  {Kuss}, {Lande}, {Larsson}, {Latronico}, {Longo}, {Loparco}, {Lovellette},
  {Lubrano}, {Mazziotta}, {McEnery}, {Mehault}, {Michelson}, {Mizuno}, {Monte},
  {Monzani}, {Morselli}, {Moskalenko}, {Murgia}, {Tramacere}, {Nuss},
  {Greiner}, {Ohno}, {Ohsugi}, {Omodei}, {Orienti}, {Orlando}, {Ormes},
  {Paneque}, {Perkins}, {Pesce-Rollins}, {Piron}, {Pivato}, {Porter},
  {Rain{\`o}}, {Rando}, {Razzano}, {Razzaque}, {Reimer}, {Reimer}, {Reyes},
  {Ritz}, {Rau}, {Romoli}, {Roth}, {S{\'a}nchez-Conde}, {Sanchez}, {Scargle},
  {Sgr{\`o}}, {Siskind}, {Spandre}, {Spinelli}, {Stawarz}, {Suson},
  {Takahashi}, {Tanaka}, {Thayer}, {Thompson}, {Tibaldo}, {Tinivella},
  {Torres}, {Tosti}, {Troja}, {Usher}, {Vandenbroucke}, {Vasileiou},
  {Vianello}, {Vitale}, {Waite}, {Winer}, {Wood}, \& {Wood}}]{Ackermann2012Sci}
{Ackermann}, M., {Ajello}, M., {Allafort}, A., {et~al.} 2012, Science, 338,
  1190, \dodoi{10.1126/science.1227160}

\bibitem[{{Atwood} {et~al.}(2009){Atwood}, {Abdo}, {Ackermann}, {Althouse},
  {Anderson}, {Axelsson}, {Baldini}, {Ballet}, {Band}, {Barbiellini},
  {Bartelt}, {Bastieri}, {Baughman}, {Bechtol}, {B{\'e}d{\'e}r{\`e}de},
  {Bellardi}, {Bellazzini}, {Berenji}, {Bignami}, {Bisello}, {Bissaldi},
  {Blandford}, {Bloom}, {Bogart}, {Bonamente}, {Bonnell}, {Borgland},
  {Bouvier}, {Bregeon}, {Brez}, {Brigida}, {Bruel}, {Burnett}, {Busetto},
  {Caliandro}, {Cameron}, {Caraveo}, {Carius}, {Carlson}, {Casandjian},
  {Cavazzuti}, {Ceccanti}, {Cecchi}, {Charles}, {Chekhtman}, {Cheung},
  {Chiang}, {Chipaux}, {Cillis}, {Ciprini}, {Claus}, {Cohen-Tanugi},
  {Condamoor}, {Conrad}, {Corbet}, {Corucci}, {Costamante}, {Cutini}, {Davis},
  {Decotigny}, {DeKlotz}, {Dermer}, {de Angelis}, {Digel}, {do Couto e Silva},
  {Drell}, {Dubois}, {Dumora}, {Edmonds}, {Fabiani}, {Farnier}, {Favuzzi},
  {Flath}, {Fleury}, {Focke}, {Funk}, {Fusco}, {Gargano}, {Gasparrini},
  {Gehrels}, {Gentit}, {Germani}, {Giebels}, {Giglietto}, {Giommi}, {Giordano},
  {Glanzman}, {Godfrey}, {Grenier}, {Grondin}, {Grove}, {Guillemot}, {Guiriec},
  {Haller}, {Harding}, {Hart}, {Hays}, {Healey}, {Hirayama}, {Hjalmarsdotter},
  {Horn}, {Hughes}, {J{\'o}hannesson}, {Johansson}, {Johnson}, {Johnson},
  {Johnson}, {Johnson}, {Kamae}, {Katagiri}, {Kataoka}, {Kavelaars}, {Kawai},
  {Kelly}, {Kerr}, {Klamra}, {Kn{\"o}dlseder}, {Kocian}, {Komin}, {Kuehn},
  {Kuss}, {Landriu}, {Latronico}, {Lee}, {Lee}, {Lemoine-Goumard}, {Lionetto},
  {Longo}, {Loparco}, {Lott}, {Lovellette}, {Lubrano}, {Madejski}, {Makeev},
  {Marangelli}, {Massai}, {Mazziotta}, {McEnery}, {Menon}, {Meurer},
  {Michelson}, {Minuti}, {Mirizzi}, {Mitthumsiri}, {Mizuno}, {Moiseev},
  {Monte}, {Monzani}, {Moretti}, {Morselli}, {Moskalenko}, {Murgia},
  {Nakamori}, {Nishino}, {Nolan}, {Norris}, {Nuss}, {Ohno}, {Ohsugi}, {Omodei},
  {Orlando}, {Ormes}, {Paccagnella}, {Paneque}, {Panetta}, {Parent}, {Pearce},
  {Pepe}, {Perazzo}, {Pesce-Rollins}, {Picozza}, {Pieri}, {Pinchera}, {Piron},
  {Porter}, {Poupard}, {Rain{\`o}}, {Rando}, {Rapposelli}, {Razzano}, {Reimer},
  {Reimer}, {Reposeur}, {Reyes}, {Ritz}, {Rochester}, {Rodriguez}, {Romani},
  {Roth}, {Russell}, {Ryde}, {Sabatini}, {Sadrozinski}, {Sanchez}, {Sander},
  {Sapozhnikov}, {Parkinson}, {Scargle}, {Schalk}, {Scolieri}, {Sgr{\`o}},
  {Share}, {Shaw}, {Shimokawabe}, {Shrader}, {Sierpowska-Bartosik}, {Siskind},
  {Smith}, {Smith}, {Spandre}, {Spinelli}, {Starck}, {Stephens}, {Strickman},
  {Strong}, {Suson}, {Tajima}, {Takahashi}, {Takahashi}, {Tanaka}, {Tenze},
  {Tether}, {Thayer}, {Thayer}, {Thompson}, {Tibaldo}, {Tibolla}, {Torres},
  {Tosti}, {Tramacere}, {Turri}, {Usher}, {Vilchez}, {Vitale}, {Wang},
  {Watters}, {Winer}, {Wood}, {Ylinen}, \& {Ziegler}}]{2009ApJ...697.1071A}
{Atwood}, W.~B., {Abdo}, A.~A., {Ackermann}, M., {et~al.} 2009, \apj, 697,
  1071, \dodoi{10.1088/0004-637X/697/2/1071}

\bibitem[{{B{\l}a{\.z}ejowski} {et~al.}(2000){B{\l}a{\.z}ejowski}, {Sikora},
  {Moderski}, \& {Madejski}}]{2000ApJ...545..107B}
{B{\l}a{\.z}ejowski}, M., {Sikora}, M., {Moderski}, R., \& {Madejski}, G.~M.
  2000, \apj, 545, 107, \dodoi{10.1086/317791}

\bibitem[{{Bloom} \& {Marscher}(1996)}]{1996ApJ...461..657B}
{Bloom}, S.~D., \& {Marscher}, A.~P. 1996, \apj, 461, 657,
  \dodoi{10.1086/177092}

\bibitem[{{Celotti} \& {Ghisellini}(2008)}]{2008MNRAS.385..283C}
{Celotti}, A., \& {Ghisellini}, G. 2008, \mnras, 385, 283,
  \dodoi{10.1111/j.1365-2966.2007.12758.x}

\bibitem[{{Chen} {et~al.}(2023){Chen}, {Gu}, {Fan}, {Yu}, {Zhong}, {Liu},
  {Ding}, {Xiong}, \& {Guo}}]{2023ApJS..268....6C}
{Chen}, Y., {Gu}, Q., {Fan}, J., {et~al.} 2023, \apjs, 268, 6,
  \dodoi{10.3847/1538-4365/ace444}

\bibitem[{{Chiang} {et~al.}(2010){Chiang}, {Done}, {Still}, \&
  {Godet}}]{2010MNRAS.403.1102C}
{Chiang}, C.~Y., {Done}, C., {Still}, M., \& {Godet}, O. 2010, \mnras, 403,
  1102, \dodoi{10.1111/j.1365-2966.2009.16129.x}

\bibitem[{{Costamante} {et~al.}(2018){Costamante}, {Cutini}, {Tosti},
  {Antolini}, \& {Tramacere}}]{2018MNRAS.477.4749C}
{Costamante}, L., {Cutini}, S., {Tosti}, G., {Antolini}, E., \& {Tramacere}, A.
  2018, \mnras, 477, 4749, \dodoi{10.1093/mnras/sty887}

\bibitem[{{Dermer} \& {Schlickeiser}(2002)}]{2002ApJ...575..667D}
{Dermer}, C.~D., \& {Schlickeiser}, R. 2002, \apj, 575, 667,
  \dodoi{10.1086/341431}

\bibitem[{{Desai} {et~al.}(2019){Desai}, {Helgason}, {Ajello}, {Paliya},
  {Dom{\'\i}nguez}, {Finke}, \& {Hartmann}}]{Desai2019ApJ}
{Desai}, A., {Helgason}, K., {Ajello}, M., {et~al.} 2019, \apjl, 874, L7,
  \dodoi{10.3847/2041-8213/ab0c10}

\bibitem[{{Fan} {et~al.}(2023){Fan}, {Xiao}, {Yang}, {Zhang}, {Strigachev},
  {Bachev}, \& {Yang}}]{2023ApJS..268...23F}
{Fan}, J., {Xiao}, H., {Yang}, W., {et~al.} 2023, \apjs, 268, 23,
  \dodoi{10.3847/1538-4365/ace7c8}

\bibitem[{{Fermi-LAT Collaboration} {et~al.}(2018){Fermi-LAT Collaboration},
  {Abdollahi}, {Ackermann}, {Ajello}, {Atwood}, {Baldini}, {Ballet},
  {Barbiellini}, {Bastieri}, {Becerra Gonzalez}, {Bellazzini}, {Bissaldi},
  {Blandford}, {Bloom}, {Bonino}, {Bottacini}, {Buson}, {Bregeon}, {Bruel},
  {Buehler}, {Cameron}, {Caputo}, {Caraveo}, {Cavazzuti}, {Charles}, {Chen},
  {Cheung}, {Chiaro}, {Ciprini}, {Cohen-Tanugi}, {Cominsky}, {Conrad},
  {Costantin}, {Cutini}, {D'Ammando}, {de Palma}, {Desai}, {Digel}, {Di Lalla},
  {Di Mauro}, {Di Venere}, {Dom{\'\i}nguez}, {Favuzzi}, {Fegan}, {Finke},
  {Franckowiak}, {Fukazawa}, {Funk}, {Fusco}, {Gallardo Romero}, {Gargano},
  {Gasparrini}, {Giglietto}, {Giordano}, {Giroletti}, {Green}, {Grenier},
  {Guillemot}, {Guiriec}, {Hartmann}, {Hays}, {Helgason}, {Horan},
  {J{\'o}hannesson}, {Kocevski}, {Kuss}, {Larsson}, {Latronico}, {Li}, {Longo},
  {Loparco}, {Lott}, {Lovellette}, {Lubrano}, {Madejski}, {Magill}, {Maldera},
  {Manfreda}, {Marcotulli}, {Mazziotta}, {McEnery}, {Meyer}, {Michelson},
  {Mizuno}, {Monzani}, {Morselli}, {Moskalenko}, {Negro}, {Nuss}, {Ojha},
  {Omodei}, {Orienti}, {Orlando}, {Ormes}, {Palatiello}, {Paliya}, {Paneque},
  {Perkins}, {Persic}, {Pesce-Rollins}, {Petrosian}, {Piron}, {Porter},
  {Primack}, {Principe}, {Rain{\`o}}, {Rando}, {Razzano}, {Razzaque}, {Reimer},
  {Reimer}, {Saz Parkinson}, {Sgr{\`o}}, {Siskind}, {Spandre}, {Spinelli},
  {Suson}, {Tajima}, {Takahashi}, {Thayer}, {Tibaldo}, {Torres}, {Torresi},
  {Tosti}, {Tramacere}, {Troja}, {Valverde}, {Vianello}, {Vogel}, {Wood}, \&
  {Zaharijas}}]{Fermi2018Sci}
{Fermi-LAT Collaboration}, {Abdollahi}, S., {Ackermann}, M., {et~al.} 2018,
  Science, 362, 1031, \dodoi{10.1126/science.aat8123}

\bibitem[{{Finke} {et~al.}(2022){Finke}, {Ajello}, {Dom{\'\i}nguez}, {Desai},
  {Hartmann}, {Paliya}, \& {Saldana-Lopez}}]{Finke2022ApJ}
{Finke}, J.~D., {Ajello}, M., {Dom{\'\i}nguez}, A., {et~al.} 2022, \apj, 941,
  33, \dodoi{10.3847/1538-4357/ac9843}

\bibitem[{{Finke} {et~al.}(2010){Finke}, {Razzaque}, \&
  {Dermer}}]{Finke2010ApJ}
{Finke}, J.~D., {Razzaque}, S., \& {Dermer}, C.~D. 2010, \apj, 712, 238,
  \dodoi{10.1088/0004-637X/712/1/238}

\bibitem[{{Foreman-Mackey} {et~al.}(2013){Foreman-Mackey}, {Hogg}, {Lang}, \&
  {Goodman}}]{2013PASP..125..306F}
{Foreman-Mackey}, D., {Hogg}, D.~W., {Lang}, D., \& {Goodman}, J. 2013, \pasp,
  125, 306, \dodoi{10.1086/670067}

\bibitem[{{Fossati} {et~al.}(1998){Fossati}, {Maraschi}, {Celotti}, {Comastri},
  \& {Ghisellini}}]{Fossati1998MNRAS}
{Fossati}, G., {Maraschi}, L., {Celotti}, A., {Comastri}, A., \& {Ghisellini},
  G. 1998, \mnras, 299, 433, \dodoi{10.1046/j.1365-8711.1998.01828.x}

\bibitem[{{Franceschini} {et~al.}(2008){Franceschini}, {Rodighiero}, \&
  {Vaccari}}]{2008A&A...487..837F}
{Franceschini}, A., {Rodighiero}, G., \& {Vaccari}, M. 2008, \aap, 487, 837,
  \dodoi{10.1051/0004-6361:200809691}

\bibitem[{{Ghisellini} {et~al.}(1989){Ghisellini}, {George}, \&
  {Done}}]{1989MNRAS.241P..43G}
{Ghisellini}, G., {George}, I.~M., \& {Done}, C. 1989, \mnras, 241, 43P,
  \dodoi{10.1093/mnras/241.1.43P}

\bibitem[{{Ghisellini} {et~al.}(2013){Ghisellini}, {Haardt}, {Della Ceca},
  {Volonteri}, \& {Sbarrato}}]{Ghisellini2013MNRAS}
{Ghisellini}, G., {Haardt}, F., {Della Ceca}, R., {Volonteri}, M., \&
  {Sbarrato}, T. 2013, \mnras, 432, 2818, \dodoi{10.1093/mnras/stt637}

\bibitem[{{Ghisellini} {et~al.}(1985){Ghisellini}, {Maraschi}, \&
  {Treves}}]{1985A&A...146..204G}
{Ghisellini}, G., {Maraschi}, L., \& {Treves}, A. 1985, \aap, 146, 204

\bibitem[{{Ghisellini} \& {Tavecchio}(2008)}]{2008MNRAS.387.1669G}
{Ghisellini}, G., \& {Tavecchio}, F. 2008, \mnras, 387, 1669,
  \dodoi{10.1111/j.1365-2966.2008.13360.x}

\bibitem[{{Ghisellini} \& {Tavecchio}(2009)}]{2009MNRAS.397..985G}
---. 2009, \mnras, 397, 985, \dodoi{10.1111/j.1365-2966.2009.15007.x}

\bibitem[{{Ghisellini} {et~al.}(2010{\natexlab{a}}){Ghisellini}, {Tavecchio},
  {Foschini}, {Ghirlanda}, {Maraschi}, \& {Celotti}}]{2010MNRAS.402..497G}
{Ghisellini}, G., {Tavecchio}, F., {Foschini}, L., {et~al.} 2010{\natexlab{a}},
  \mnras, 402, 497, \dodoi{10.1111/j.1365-2966.2009.15898.x}

\bibitem[{{Ghisellini} {et~al.}(2014){Ghisellini}, {Tavecchio}, {Maraschi},
  {Celotti}, \& {Sbarrato}}]{2014Natur.515..376G}
{Ghisellini}, G., {Tavecchio}, F., {Maraschi}, L., {Celotti}, A., \&
  {Sbarrato}, T. 2014, \nat, 515, 376, \dodoi{10.1038/nature13856}

\bibitem[{{Ghisellini} {et~al.}(2010{\natexlab{b}}){Ghisellini}, {Della Ceca},
  {Volonteri}, {Ghirlanda}, {Tavecchio}, {Foschini}, {Tagliaferri}, {Haardt},
  {Pareschi}, \& {Grindlay}}]{2010MNRAS.405..387G}
{Ghisellini}, G., {Della Ceca}, R., {Volonteri}, M., {et~al.}
  2010{\natexlab{b}}, \mnras, 405, 387,
  \dodoi{10.1111/j.1365-2966.2010.16449.x}

\bibitem[{{Ghisellini} {et~al.}(2011){Ghisellini}, {Tagliaferri}, {Foschini},
  {Ghirlanda}, {Tavecchio}, {Della Ceca}, {Haardt}, {Volonteri}, \&
  {Gehrels}}]{2011MNRAS.411..901G}
{Ghisellini}, G., {Tagliaferri}, G., {Foschini}, L., {et~al.} 2011, \mnras,
  411, 901, \dodoi{10.1111/j.1365-2966.2010.17723.x}

\bibitem[{{Iyida} {et~al.}(2020){Iyida}, {Odo}, {Chukwude}, \&
  {Ubachukwu}}]{2020OAst...29..168I}
{Iyida}, E.~U., {Odo}, F.~C., {Chukwude}, A.~E., \& {Ubachukwu}, A.~A. 2020,
  Open Astronomy, 29, 168, \dodoi{10.1515/astro-2020-0015}

\bibitem[{{Jones}(1968)}]{1968PhRv..167.1159J}
{Jones}, F.~C. 1968, Physical Review, 167, 1159,
  \dodoi{10.1103/PhysRev.167.1159}

\bibitem[{{Keshet} \& {Waxman}(2005)}]{2005PhRvL..94k1102K}
{Keshet}, U., \& {Waxman}, E. 2005, \prl, 94, 111102,
  \dodoi{10.1103/PhysRevLett.94.111102}

\bibitem[{{Landau} {et~al.}(1986){Landau}, {Golisch}, {Jones}, {Jones},
  {Pedelty}, {Rudnick}, {Sitko}, {Kenney}, {Roellig}, {Salonen}, {Urpo},
  {Schmidt}, {Neugebauer}, {Matthews}, {Elias}, {Impey}, {Clegg}, \&
  {Harris}}]{1986ApJ...308...78L}
{Landau}, R., {Golisch}, B., {Jones}, T.~J., {et~al.} 1986, \apj, 308, 78,
  \dodoi{10.1086/164480}

\bibitem[{{Luna-Cervantes} {et~al.}(2024){Luna-Cervantes}, {Tramacere}, \&
  {Ben{\'\i}tez}}]{2024MNRAS.532.3729L}
{Luna-Cervantes}, J., {Tramacere}, A., \& {Ben{\'\i}tez}, E. 2024, \mnras, 532,
  3729, \dodoi{10.1093/mnras/stae1687}

\bibitem[{{Maraschi} \& {Tavecchio}(2003)}]{2003ApJ...593..667M}
{Maraschi}, L., \& {Tavecchio}, F. 2003, \apj, 593, 667, \dodoi{10.1086/342118}

\bibitem[{{Marcotulli} {et~al.}(2020){Marcotulli}, {Paliya}, {Ajello}, {Kaur},
  {Marchesi}, {Rajagopal}, {Hartmann}, {Gasparrini}, {Ojha}, \&
  {Madejski}}]{Marcotulli2020ApJ}
{Marcotulli}, L., {Paliya}, V., {Ajello}, M., {et~al.} 2020, \apj, 889, 164,
  \dodoi{10.3847/1538-4357/ab65f5}

\bibitem[{{Massaro} {et~al.}(2006){Massaro}, {Tramacere}, {Perri}, {Giommi}, \&
  {Tosti}}]{2006A&A...448..861M}
{Massaro}, E., {Tramacere}, A., {Perri}, M., {Giommi}, P., \& {Tosti}, G. 2006,
  \aap, 448, 861, \dodoi{10.1051/0004-6361:20053644}

\bibitem[{{Narayan} \& {Raymond}(1999)}]{1999ApJ...515L..69N}
{Narayan}, R., \& {Raymond}, J. 1999, \apjl, 515, L69, \dodoi{10.1086/311973}

\bibitem[{{Narayan} {et~al.}(2017){Narayan}, {Sa{\`I}{\textsection}dowski}, \&
  {Soria}}]{2017MNRAS.469.2997N}
{Narayan}, R., {Sa{\`I}{\textsection}dowski}, A., \& {Soria}, R. 2017, \mnras,
  469, 2997, \dodoi{10.1093/mnras/stx1027}

\bibitem[{{Nolan} {et~al.}(2012){Nolan}, {Abdo}, {Ackermann}, {Ajello},
  {Allafort}, {Antolini}, {Atwood}, {Axelsson}, {Baldini}, {Ballet},
  {Barbiellini}, {Bastieri}, {Bechtol}, {Belfiore}, {Bellazzini}, {Berenji},
  {Bignami}, {Blandford}, {Bloom}, {Bonamente}, {Bonnell}, {Borgland},
  {Bottacini}, {Bouvier}, {Brandt}, {Bregeon}, {Brigida}, {Bruel}, {Buehler},
  {Burnett}, {Buson}, {Caliandro}, {Cameron}, {Campana}, {Ca{\~n}adas},
  {Cannon}, {Caraveo}, {Casandjian}, {Cavazzuti}, {Ceccanti}, {Cecchi},
  {{\c{C}}elik}, {Charles}, {Chekhtman}, {Cheung}, {Chiang}, {Chipaux},
  {Ciprini}, {Claus}, {Cohen-Tanugi}, {Cominsky}, {Conrad}, {Corbet}, {Cutini},
  {D'Ammando}, {Davis}, {de Angelis}, {DeCesar}, {DeKlotz}, {De Luca}, {den
  Hartog}, {de Palma}, {Dermer}, {Digel}, {Silva}, {Drell}, {Drlica-Wagner},
  {Dubois}, {Dumora}, {Enoto}, {Escande}, {Fabiani}, {Falletti}, {Favuzzi},
  {Fegan}, {Ferrara}, {Focke}, {Fortin}, {Frailis}, {Fukazawa}, {Funk},
  {Fusco}, {Gargano}, {Gasparrini}, {Gehrels}, {Germani}, {Giebels},
  {Giglietto}, {Giommi}, {Giordano}, {Giroletti}, {Glanzman}, {Godfrey},
  {Grenier}, {Grondin}, {Grove}, {Guillemot}, {Guiriec}, {Gustafsson},
  {Hadasch}, {Hanabata}, {Harding}, {Hayashida}, {Hays}, {Hill}, {Horan},
  {Hou}, {Hughes}, {Iafrate}, {Itoh}, {J{\'o}hannesson}, {Johnson}, {Johnson},
  {Johnson}, {Johnson}, {Kamae}, {Katagiri}, {Kataoka}, {Katsuta}, {Kawai},
  {Kerr}, {Kn{\"o}dlseder}, {Kocevski}, {Kuss}, {Lande}, {Landriu},
  {Latronico}, {Lemoine-Goumard}, {Lionetto}, {Llena Garde}, {Longo},
  {Loparco}, {Lott}, {Lovellette}, {Lubrano}, {Madejski}, {Marelli}, {Massaro},
  {Mazziotta}, {McConville}, {McEnery}, {Mehault}, {Michelson}, {Minuti},
  {Mitthumsiri}, {Mizuno}, {Moiseev}, {Mongelli}, {Monte}, {Monzani},
  {Morselli}, {Moskalenko}, {Murgia}, {Nakamori}, {Naumann-Godo}, {Norris},
  {Nuss}, {Nymark}, {Ohno}, {Ohsugi}, {Okumura}, {Omodei}, {Orlando}, {Ormes},
  {Ozaki}, {Paneque}, {Panetta}, {Parent}, {Perkins}, {Pesce-Rollins},
  {Pierbattista}, {Pinchera}, {Piron}, {Pivato}, {Porter}, {Racusin},
  {Rain{\`o}}, {Rando}, {Razzano}, {Razzaque}, {Reimer}, {Reimer}, {Reposeur},
  {Ritz}, {Rochester}, {Romani}, {Roth}, {Rousseau}, {Ryde}, {Sadrozinski},
  {Salvetti}, {Sanchez}, {Saz Parkinson}, {Sbarra}, {Scargle}, {Schalk},
  {Sgr{\`o}}, {Shaw}, {Shrader}, {Siskind}, {Smith}, {Spandre}, {Spinelli},
  {Stephens}, {Strickman}, {Suson}, {Tajima}, {Takahashi}, {Takahashi},
  {Tanaka}, {Thayer}, {Thayer}, {Thompson}, {Tibaldo}, {Tibolla}, {Tinebra},
  {Tinivella}, {Torres}, {Tosti}, {Troja}, {Uchiyama}, {Vandenbroucke}, {Van
  Etten}, {Van Klaveren}, {Vasileiou}, {Vianello}, {Vitale}, {Waite},
  {Wallace}, {Wang}, {Werner}, {Winer}, {Wood}, {Wood}, {Wood}, {Yang}, \&
  {Zimmer}}]{2012ApJS..199...31N}
{Nolan}, P.~L., {Abdo}, A.~A., {Ackermann}, M., {et~al.} 2012, \apjs, 199, 31,
  \dodoi{10.1088/0067-0049/199/2/31}

\bibitem[{{Ostrowski} \& {Bednarz}(2002)}]{2002A&A...394.1141O}
{Ostrowski}, M., \& {Bednarz}, J. 2002, \aap, 394, 1141,
  \dodoi{10.1051/0004-6361:20021173}

\bibitem[{{Paliya} {et~al.}(2020){Paliya}, {Ajello}, {Cao}, {Giroletti},
  {Kaur}, {Madejski}, {Lott}, \& {Hartmann}}]{Paliya2020ApJ}
{Paliya}, V.~S., {Ajello}, M., {Cao}, H.~M., {et~al.} 2020, \apj, 897, 177,
  \dodoi{10.3847/1538-4357/ab9c1a}

\bibitem[{{Paliya} {et~al.}(2021){Paliya}, {Dom{\'\i}nguez}, {Ajello},
  {Olmo-Garc{\'\i}a}, \& {Hartmann}}]{2021ApJS..253...46P}
{Paliya}, V.~S., {Dom{\'\i}nguez}, A., {Ajello}, M., {Olmo-Garc{\'\i}a}, A., \&
  {Hartmann}, D. 2021, \apjs, 253, 46, \dodoi{10.3847/1538-4365/abe135}

\bibitem[{{Paliya} {et~al.}(2017){Paliya}, {Marcotulli}, {Ajello}, {Joshi},
  {Sahayanathan}, {Rao}, \& {Hartmann}}]{2017ApJ...851...33P}
{Paliya}, V.~S., {Marcotulli}, L., {Ajello}, M., {et~al.} 2017, \apj, 851, 33,
  \dodoi{10.3847/1538-4357/aa98e1}

\bibitem[{{Potter} \& {Cotter}(2013)}]{2013MNRAS.431.1840P}
{Potter}, W.~J., \& {Cotter}, G. 2013, \mnras, 431, 1840,
  \dodoi{10.1093/mnras/stt300}

\bibitem[{{Prandini} \& {Ghisellini}(2022)}]{Prandini2022Galax}
{Prandini}, E., \& {Ghisellini}, G. 2022, Galaxies, 10, 35,
  \dodoi{10.3390/galaxies10010035}

\bibitem[{{Rajguru} \& {Chatterjee}(2022)}]{2022PhRvD.106f3001R}
{Rajguru}, G., \& {Chatterjee}, R. 2022, \prd, 106, 063001,
  \dodoi{10.1103/PhysRevD.106.063001}

\bibitem[{{Rawlings} \& {Saunders}(1991)}]{1991Natur.349..138R}
{Rawlings}, S., \& {Saunders}, R. 1991, \nat, 349, 138,
  \dodoi{10.1038/349138a0}

\bibitem[{{Rees} {et~al.}(1982){Rees}, {Begelman}, {Blandford}, \&
  {Phinney}}]{1982Natur.295...17R}
{Rees}, M.~J., {Begelman}, M.~C., {Blandford}, R.~D., \& {Phinney}, E.~S. 1982,
  \nat, 295, 17, \dodoi{10.1038/295017a0}

\bibitem[{{Sahakyan}(2020)}]{2020MNRAS.496.5518S}
{Sahakyan}, N. 2020, \mnras, 496, 5518, \dodoi{10.1093/mnras/staa1893}

\bibitem[{{Sahakyan} {et~al.}(2020){Sahakyan}, {Israyelyan}, {Harutyunyan},
  {Khachatryan}, \& {Gasparyan}}]{2020MNRAS.498.2594S}
{Sahakyan}, N., {Israyelyan}, D., {Harutyunyan}, G., {Khachatryan}, M., \&
  {Gasparyan}, S. 2020, \mnras, 498, 2594, \dodoi{10.1093/mnras/staa2477}

\bibitem[{{Shakura} \& {Sunyaev}(1973)}]{1973A&A....24..337S}
{Shakura}, N.~I., \& {Sunyaev}, R.~A. 1973, \aap, 24, 337

\bibitem[{{Sharma} {et~al.}(2024){Sharma}, {Kamaram}, {Prince}, {Khatoon}, \&
  {Bose}}]{2024MNRAS.527.2672S}
{Sharma}, A., {Kamaram}, S.~R., {Prince}, R., {Khatoon}, R., \& {Bose}, D.
  2024, \mnras, 527, 2672, \dodoi{10.1093/mnras/stad3399}

\bibitem[{{Sikora} {et~al.}(1994){Sikora}, {Begelman}, \&
  {Rees}}]{1994ApJ...421..153S}
{Sikora}, M., {Begelman}, M.~C., \& {Rees}, M.~J. 1994, \apj, 421, 153,
  \dodoi{10.1086/173633}

\bibitem[{{Sikora} {et~al.}(2009){Sikora}, {Stawarz}, {Moderski}, {Nalewajko},
  \& {Madejski}}]{2009ApJ...704...38S}
{Sikora}, M., {Stawarz}, {\L}., {Moderski}, R., {Nalewajko}, K., \& {Madejski},
  G.~M. 2009, \apj, 704, 38, \dodoi{10.1088/0004-637X/704/1/38}

\bibitem[{{Stecker} {et~al.}(2007){Stecker}, {Baring}, \&
  {Summerlin}}]{2007ApJ...667L..29S}
{Stecker}, F.~W., {Baring}, M.~G., \& {Summerlin}, E.~J. 2007, \apjl, 667, L29,
  \dodoi{10.1086/522005}

\bibitem[{{Stickel} {et~al.}(1991){Stickel}, {Padovani}, {Urry}, {Fried}, \&
  {Kuehr}}]{1991ApJ...374..431S}
{Stickel}, M., {Padovani}, P., {Urry}, C.~M., {Fried}, J.~W., \& {Kuehr}, H.
  1991, \apj, 374, 431, \dodoi{10.1086/170133}

\bibitem[{{Sulentic} {et~al.}(2000){Sulentic}, {Marziani}, \&
  {Dultzin-Hacyan}}]{2000ARA&A..38..521S}
{Sulentic}, J.~W., {Marziani}, P., \& {Dultzin-Hacyan}, D. 2000, \araa, 38,
  521, \dodoi{10.1146/annurev.astro.38.1.521}

\bibitem[{{Taam} {et~al.}(2012){Taam}, {Liu}, {Yuan}, \&
  {Qiao}}]{2012ApJ...759...65T}
{Taam}, R.~E., {Liu}, B.~F., {Yuan}, W., \& {Qiao}, E. 2012, \apj, 759, 65,
  \dodoi{10.1088/0004-637X/759/1/65}

\bibitem[{{Thorne}(1974)}]{1974ApJ...191..507T}
{Thorne}, K.~S. 1974, \apj, 191, 507, \dodoi{10.1086/152991}

\bibitem[{{Tramacere} {et~al.}(2009){Tramacere}, {Giommi}, {Perri},
  {Verrecchia}, \& {Tosti}}]{2009A&A...501..879T}
{Tramacere}, A., {Giommi}, P., {Perri}, M., {Verrecchia}, F., \& {Tosti}, G.
  2009, \aap, 501, 879, \dodoi{10.1051/0004-6361/200810865}

\bibitem[{{Tramacere} {et~al.}(2011){Tramacere}, {Massaro}, \&
  {Taylor}}]{2011ApJ...739...66T}
{Tramacere}, A., {Massaro}, E., \& {Taylor}, A.~M. 2011, \apj, 739, 66,
  \dodoi{10.1088/0004-637X/739/2/66}

\bibitem[{{Urry} \& {Padovani}(1995)}]{1995PASP..107..803U}
{Urry}, C.~M., \& {Padovani}, P. 1995, \pasp, 107, 803, \dodoi{10.1086/133630}

\bibitem[{{Virtanen} \& {Vainio}(2005)}]{2005ApJ...621..313V}
{Virtanen}, J. J.~P., \& {Vainio}, R. 2005, \apj, 621, 313,
  \dodoi{10.1086/427324}

\bibitem[{{Volonteri} {et~al.}(2011){Volonteri}, {Haardt}, {Ghisellini}, \&
  {Della Ceca}}]{2011MNRAS.416..216V}
{Volonteri}, M., {Haardt}, F., {Ghisellini}, G., \& {Della Ceca}, R. 2011,
  \mnras, 416, 216, \dodoi{10.1111/j.1365-2966.2011.19024.x}

\bibitem[{{Wielgus} {et~al.}(2022){Wielgus}, {Lan{\v{c}}ov{\'a}}, {Straub},
  {Klu{\'z}niak}, {Narayan}, {Abarca}, {R{\'o}{\.z}a{\'n}ska}, {Vincent},
  {T{\"o}r{\"o}k}, \& {Abramowicz}}]{2022MNRAS.514..780W}
{Wielgus}, M., {Lan{\v{c}}ov{\'a}}, D., {Straub}, O., {et~al.} 2022, \mnras,
  514, 780, \dodoi{10.1093/mnras/stac1317}

\bibitem[{{Wu} {et~al.}(2024){Wu}, {Hu}, \& {Dai}}]{2024ApJ...972..183W}
{Wu}, F., {Hu}, W., \& {Dai}, B. 2024, \apj, 972, 183,
  \dodoi{10.3847/1538-4357/ad5f8a}

\bibitem[{{Xiao} {et~al.}(2022){Xiao}, {Ouyang}, {Zhang}, {Fu}, {Zhang},
  {Zeng}, \& {Fan}}]{2022ApJ...925...40X}
{Xiao}, H., {Ouyang}, Z., {Zhang}, L., {et~al.} 2022, \apj, 925, 40,
  \dodoi{10.3847/1538-4357/ac36da}

\bibitem[{{Xie} {et~al.}(2024){Xie}, {Ouyang}, {Wu}, {Xiao}, {Zhang}, {Chen},
  {Luo}, \& {Fan}}]{2024ApJ...976...78X}
{Xie}, S., {Ouyang}, Z., {Wu}, J., {et~al.} 2024, \apj, 976, 78,
  \dodoi{10.3847/1538-4357/ad8353}

\bibitem[{{Xiong} \& {Zhang}(2014)}]{2014MNRAS.441.3375X}
{Xiong}, D.~R., \& {Zhang}, X. 2014, \mnras, 441, 3375,
  \dodoi{10.1093/mnras/stu755}

\bibitem[{{Yuan} \& {Narayan}(2004)}]{2004ApJ...612..724Y}
{Yuan}, F., \& {Narayan}, R. 2004, \apj, 612, 724, \dodoi{10.1086/422802}

\bibitem[{{Yuan} \& {Narayan}(2014)}]{2014ARA&A..52..529Y}
---. 2014, \araa, 52, 529, \dodoi{10.1146/annurev-astro-082812-141003}

\bibitem[{{Zhang} {et~al.}(2024){Zhang}, {Chen}, {He}, {Nie}, {Tang}, {Huang},
  {Chen}, \& {Fan}}]{2024ApJS..271...27Z}
{Zhang}, L., {Chen}, X., {He}, S., {et~al.} 2024, \apjs, 271, 27,
  \dodoi{10.3847/1538-4365/ad20c8}

\bibitem[{{Zhu} {et~al.}(2024){Zhu}, {Chen}, \& {Zhang}}]{2024ApJS..275...41Z}
{Zhu}, K.~R., {Chen}, J.~M., \& {Zhang}, L. 2024, \apjs, 275, 41,
  \dodoi{10.3847/1538-4365/ad8639}

\end{thebibliography}
\bibliographystyle{aasjournal}
\end{document}